\documentclass[10pt,epsfig]{article} 
\usepackage{enumerate}
\usepackage{float}
\usepackage{subfig}
\usepackage{color}
\usepackage{slashed}
\usepackage{amssymb, amsmath,mathrsfs}
\usepackage{graphicx}
\usepackage{multirow}
\usepackage{cite}
\usepackage{chngcntr}
\usepackage[colorlinks=true,
            linkcolor=red,
            urlcolor=blue,
            citecolor=blue]{hyperref}
\usepackage{dcolumn}
\usepackage{bm}
\usepackage{color}
\long\def\rpl#1!!#2!!{\textcolor{red}{#1} \textcolor{blue}{#2}}
\def\baselinestretch{1.3}
\newcommand{\eq}[1]{\begin{equation}\begin{split}#1\end{split}\end{equation}}

\newcommand{\beq}{\begin {equation}}  
\newcommand{\eeq}{\end   {equation}} 
\newcommand{\bea}{\begin {eqnarray}} 
\newcommand{\eea}{\end   {eqnarray}}  
\newcommand{\baa}{\begin {array}   } 
\newcommand{\eaa}{\end   {array}   }     
\newcommand{\bit}{\begin {itemize} }
\newcommand{\eit}{\end   {itemize} }
\newcommand{\be }{\begin {equation}} 
\newcommand{\ee }{\end   {equation}}
\newcommand{\nn }{\nonumber        }

\newcommand{\tbox}[1]{\mbox{\tiny #1}}

\newcommand{\eff}{\textrm{eff}}
\newcommand{\tree}{\textrm{tree}}
\newcommand{\Tr}{\textbf{Tr}}
\newcommand{\STr}{\textbf{STr}}

\newcommand\matTwo[2]{\ensuremath{\begin{pmatrix} #1  \\ #2\end{pmatrix}}}
\newcommand\matTwoT[2]{\ensuremath{\begin{pmatrix} #1 & #2 \end{pmatrix}}}
\newcommand\matFour[4]{\ensuremath{\begin{pmatrix} #1 & #2 \\ #3 & #4\end{pmatrix}}}

\allowdisplaybreaks
\begin{document}




%

\begin{flushright}
{UTTG-15-15 \\
ACFI-T15-16 }
\end{flushright}

\begin{center}

{\Large \textbf {Electroweak Baryogenesis in a Scalar-Assisted Vector-Like Fermion Model}}\\[10mm]

Ming-Lei Xiao$^{*}$\footnote{mingleix@utexas.edu}, Jiang-Hao Yu$^{*,\dagger}$\footnote{jhyu@physics.umass.edu}  \\
$^{*}${\em Theory Group, Department of Physics and Texas Cosmology Center,
\\The University of Texas at Austin,  Austin, TX 78712 U.S.A.}\\
$^{\dagger}${\em Amherst Center for Fundamental Interactions, Physics Department, \\
University of Massachusetts Amherst, Amherst, MA 01003, USA}\\[5mm] 

\end{center}


\begin{abstract} 

We extend the standard model to a scalar-assisted vector-like fermion model to realize electroweak baryogenesis. 
The extended Cabbibo-Kobayashi-Maskawa matrix, due to the mixing among the vector-like quark and the standard model quarks,
provides additional sources of the CP violation. 
Together with the enhancement from large vector-like quark mass, a large enough baryon-to-photon ratio could be obtained.
The strongly first-order phase transition could be realized via the potential barrier which separate the broken minimum and the symmetric minimum in the scalar potential.
We investigate in detail the one-loop temperature-dependent effective potential, 
and perform a random parameter scan to study the allowed parameter region that satisfies the strongly first order phase transition criteria $v_c \ge T_c$.
Several distinct patterns of phase transition are classified and discussed.
Among these patterns, large trilinear mass term between the Higgs boson and the scalar is preferred, for it controls the width of the potential barrier.
Our results indicate large quartic scalar couplings, and moderate mixing angle between
the Higgs boson and the new scalar. 
This parameter region could be further explored at the Run 2 LHC. 

%
%

%
%
\end{abstract}

\newpage
\setcounter{footnote}{0}

\def\baselinestretch{1.5}
\counterwithin{equation}{section}


\section{Introduction}
\label{sec:intro}

The baryonic matter that remains after the baryon-antibaryon annihilation,
makes up around 5\% of the total energy density of the universe. 
It is puzzling that the universe does not have equal amounts of matter and antimatter. 
We can characterize the asymmetry between matter and antimatter in terms of the baryon-to-photon ratio
\bea
	\eta \equiv \frac{n_B}{n_\gamma},
\eea
where $n_B = n_{b} - n_{\bar{b}}$ is the difference between the number density of baryons and antibaryons, and 
$n_\gamma$ is the number density of photon. 
The $n_\gamma$ is introduced to prevent the parameters $\eta$ from diluting during the expansion of the universe after nucleosynthesis.
The baryonic matter desity $n_B$ at present time has been consistently measured by the big bang nucleosynthesis 
and the fluctuations of the cosmic microwave background. 
The Planck result for the cosmological density parameter~\cite{Ade:2015xua} 
\bea
\Omega_B h^2 = 0.02226 \pm 0.00016,
\eea 
translates to the baryon-to-photon ratio
\bea
	\eta = (6.05 \pm 0.07) \times 10^{-10}.
\eea

Explaining the observed baryon asymmetry has been one of the greatest challenges of particle physics and cosmology. 
As the entropy production during inflation could greatly dilute and thus wash out any existing baryon asymmetry, it is reasonable to assume a zero baryon number density after the inflation. Later, the asymmetry is generated dynamically through the so-called "baryogenesis".
It has been suggested by Sakharov~\cite{Sakharov:1967dj} long time ago that 
the general baryogenesis has three necessary conditions: 
baryon number violation, sufficient $C$ and $CP$ violation, and departure from 
thermal equilibrium.
Hence, we look forward to a mechanism in which these three conditions are satisfied and could provide the observed baryon asymmetry.

Electroweak baryogenesis~\cite{Kuzmin:1985mm, Shaposhnikov:1986jp, Shaposhnikov:1987tw} (EWBG) offers a theoretically attractive and experimentally testable mechanism  
to realize baryogenesis. 
The great attraction of this mechanism is that the baryogenesis took place at or near the electroweak scale, suggesting that it might be probed in the near future by experiments at the accelerators.
The EWBG proceeds as follows (see \cite{Cohen:1993nk, Trodden:1998ym, Quiros:1999jp, Bernreuther:2002uj, Cline:2006ts, Morrissey:2012db} for reviews).
At temperatures far above the electroweak scale, the electroweak symmetry is manifest, which implied a high sphaleron rate that preserves baryon symmetry in thermal equilibrium.
As the universe cools down to near the electroweak phase transition scale, 
bubbles of the symmetry-broken vacuum began to emerge and grow. 
CP violating processes involving the electroweak sector were triggered at the expanding wall of the bubbles, leaving baryons inside the bubbles and antibaryons outside.
%
Through the rapid sphaleron transitions in the unbroken phase, the excess of antibaryons are washed out. Meanwhile, if the sphaleron rate in the broken phase could be suppressed enough, the excess of baryons inside the bubbles could survive. 
We can easily realize a Boltzmann suppression of the sphaleron rate, because the sphaleron has an excitation energy $E_{sph}$ that is related to the Higgs vacuum expectation value (vev) $v$. 
It has been shown~\cite{Bochkarev:1990gb} that the suppression is strong enough when
\bea
E_{sph}/T_c \ge 45,
\eea
which serves as the condition for a \emph{strong first-order phase transition} (SFOPT) in the context of electroweak baryogenesis.

The standard model (SM) contains all the necessary ingredients to realize electroweak baryogenesis: baryon number is violated by sphaleron processes; CP violation comes from the Cabbibo-Kobayashi-Maskawa (CKM) matrix; departure from equilibrium is realized by the bubble nucleation and expansion during the first-order electroweak phase transition (EWPT).
However, given the observed Higgs boson mass $M_H = 125 $ GeV, the EWPT is not strong enough to suppress the sphaleron rate inside the bubbles~\cite{Bochkarev:1987wf}. 
Also, the CP violation in the CKM matrix is not large enough to generate the expected asymmetry.
Therefore a successful electroweak baryogenesis needs new physics beyond the Standard Model.
The new physics should provide new sources of CP violation that can be manifested by the advancing bubble walls, and also provide strong enough first order EWPT. Both conditions require the existence of new physics at around the electroweak scale that directly couples to the SM Higgs sector. 
A simple and economic way to realize the strong first order EWPT is to add a new scalar which couples to the Higgs boson, such as the singlet extended standard model, etc~\cite{Choi:1993cv, Cline:1996mga, Ham:2004cf, Ahriche:2007jp, Profumo:2007wc, Espinosa:2011ax, Li:2014wia, Fuyuto:2014yia, Profumo:2014opa}.
Moreover, if the scalar is a real singlet~\cite{Profumo:2007wc, Profumo:2014opa}, the cubic terms could exist in the potential at tree-level, and therefore the phase transition gets stronger without the need of the thermally induced barrier. 

We consider the electroweak baryogenesis in a scalar-assisted vectorlike fermion model~\cite{Xiao:2014kba}, in which a singlet scalar and vectorlike fermions are added to the SM particle content. 
Originally, the model is motivated by the possible instability of the vacuum structure in the vector-like fermion model. 
The singlet scalar is added to the scalar sector and couples to the vector-like fermion. 
This model solves the vacuum stability problem in vectorlike fermion model and possible perturbativity issues in singlet scalar extended standard model.
Recently this model attracts lots of attentions because it could naturally explain the diphoton excess 
observed at both ATLAS and CMS~\cite{ATLAS13, CMS13}. 
The diphoton signature of this model and its extensions has been considered in Ref.~\cite{Franceschini:2015kwy}.
Due to constraints from other $WW$ and $ZZ$ channels, a 750 GeV scalar singlet could accommodate the observed diphoton excess more readily than the $SU(2)_L \times U(1)_Y$ scalar multiplets.
The diphoton signature is produced via the gluon fusion and subsequent diphoton decay with vector-like fermion running in the loop.

In this work, we consider that this model realize the ectroweak baryogenesis. 
The vector-like fermion mixes with the SM quarks, extending the $3 \times 3$ CKM matrix to a $3 \times 4$ matrix, which provides additional sources of CP violation. 
Due to the coupling between Higgs and the new scalar, the phase transition happens in an extended scalar space, which leads to more possibilities on phase transition. 
We will discuss the scalar potential in detail, perform numerial calculations, and investigate how the extended scalar sector provides us the SFOPT.  
Furthermore, we classify the phase transition patterns and explore the parameter preferences in each pattern using the shape of the derivatives of the scalar potential. 
Finally, we explore the discovery potential of the parameter space favored by SFOPT at the LHC.

The organization of presentation is as follows. We begin with the description of the new physics model. In Sec. 3, we discuss the CP violation in this model. In Sec. 4, we present the effective potential and the shape of the scalar potential. In Sec. 5, we discuss the phase transition pattern and explain our numerical results. In sec. 6, the LHC discovery potential of the favored parameter region is discussed. We then make our conclusion.


\section{Scalar-assisted Vectorlike Fermion Model}
\label{sec:model}

In our setup, we consider an extension of the SM in which a vector-like fermion $U$ and a real singlet scalar $s$ are added to the particle content~\cite{Xiao:2014kba}.
The vector-like fermion $U$ transforms as $(3, 1)_{\frac23}$ under the SM gauge symmetry $SU(3)_C \times SU(2)_L \times U(1)_Y$. 
Due to the same quantum number, its right-handed component mixes with the SM right-handed up-type quarks. 
%
%
As is known~\cite{Xiao:2014kba}, the vector-like fermion model encounters the vacuum instability problem. 
To have a stabilized scalar potential, 
a singlet scalar $s$ is introduced. 
The scalar mixes with the SM Higgs boson and couples to the vector-like fermion. 
Here, we assume no $Z_2$ symmetry for the new scalar, so that it has a non-zero vev in general. 
We refer this model as the scalar-assisted vector-like fermion model. 
Let us discuss the quark sector and the scalar sector in detail.

In the quark sector, a SM family contains a doublet $Q_L = (u_L, d_L)^T \sim (3, 2)_{1/6}$ and two singlets $u_R \sim (3, 1)_{2/3}$ and $d_R \sim (3, 1)_{-1/3}$ that couple to each other via the Higgs doublet $\Phi \sim (1, 2)_{1/2}$ .
Due to the flavor experiment constraints, we only allow the new vector-like fermion to have significant mixing with the third generation quarks, so it's also called a top partner.
Vector-like top partner is well motivated in the little Higgs~\cite{ArkaniHamed:2002qy}, composite Higgs~\cite{Kaplan:1983sm}, and extra dimension models~\cite{Randall:1999ee}, etc.
Let us first write down the Lagrangian for one flavor mixing between the new fermion and the third generation quarks, and then extend to the three flavor mixing.
We could write down the following new Yukawa couplings:
\bea
	{\mathcal L}_{\rm Yuk} = y_t \overline{Q}_{L3} \tilde{\Phi} \,u_{R3}  - y'  \overline{Q}_{L3} \tilde{\Phi} \,U_{R} - y_s s \overline{U}_L U_R - M \overline{U}_L U_R + {\rm h.c.},
	\label{eq:yuktop}
\eea
where ${Q}_{L3}$ and $u_{R3}$ are the left-handed quark doublet and the right-handed up-type quark in the third generation. 
The vacuum expectation values of the two scalars are denoted as
\bea
	v \equiv \langle \phi \rangle, \quad u \equiv \langle s \rangle.
\eea
The mass term becomes
\bea
	{\mathcal L}_{\rm mass} = -   \matTwoT{\overline{u}_{L3}}{\overline{U}_L}  
	\matFour{v y_t}{v y' }{0}{M + y_s u} 
	\matTwo{u_{R}}{U_R}  + {\rm h.c.}.
\eea
To get the mass eigenstates $(t, T)$, we diagonalize the fermion mass matrix
\bea
	\matFour{v y_t}{v y' }{0}{M + y_s u} 
	= \matFour{\cos\theta_{L}}{\sin\theta_{L}}{-\sin\theta_{L}}{\cos\theta_{L}}
		\matFour{m_t}{0}{0}{m_T}
	 	\matFour{\cos\theta_{R}}{-\sin\theta_{R}}{\sin\theta_{R}}{\cos\theta_{R}}.
\eea
Note that the two mixing angles are not independent parameters, 
\bea
	\tan\theta_R = \frac{m_t}{m_T} \tan \theta_L.
\eea

Despite the tight constraints on the flavor mixing between the new vector-like fermion and the first two generations,
these mixings are still essential for the new CP violation.
If we consider the three families of the quarks in the SM, the Yukawa couplings $y_t$ and $y'$ in Eq.~\ref{eq:yuktop} becomes matrix $Y^{u}_{ij}$ and vector $Y'_i$ in the flavor space. 
With explicit flavor indices, the Yukawa Lagrangian becomes
\bea
	{\mathcal L}_{\rm Yuk} &=& - Y^d_{ij} \overline{Q}_{Li} \Phi \,d_{Rj} - Y^u_{ij} \overline{Q}_{Li} \tilde{\Phi} \,u_{Rj} \nn \\ 
	&-& Y'_i  \overline{Q}_{Li} \tilde{\Phi} \,U_{R} - y_s s \overline{U}_L U_R - M \overline{U}_L U_R + {\rm h.c.}
\eea
The mass term of the fermion sector is
\bea
	{\mathcal L}_{\rm mass} = -   \matTwoT{\overline{u}_{Li}}{\overline{U}_L}  
	\matFour{v Y^u_{ij}}{v Y'_i}{0}{M + y_s u} 
	\matTwo{u_{R}^j}{U_R} 
	- v Y^d_{ij} \overline{d}_{Li}  d_{R}^j + {\rm h.c.}.
\eea
Hereafter we identify
\bea
	\mathcal{M}^{u}_{IJ} = \matFour{v Y^u_{ij}}{v Y'_i}{0}{M + y_s u}, \quad
	\mathcal{M}^{d}_{ij} = v  Y^d_{ij},
\eea
where $I$ and $J$ run over $1$ to $4$. 
Using bidiagonalisation, the mass matrix transforms as
\bea
	{\mathcal U}^{\dagger}_L  \mathcal{M}^{u}{\mathcal U}_R &=& {\rm Diag} (m_u, m_c, m_t, m_T),\\
	{\mathcal D}^{\dagger}_L  \mathcal{M}^{d}{\mathcal D}_R &=& {\rm Diag} (m_d, m_s, m_b),
\eea
through rotations of the quark flavor basis
\bea
	u_L^I = \mathcal{U}^{I}_{\ J}u_L^J , \quad d_L^i = \mathcal{D}^{i}_{\ j}d_L^j,
\eea
where $\mathcal{U}^{I}_{ J}$ and $\mathcal{D}^{i}_{ j}$ are $4\times 4$ unitary matrix, and $3\times 3$ unitary matrix, respectively.

The new CKM matrix $V'_{\rm CKM}$ is obtained from the charged current. In weak eigenstates
\bea
	{\mathcal L}_{\rm CC} =  \frac{g_2}{\sqrt{2}} \overline{u}_{Li} \gamma^{\mu} d_L^i W_\mu + {\rm h.c.}.
\eea
Rotating into the mass eigenstates, we get
\bea
	{\mathcal L}_{\rm CC} =  \frac{g_2}{\sqrt{2}} \overline{u}_{LI} \gamma^{\mu} \left({\mathcal U}^{\dagger}_L \right)^I_{\ i} \left({\mathcal D}_L \right)^i_{\ j} d_L^j W_\mu + {\rm h.c.}.
\eea
$V'_{\rm CKM}$ is defined as a $4\times 3$ matrix
\bea
	V'_{\textrm{CKM}} = \left({\mathcal U}^{\dagger}_L \right)^I_{\ i} \left({\mathcal D}_L \right)^i_{\ j} \equiv \mathcal{U}^{\dagger}E\mathcal{D},
\eea
where the explicit form of the new CKM matrix are expressed as
\bea
	E = \begin{pmatrix}
		1	&	0	&	0	\\
		0	&	1	&	0	\\
		0	&	0	&	1	\\
		0	&	0	&	0
	\end{pmatrix},
	\quad
	V'_{\rm CKM} = \begin{pmatrix}
		V_{ud}	&	V_{us}	&	V_{ub}	\\
		V_{cd}	&	V_{cs}	&	V_{cb}	\\
		V_{td}	&	V_{ts}	&	V_{tb}	\\
		V_{Td}	&	V_{Ts}	&	V_{Tb}
	\end{pmatrix}.
\eea

The unitarity of the transformation matrices $\mathcal{U}$ and $\mathcal{D}$ implies
\begin{equation}
	\label{uniCKM}
	V_{\textrm{CKM}}^{\prime\dagger}V'_{\textrm{CKM}} = \mathcal{D}^{\dagger}E^{\dagger}\mathcal{U}\mathcal{U}^{\dagger}E\mathcal{D} = \mathbf{1}_{3\times3},
\end{equation}
which means that the 3 columns of $V'_{\rm CKM}$ are orthonormal to each other. For future use, we can complement one column to make up a unitary $4 \times 4$ matrix expressed as
\eq{
\bar{V}_{\rm CKM} = \begin{pmatrix}
		V_{ud}	&	V_{us}	&	V_{ub}	&	V_{u4}	\\
		V_{cd}	&	V_{cs}	&	V_{cb}	&	V_{c4}	\\
		V_{td}	&	V_{ts}	&	V_{tb}	&	V_{t4}	\\
		V_{Td}	&	V_{Ts}	&	V_{Tb}	&	V_{T4}
	\end{pmatrix}.
}

In the scalar sector, the new scalar couples to the SM Higgs boson. 
The general scalar potential is
\bea
		V_{\rm tree} & = & -\frac12 \mu_{\phi}^2 \phi^2 + \frac14\lambda_{\phi} \phi^4 - \frac12\mu_s^2 s^2 
		+ \mu_1^3 s + \frac13 \mu_3 s^3 + \frac14\lambda_{s} s^4 + \frac12 \mu_{s\phi} \phi^2 s + \frac14\lambda_{s\phi}\phi^2 s^2.\nn\\  
\eea
The parameter $\mu_1$ can be eliminated by a redefinition of the scalar field $s \to s + \sigma$. 
The minimization conditions at the vacuum $(v,u)$ are used to eliminate the quadratic coefficients
\bea
 	\mu_{\phi}^2 &=& \lambda_\phi v^2 + \mu_{s\phi} u + \frac12\lambda_{s\phi}u^2 , \\
    \mu_s^2 &=& \lambda_s u^2 + \mu_3 u + \frac12\lambda_{s\phi}v^2 + \frac{\mu_{s\phi} v^2}{2 u},
\eea
The second derivatives of the tree-level potential 
describe the mass squared matrix of $\phi$ and $s$:
\bea
\label{MMs_tree}
	{\cal M}^2_{\rm scalar} 
		\equiv \left( \begin{array}{cc}
		m^2_{\phi\phi}	&	m^2_{\phi s}	\\
		m^2_{s\phi }	&	m^2_{s s}
	\end{array} \right) 
	= \left( \begin{array}{cc}
		2\lambda_\phi v^2	&	\mu_{s\phi} v + \lambda_{s\phi}uv	\\
		\mu_{s\phi} v + \lambda_{s\phi}uv	&	2\lambda_s u^2 + \mu_3 u - \frac{\mu_{s\phi}v^2}{2u}
	\end{array}\right).  
\eea
Diagonalizing the above matrix, we obtain the mass squared eigenvalues
\bea
	m^2_{h,S} &=&  \frac{1}{2} \left(m^2_{\phi\phi} + m^2_{ss}\right) \mp 
	\frac12\sqrt{  \left(m^2_{\phi\phi} - m^2_{ss}\right)^2 +4 m^4_{s\phi } },
		  \label{eq:scalarmass}
\eea
and the eigenvectors
\bea
	\left( \begin{array}{c} h \\ S \end{array}\right) = 
	\left( \begin{array}{cc}
	\cos\varphi & -\sin\varphi\\
	\sin\varphi & \cos\varphi
	\end{array}\right)
	\left( \begin{array}{c} \phi \\ s \end{array}\right), 
\eea
where the mixing angle $\varphi$ is given by
\bea
	\tan 2\varphi = \frac{ 2 m_{s\phi}^2 }{ m^2_{ss} - m^2_{\phi\phi} }.
	\label{eq:scalarangle}
\eea


\section{Sources of CP Violation}
\label{sec:cpv}

In the SM, the CP violation is characterized by the quark-rephasing invariant quantity, the Jarlskog invariant~\cite{Jarlskog:1985ht}
\begin{equation}
J_{CP} = (m_t^2-m_c^2)(m_t^2-m_u^2)(m_c^2-m_u^2)(m_b^2-m_s^2)(m_b^2-m_d^2)(m_s^2-m_d^2)A, 
\end{equation}
where
\begin{equation}
	A = \textrm{Im} V_{ud}V_{cb}V^*_{ub}V^*_{cd} 
\end{equation}	
is twice the area of the unitarity triangle of the CKM matrix. 
As the SM CKM matrix only has 1 independent CP phase, the three unitarity conditions give the same area, which is the only CP violating source.
This quantity can also be written as~\cite{Bernabeu:1986fc}
\begin{equation}
	\label{JarlskogM}
J_{CP} = -\frac{i}{2}\textrm{det}[H_u,H_d],
\end{equation}
where $H_u = M_uM_u^{\dagger}$ and $H_d = M_dM_d^{\dagger}$ are the building blocks of rephasing invariants. 
In the picture of electroweak baryogenesis, this quantity provides a dimensionless CP violation strength $J_{CP}/T_c^{12}\sim 10^{-20}$, 
which is too small compared to the typical strength of Baryogenesis $\eta\sim 10^{-10}$.
We observe that both the fermion masses and the unitarity triangle $A$ suppress the CP violation.
Thus we expect that adding heavy quarks could enhance the CP violating effect via the large fermion mass. 
In general, this heavy quark can be chiral or vector-like.
A lot of efforts were performed for both the model of 4th generation quarks~\cite{Hou:2008xd} and vector-like bottom quark~\cite{delAguila:1997vn, Branco:1998yk}, among which was the study of the enhancement of CP violation. 
%
%
However, the simplest 4th generation model was ruled out by the experiment data~\cite{Djouadi:2012ae, Eberhardt:2012gv}. 
On the other hand, the vector-like quark is still alive and provides enhancement of CP violation  in a similar way.

Let us discuss the CP violation strength in our model. 
First, let's count how many independent CP phases are there in the model. 
The unitary condition Eq.~\ref{uniCKM} sets 9 constraints to the CKM elements. There are also $6$ rephasing redundencies, coming from the 7 involved quark fields modular the total baryon phase. Finally, the number of independent matrix elements in the CKM matrix is
\begin{equation}
	12\times2 - 9 - 6 = 9,
\end{equation}
among which 6 degrees of freedom attribute to real rotations, and the other 3 are independent CP phases.

The 3 phases can be parameterized as the following 3 rephrasing invariants
\begin{align}
	B_1 &= \mathrm{Im}V_{cb}V^*_{Tb}V_{T4}V^*_{c4},\\
	B_2 &= \mathrm{Im}V_{tb}V^*_{Tb}V_{T4}V^*_{t4},\\
	B_3 &= \mathrm{Im}V_{cb}V^*_{tb}V_{t4}V^*_{c4},
\end{align}
all of which represents the area of a subtriangle of the unitarity quadrangle formed by the complex numbers $V^*_{Tb}V_{T4}$, $V^*_{tb}V_{t4}$, $V^*_{cb}V_{c4}$, $V^*_{ub}V_{u4}$. In the chiral limit where $u$ and $c$ are supposed to be massless, $m_u=m_c=0$, the $B$'s that involve $c$ and $u$ are not observable, hence we only care about the quantity $B_2$.

Similar to the SM case, those unitarity areas are not the only quantities which appear in the CP violation  processes. In the case that we are interested in, where CP violation occurs simply via the evolution described by the Dirac equation~\cite{Bernreuther:2002uj}, the mass matrices also play the game. Therefore, dimensionful quantities like Eq.~\ref{JarlskogM} should be used to characterize the CP violation strength.
In the gauge basis, the mass terms are
\begin{equation}
	- \mathcal{L}_{\rm mass} = \bar{u}_{LI} \mathcal{M}^u_{IJ} u_{RJ} + \bar{d}_{Li}\mathcal{M}^d_{ij}d_{Rj}.
\end{equation}
We decompose the up-type mass matrix as 
\bea
\mathcal{M}^u = \begin{pmatrix}M^u \\ m^u \end{pmatrix},
\eea
where 
$M^u$ and $m^u$ are
submatrices with dimension $3 \times 4$ and $1 \times 4$, respectively.  
Then we can define $H_u = M_uM_u^{\dagger}$ and $H_d = M_dM_d^{\dagger}$ as before. We can also define another building block of rephasing invariants $h_u = M^{u}m^{u\dagger}$.

It was investigated that in the vector-like bottom partner model~\cite{delAguila:1997vn}, the CP violation is characterized by 7 Jarlskog-like invariants (J-invariants). The top partner model should be similar. 
In the simple case of chiral limit $m_u=m_d=m_s=m_c=0$, only one of them is  independent:
\begin{equation}
\begin{split}
J& = \mathrm{Im}\ \mathbf{tr}H_dH_uh_uh_u^{\dagger}\\
& = m_b^2m_T^2m_t^2(m_T^2-m_t^2) B_2,
\end{split}
\end{equation}

In this work, we only estimate the CP asymmetry using the Jarlskog-like invariants, and leave more detailed study via transport equation for future work.
To estimate the strength of CP violation in our model, we need to look at the experimental constraints on the heavy fermion mass and the extended CKM matrix elements. 
Current experiments on flavor physics, such as K and B decay and $B-\bar{B}$ mixing were analyzed
%
%
in the literature\cite{Botella:2012ju, Alok:2015iha} by performing a global fitting on the $4\times 3$ CKM matrix elements using 68 flavor physics observables~\cite{Alok:2015iha}. 
The analysis includes the direct measurements of the CKM elements, CP violation in $K_L \to \pi \pi$, branching fraction of the decay
$K^+ \to \pi^+ \nu \bar{\nu}$, branching fraction of the decay $K_L \to \mu^+ \mu^-$, $Z \to b \bar{b}$ decay, $B^0_q-\bar{B}^0_q$ mixing
($q = d, s$), indirect CP violation in $B^0_d \to J/\psi K_S$ and $B^0_s \to J/\psi \phi $, CKM angle $\gamma$, 
branching ratio of $B \to X_s \ell^+ \ell^-$ ($\ell = e, \mu$), branching ratio of $B \to X_s\gamma$, branching ratio of $B \to K\mu^+ \mu^-$, constraints from $B \to K^* \mu^+ \mu^-$, branching ratio of $B^+ \to \pi^+\mu^+ \mu^-$, branching ratio of $B_q \to \mu^+ \mu^-$ ($q = s, d$), branching ratio of $B \to \tau \bar{\nu}$, like-sign dimuon charge asymmetry $A_{SL}^b$, and finally the oblique parameters $S$ and $T$. 
The results of the global fitting are shown in Table 5 and 6 of Ref.~\cite{Alok:2015iha} for $m_T = 800 $ GeV and $1200 $ GeV.
The results suggest that $B_2$ could be as large as $10^{-6}$ for a TeV top partner
\footnote{ Using the results in table 6 of Ref.~\cite{Alok:2015iha} where the moduli of the $V^{\dagger}V$ elements are estimated by mixing and decay of B and K mesons, the $B_i$ quantities can be estimated by $B_i \lesssim |V^{\dagger}V|\times|V^{\dagger}V|$. 
Taking imaginary parts might introduce one or two orders of magnitude smaller, but would not ruin the estimation.
}.
More importantly, the enhancement from the heavy top quark mass implies a J-invariant of order $J \lesssim 10^{11}{\rm GeV}^8$.
Assuming that the typical energy scale during the baryogenesis is the critical temperature of EW phase transition $T_c$, the dimensionless CP violation strength formulated as $J/T_c^8$ needs to be greater than the observed baryon number asymmetry. 
\eq{
\frac{J}{T_c^8} \ge \eta \sim 10^{-10},
}
which sets an upper bound for the critical temperature $T_c \lesssim 420 $TeV.  

Given the possible large CP violation effects, we also need to check the current constraints from the non-observation of the electric dipole moment (EDM) of the electron and neutron. 
The electroweak sector of the Standard Model gives an EDM for the neutron of size $|d_n| \sim 10^{-32} - 10^{-31} e $cm.
The model with an extended quark sector typically gives rise to a quark EDM at the two loop level, thus contributes to the neutron EDM. 
The current experimental limit on the EDM of the neutron is $|d_n| < 2.9 \times 10^{-26} e $cm (90\% CL)~\cite{Baker:2006ts}. 
It is shown in the literature~\cite{Liao:2000re} that the EDM in the model with an extra down-type singlet quark and found that the induced neutron EDM is of order $10^{-29} e $cm. 
Ref.~\cite{Liao:2000re} also comments that
in the model with an extra up-type singlet quark, the down-type quark EDM's vanish identically at two loop order.
Since all down-type quarks are light in this case, the leading terms for the $u_e$ EDM is proportional to
$m_{u_e} m^2_{d_i}$. 
The contributions from the vector-like up-type quark to the neutron EDM are thus even smaller compared to the extra down-type  quark.


\section{The Scalar Potential}
\label{sec:potential}

\subsection{A Brief Review of Effective Scalar Potential}

To study the phase transition, we consider the potential of the two scalar fields
at finite temperature (see Ref.~\cite{Quiros:1999jp} for review). 
At the one-loop order, the zero-temperature effective potential in the Landau gauge
\footnote{While the effective potential in the Landau gauge is not gauge invariant, the potential at its minimum is well-defined. For concerns about the gauge invariance and a treatment of the gauge invariant effective potential, see Ref.~\cite{Patel:2011th}.} 
has the form
\bea
	V_{\rm CW}(\phi, s) = V_{\rm tree}(\phi, s) + \sum_i \frac{n_i}{64 \pi^2} m_i^4(\phi, s) \left(\log \frac{m_i^2(\phi,s)}{Q^2} - c_i \right),
\eea
where $n_i$ is the number of degrees of freedom of the particle $i$ running in the loop, with negative sign for fermions, and
$m^2_i(\phi, s)$ is the corresponding field-dependent squared mass, defined in Appendix A.3.
Here $c_i$ are constants that depend on the renormalization scheme,  
and $Q$ is the renormalization scale.
For convenience, counter-terms $V_{\rm CT}$ are chosen to preserve the input parameters, like the vacuum expectation values (vev) and the masses, from loop; corrections:
\bea
	&& \frac{\partial (V_{\rm CW} + V_{\rm CT})}{\partial \phi_i } \Big|_{\phi_i =  \langle \phi_i\rangle} = 0, 
	\label{eq:cwvev}\\
	&& \frac{\partial^2 (V_{\rm CW} + V_{\rm CT})}{\partial \phi_i \partial \phi_j } \Big|_{\phi_i = \langle \phi_i\rangle} = m_{ij}^2,
	\label{eq:cwmass}
\eea
where $\langle \phi_i \rangle = v, u$ and $m_{ij}^2$ are tree level vev and mass squared matrix defined in Sec.~\ref{sec:model}. 
This naive treatment fails when we consider the Goldstone contribution.
The Goldstone boson contribution to the scalar masses in Eq.~\ref{eq:cwmass} is infrared log-divergent due to its zero pole mass.
This indicates that renormalizing the scalar potential at zero external momenta, as is done in the effective potential calculation, is not a well-defined procedure when Goldstone bosons are involved.
An alternative on-shell renormalization procedure was proposed~\cite{Anderson:1991zb, Cline:1996mga, Delaunay:2007wb} to cure the problem, as described in Appendix A.1 and A.2 in detail. We extend the results of the Ref.~\cite{Delaunay:2007wb} to the effective potential with mixture of the Higgs boson and new scalar. Here we list the final expression of the zero temperature one-loop effective potential:
\bea
	V^{\rm on-shell}_{\rm CW}(\phi, s) &=& V_{\rm tree}(\phi, s) 
	+ \sum_{i\neq G} \frac{n_i}{64 \pi^2} \left[m_i^4(\phi, s) \left(\log \frac{m_i^2(\phi,s)}{m_i^2(v, u)} - \frac23 \right)
	+ 2 m^2_i(v,u) m_i^2 (\phi, s) \right] \nn\\
	&+& \frac{3}{64\pi^2}m_G^4(\phi,s)\ln \frac{m_G^2(v,u)}{m_H^2}. 
\eea

The one-loop thermal corrections to the effective potential at
finite temperature $T$ is
\bea
 V_{\rm thermal}(\phi, s, T) = \sum_i \frac{n_i T^4}{2 \pi^2} J_{\rm B,F}
\left(\frac{m_i^2(\phi, s)}{T^2}\right),
\label{VT}
\eea
where
\bea
J_{\rm B,F}(y) = \int_0^\infty dx\ x^2 \log\left[1 \mp
e^{-\sqrt{x^2+y}} \right].
\label{Ibf} 
\eea
with the sign $-$ for bosons and $+$ for fermions.

The finite-temperature potential needs to be corrected, due to the infrared divergences, generated by bosonic long-range fluctuations called Matsubara zero modes. 
%
%
This can be solved schematically by resumming over all diagrams with bubbles attached to the big loop~\cite{Parwani:1991gq, Arnold:1992rz, Sanchez:2006tt}, which are called the "ring diagrams".  
This leads to a shift of the bosonic field-dependent masses $m_i^2(\phi,s)$ to the thermal field-dependent masses  
\bea
	m_i^2(\phi,s, T) \equiv m_i^2(\phi,s)+ \Pi_i (T),
\eea
where the thermal shifts $\Pi_i$ are defined in Appendix A.3.  
After resummation, the ring-diagram contribution to the effective potential reads 
\bea
	V_{\rm ring} &=& - \frac{T}{12\pi} \sum_{i = {\rm B}} n_i\left(\left[m_i^{2}(\phi,s,T)\right]^{3/2} -  \left[m_i^{2}(\phi,s)\right]^{3/2} \right). 
\eea
%

In the SM, this cubic term from the Matsubara zero mode is the only source to induce a thermal barrier between a symmetric minimum and a symmetry-broken minimum in the effective potential. It was because that all the other terms in SM Higgs sector are quadratic or quartic, which can't create such degenerate minima in one-dimensional scalar space. However, in our model, as shown later, the new dimension in the scalar space greatly enriches the possibility, and hence the ring diagram contribution is much less important.

The total effective potential at finite temperature is the sum of the above terms
\beq
V_{\mathrm{eff}}(\phi,s,T) = V^{\rm on-shell}_{\rm CW}(\phi, s) + 
V_{\rm thermal}(\phi, s, T) +V_{\rm ring}(\phi, s, T).
 \label{Veff}  
\eeq
For part of the field space, the field-dependent masses of the scalars and the Goldstone bosons can be negative, and the non-convexity of the potential would induce an imaginary part that indicates a vacuum decay rate per unit volume.
However, the real part can still be interpreted as the expectation value of the energy density. Therefore we only take the real part of the potential to do the numerical analysis.

%

\subsection{Approximate Analysis of the Scalar Potential}

In the next section, we perform a numerical study on the full effective potential based on eq.~(\ref{Veff}), and scan the parameter space for strong first order phase transition. 
But to understand the numerical results, we need some approximate methods to analyse the complicated potential function. 
In the high temperature limit, the effective potential can be simplified as
\bea
\label{Full_highT}
	V_{\mathrm{eff}}(\phi,s,T)  &\simeq& V_{\rm tree} 
	+ \frac{T}{12\pi} \sum_{i = {\rm B}} n_i \left[m_i^{2}(\phi,s,T)\right]^{3/2}  
	+ \sum_{i = {\rm B, F}}  \frac{|n_i|}{24(1+\delta_{i{\rm F}})} m_i^2(\phi,s)T^2 \nn\\
	&-& \sum_{i = {\rm B, F}}  \frac{n_i}{64\pi^2} \left(m_i^4(\phi,s)\ln \frac{m_i^2(v,u)}{a'_i T^2}
	- 2 m^2_i(v,u) m_i^2 (\phi, s)\right) + \rho(T),
\eea
where $a'_{\rm B} = 16 \pi^2 \exp(-2\gamma_E)$ for bosons and $a'_{\rm F} = \pi^2 \exp(-2\gamma_E)$ for fermions. 
$\rho(T) = \frac{\pi^2}{90} n_\rho T^4$ is the Stefan-Boltzmann contributions with $n_\rho = n_B + \frac78 n_F$, which is field-independent and can be drop out for our purpose.
If we series expand the second term in $m^2_i(\phi,s) / T^2$, the remaining relevant term is another $m^2T^2$, which, combined with the third term here, gives the main temperature dependence of the full potential. The log terms can be absorbed into the running parameters that vary little within the energy scope of our discussion. Now the effective potential is simplified as
\bea
	V_{\mathrm{eff}}(\phi,s,T)  &\simeq&  V_0 + \sum_{i = {\rm B}} \frac{n_ic_i^{1/2}}{8\pi} m_i^2(\phi,s)T^2
	+ \sum_{i = {\rm B, F}}  \frac{|n_i|}{24(1+\delta_{i{\rm F}})} m_i^2(\phi,s)T^2 + \rho(T),
\eea
where $V_0$ is the zero-temperature part of the potential, being the tree-level potential plus loop suppressed corrections. 
Taking the approximation $c_i^{1/2}/8\pi \approx 1/24$ for all bosons, we obtain the temperature-dependent terms
\bea
\label{Full_Polinomial}
V_{\mathrm{eff}}(\phi,s,T) \simeq V_0  + T^2 \left[ A s + B \phi^2 + C s^2 \right],
\eea
where the coefficients are
\bea
A & = & \frac{1}{12} \left[ 4\mu_{s\phi} + 2\mu_3 + 6y_sM \right],\\
B & = & \frac{1}{12} \left[ (6 \lambda_\phi + \lambda_{s\phi}/2) + \frac{1}{4}(3g^2 + g^{'2}) + \frac{2}{3}(y_t^2 + y^{\prime 2}) \right],\\
C & = & \frac{1}{12} \left[ (3 \lambda_s  + 2\lambda_{s\phi} ) - 3y_s^2 \right].
\eea
Note that the fermion Yukawa couplings are also contributed to the configuration of the scalar potential. 
%
This simplification results in a polynomial form of the potential, which is convenient for our analysis. There are sometimes significant errors for this simplification, but we will show that the qualitative analyses based on this polynomial potential explain many key features of the numerical results we obtained.



We will analyse the scalar potential at the moment of the phase transition.
More specifically, we are only interested in the stable vacuum, i.e. the global minima of the potential, which is degenerate at the critical temperature of phase transition. They are defined as $(0,u_s)$ and $(v_c,u_b)$, representing the symmetric vauum and symmetry breaking vacuum respectively. In the following, we introduce two kinds of method to estimate the properties of these vacuum configurations.

\subsubsection{Barrier Width Estimation}
We used to express the potential in the cartetian coordinates of the field space $(\phi, s)$. 
For the analysis of the phase transition, it is also convenient to utilize the polar coordinates $(\rho, \alpha)$~\cite{Profumo:2007wc, Barger:2011vm, Profumo:2014opa}
\bea
	\rho = \sqrt{\phi^2 + (s - u)^2}, \quad \cos\alpha = \frac{\phi}{\sqrt{\phi^2 + (s - u)^2}}, 
\eea
from some shifted center $(0, u)$. When $u=u_s$ and
\eq{
\cos\alpha_0 = \frac{v_c}{\sqrt{v_c^2 + (u_b - u_s)^2}}
}
The potential has degenerate minima along the $\rho$ axis, $\rho=0$ being the symmetric one, and $\rho=\bar\rho$ being the symmetry breaking one. Hereafter we will use the notation $c_\alpha \equiv \cos\alpha_0$ and $s_\alpha \equiv \sin\alpha_0$ for short. Similar to the SM, we can employ the following form of parameterization of potential
\bea
	V_{\rm eff}(\rho, T) \simeq  \frac12 D(T^2 - T^2_0) \rho^2 
	+ {\mathcal E}\rho^3 + \frac{\bar{\lambda}}{4} \rho^4, 
	\label{eq:Veffradial}
\eea
where the coefficients are constant for a simplified polinomial potential. The coefficients $D, {\mathcal E}$ and $\bar{\lambda}$ are functions of the model parameters and $u_s, s_\alpha, c_\alpha$. We impose the following condition
\bea
&V_{\rm eff}(\rho, T_c) = \frac{\bar{\lambda}}{4} \rho^2(\rho-\bar{\rho})^2,
\eea
and obtain the non-zero vacuum value
\eq{
\bar{\rho} = -\frac{2{\cal E}}{\bar{\lambda}}.
\label{eq:rhobar}
}
If we neglect the zero-temperature loop corrections, they can be expressed as
\bea
\label{eq:Epsilon}
	{\mathcal E} &=& s_\alpha \left[ (\mu_{s\phi}+\lambda_{s\phi}u_s) c_\alpha^2/2 + (\mu_3/3 + \lambda_su_s) s_\alpha^2 \right] , \\
	\bar{\lambda} &=& \lambda_\phi c_\alpha^4 + \lambda_s s_\alpha^4 + \lambda_{s\phi} c_\alpha^2 s_\alpha^2.
\eea
The $\bar\rho$ determines the width of the barrier in the scalar potential, which is what we finally concern. But this analysis does not give estimations of $u_s, s_\alpha, c_\alpha$, which brings us to the next tool.


\subsubsection{Stationary Point Search}





%
Although the shape of scalar potential has been studied in Ref.~\cite{Espinosa:2011ax}, 
we provide a detailed and systematical recipe to describe the shape of the 2-dim potential.
We summarize our results in Table~\ref{tab:scurve}, which could be used to understand the numerical studies of the phase transition in the next section.

First, let's write the potential \ref{Full_Polinomial} as the following
\bea
V_{\mathrm{eff}}(\phi,s,T) &=& \frac{\tilde{\lambda}_{\phi}\phi^4}{4} + \frac{1}{2}\left(\frac{\tilde{\lambda}_{s\phi}}{2}s^2 + \tilde{\mu}_{s\phi}s - \tilde{\mu}_{\phi}^2\right)\phi^2 \\
&+& \frac{\tilde{\lambda}_s}{4}s^4 + \frac{\tilde{\mu}_3}{3}s^3 - \frac{\tilde{\mu}_s^2}{2}s^2 + \tilde{\chi}^3s
\eea
where all the tilded couplings are supposed to depend on temperature, logarithmically or quadratically. Since the former are negligible within the energy scope that we are interested in, we only need to focus on the quadratic temperature dependencies
\bea
\label{quadratic_T}
	\tilde{\chi}^3 &=& A T^2, \\
 	\tilde{\mu}_{\phi}^2 &=& \mu_{\phi}^2 - B T^2, \\
	\tilde{\mu}_s^2 &=&  \mu_s^2   - C T^2, 
\eea
while the other couplings are mainly their zero temperature values. We learn that a temperature around $100\sim 150$ GeV is usually smaller than the other massive parameters, therefore even the quadratic temperature dependencies are still insignificant at this range of temperature. As a result, the coefficients roughly satisfy
\eq{
|\tilde{\chi}| \ll |\tilde{\mu}_s| \sim |\mu_s|.
}
Neglecting the linear term further implies a useful corollary, that there is always a stationary point sitting around the original $(0,0)$, even until the temperature reaches $100\sim 150$ GeV, the typical values of the critical temperature in our model. Only at higher temperature when $\tilde{\chi}$ becomes important, will this stationary point gradually move away. This corollary is verified by the parameter scan, and is essential for the explanation of some of the transition patterns.

Now let's do a thorough search of the possible stationary points in the potential. First we notice that the condition for extrema consist of the following two curves:
\bea
\frac{\partial V_{\mathrm{eff}}(\phi,s,T)}{\partial \phi} = 0 , \qquad \frac{\partial V_{\mathrm{eff}}(\phi,s,T)}{\partial s} = 0,
\eea
and the vacuum must be at one of the intersections between the two curves. Let us call them the \textit{$\phi$ curve} and the \textit{$s$ curve} respectively. We will describe the shape of these two curves, and try to find some rules that the possible degenerate vacuum points should follow.

The $\phi$ curve consists of a trivial line $\phi = 0$ and a quadratic curve
\bea
\tilde{\lambda}_{\phi}\phi^2 + \frac{1}{2}\tilde{\lambda}_{s\phi} \left( s + \frac{\tilde{\mu}_{s\phi}}{\tilde{\lambda}_{s\phi}} \right)^2 =   \tilde{\mu}_{\phi}^2 + \frac{\tilde{\mu}_{s\phi}^2}{2\tilde{\lambda}_{s\phi}}.
\label{eq:curve_phi}
\eea
Obviously, symmetry-broken minimum, if there is any, must be on the quadratic curve.

For the case of $\tilde{\lambda}_{s\phi} > 0$, the quadratic curve is an ellipse, centered at point $(0,s_*\equiv-\tilde{\mu}_{s\phi}/\tilde{\lambda}_{s\phi})$, with size decreasing as $-T^2$ due to the $\tilde{\mu}_{\phi}^2$ term. At some high temperature, the ellipse shrinks to zero, and no non-zero vacuum is allowed, hence the symmetry must be restored. If $\tilde{\lambda}_{s\phi} < 0$, the quadratic curve is a hyperbola also centered at $(0,s_*)$. At high temperature, the curve will move away to infinity. As long as the potential is still bounded from below, the minimum cannot be on the hyperbola, so the symmetry must be restored.

The equation for $s$ curve is a cubic polynomial of $s$
\eq{\label{eq:s_curve}
\tilde{\lambda}_ss^3 + \tilde{\mu}_3s^2 - \left(\tilde{\mu}_s^2-\frac12\tilde{\lambda}_{s\phi}\phi^2\right)s + \left(\tilde{\chi}^3 + \frac{\tilde{\mu}_{s\phi}}{2}\phi^2\right) = 0
}
The discriminant of this polynomial turns out to be a polynomial of the $\phi$ field:
\eq{
\Delta(\phi) = a\phi^6 + b\phi^4 + c\phi^2 + d
}
where the coefficients are
\eq{
a &= -\frac{1}{2}\tilde{\lambda}_s\tilde{\lambda}_{s\phi}^3, \\
b &= \frac{1}{4}\tilde{\lambda}_{s\phi}^2\tilde{\mu}_3^2 + 3\tilde{\lambda}_s\tilde{\lambda}_{s\phi}^2\tilde{\mu}_s^2 + \frac{9}{2}\tilde{\lambda}_s\tilde{\lambda}_{s\phi}\tilde{\mu}_3\tilde{\mu}_{s\phi} - \frac{27}{4}\tilde{\lambda}_s^2\tilde{\mu}_{s\phi}^2, \\
c &= -\tilde{\lambda}_{s\phi}\tilde{\mu}_3^2\tilde{\mu}_s^2 - 6\tilde{\lambda}_s\tilde{\lambda}_{s\phi}\tilde{\mu}_s^4 - 2\tilde{\mu}_3^3\tilde{\mu}_{s\phi} - 9\tilde{\lambda}_s\tilde{\mu}_3\tilde{\mu}_s^2\tilde{\mu}_{s\phi} \\
d &= (\tilde{\mu}_3^2 + 4\tilde{\lambda}_s\tilde{\mu}_s^2)\tilde{\mu}_s^4 .
}
The insignificant $\tilde{\chi}$ has been taken to be zero as an approximation. This polynomial shows the number of points on the $s$ curve at a specific $\phi$ value: if the polynomial is positive, then there are three points at this $\phi$; if the polynomial vanishes, two of the points are degenerate; if the polynomial is negative, there is only one point at this $\phi$. Therefore, the variation of the sign of this discriminant with respect to $\phi$ could give the key features of the shape of the $s$ curve. A systematic classification is provided in Table~\ref{tab:scurve}.

\begin{table}[htbp] 
	\begin{center}
	\caption{The $s$ curve shows the first derivative of the scalar potential along the $s$ direction $\frac{\partial V}{\partial s} = 0$. $\Delta'$ is defined to be the discriminant of $\Delta(\phi)$ as a cubic polynomial of $\phi^2$. It is positive unless specified. } \label{tab:scurve}
  \hspace{0cm}
  \begin{tabular}{lll||lll} 
	  \hline \hline 
	  \shortstack{Type \\ {\small $(a, b, c, d)$} } & {\small discriminant $\Delta(\phi)$} & {\small $s$ curve {\tiny $\frac{\partial V}{\partial s} = 0$} }  & 
	  \shortstack{Type \\ {\small $(a, b, c, d)$} } & {\small discriminant $\Delta(\phi)$} & {\small $s$ curve {\tiny $\frac{\partial V}{\partial s} = 0$} }  \\
      \hline 
	  \shortstack{\textbf{A} \\ {\small $(+,+,+,+)$} }  & \parbox[c]{1em}{
  		\includegraphics[width=1in]{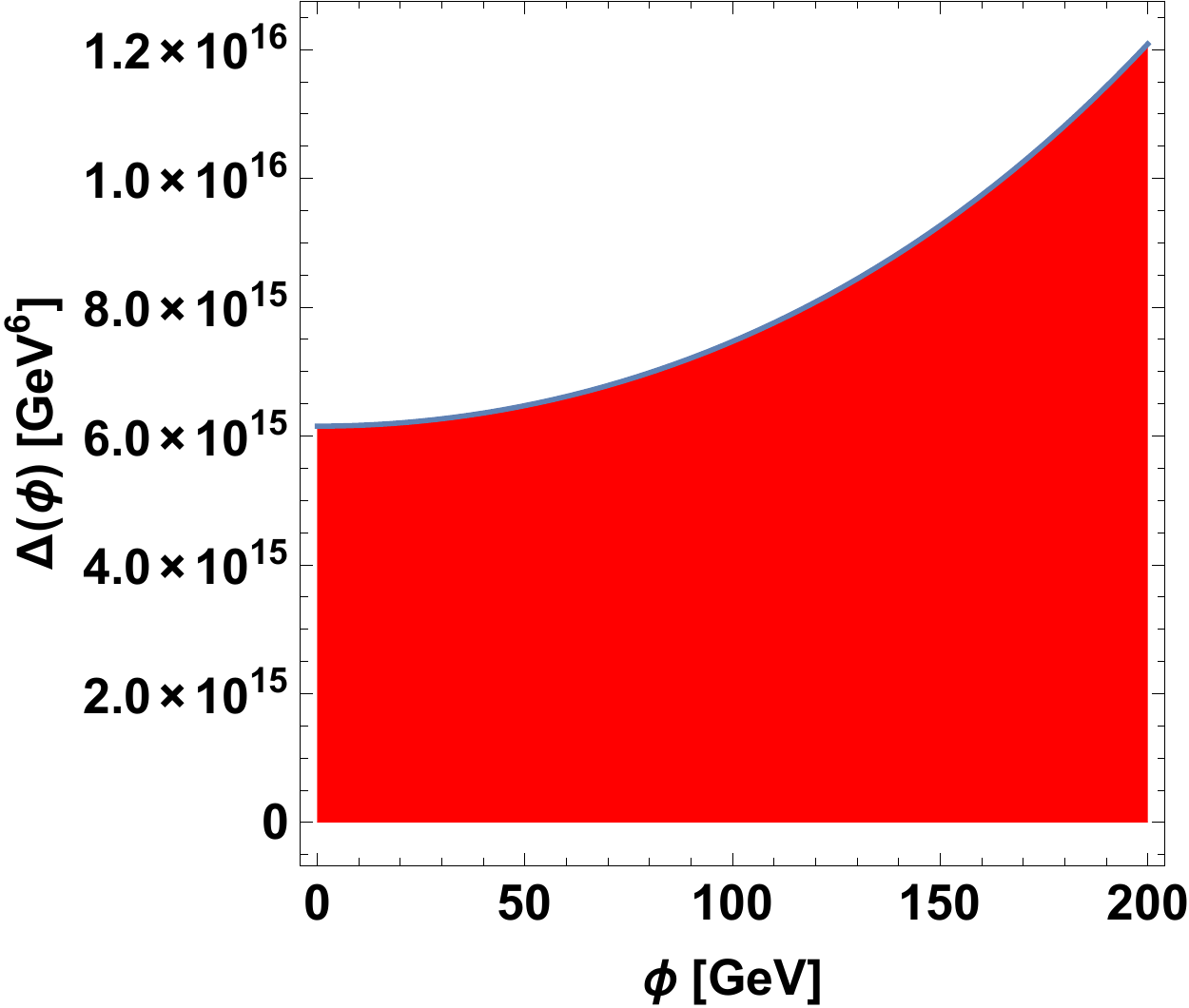}  } & \parbox[c]{1em}{
  		\includegraphics[width=1in]{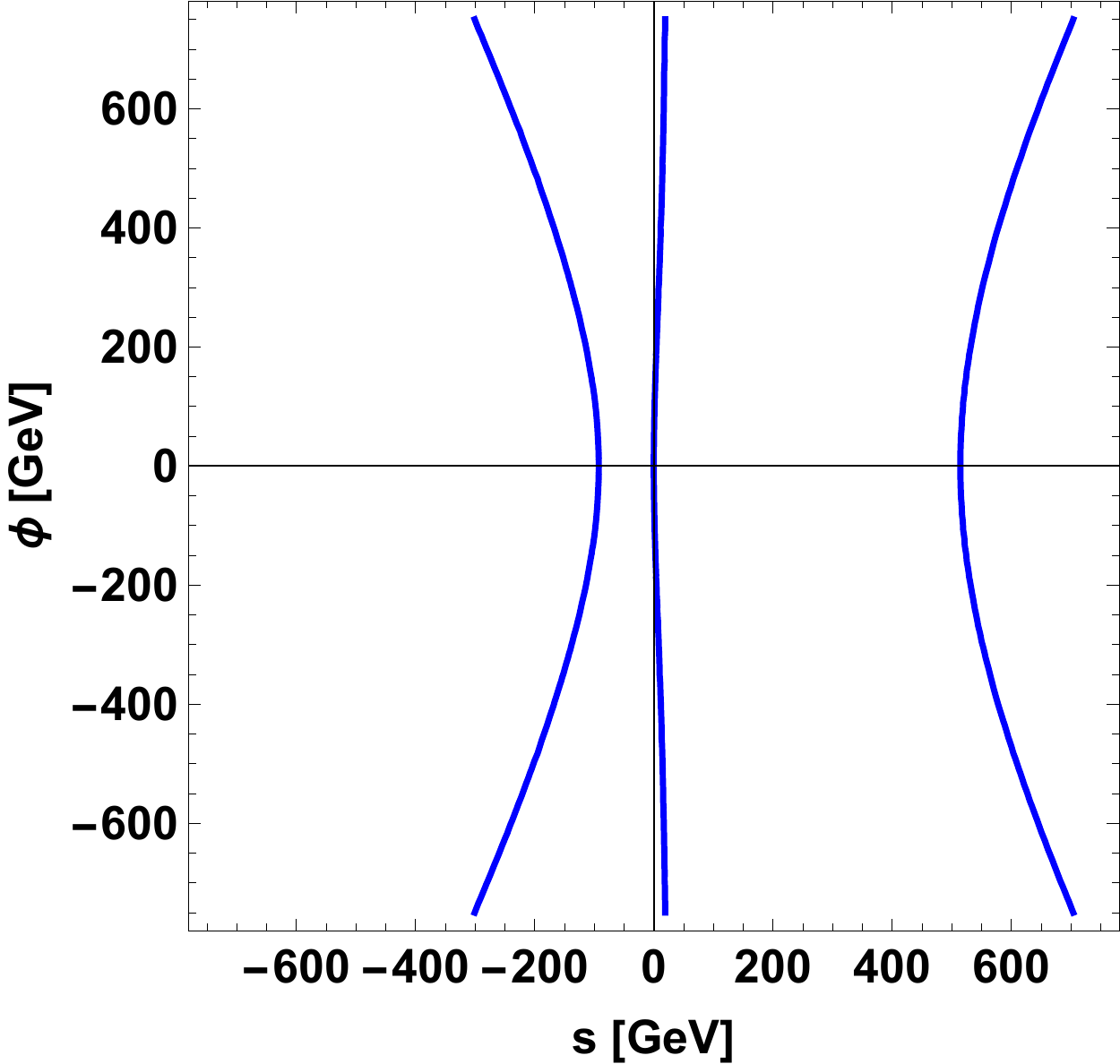}  } &
  	  \shortstack{\textbf{B} \\ {\small $(+,+,+,-)$} \\ {\small $(+,+,-,-)$} \\ {\small $(+,-,-,-)$} }  & \parbox[c]{1em}{
    		\includegraphics[width=1in]{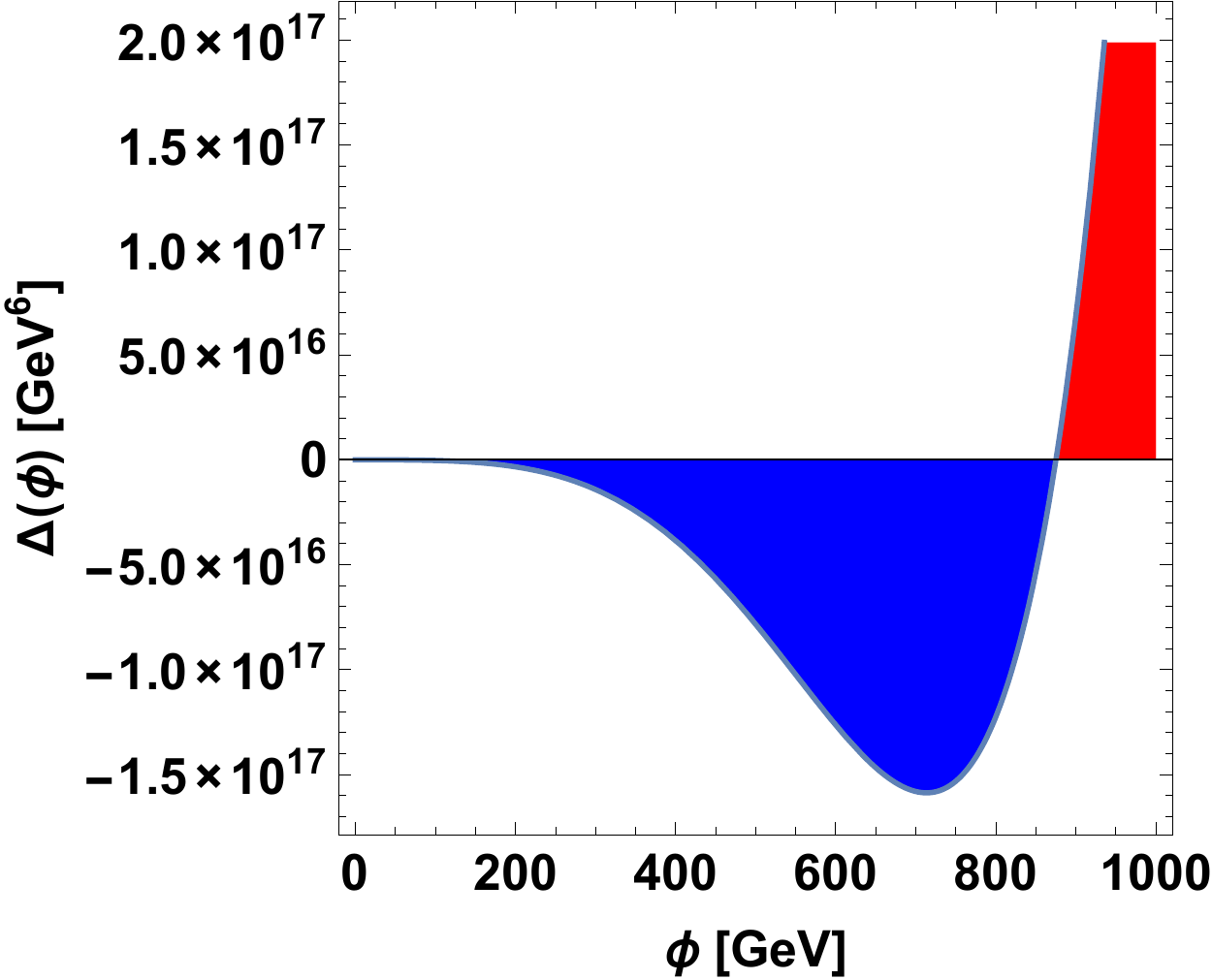}  } & \parbox[c]{1em}{
    		\includegraphics[width=1in]{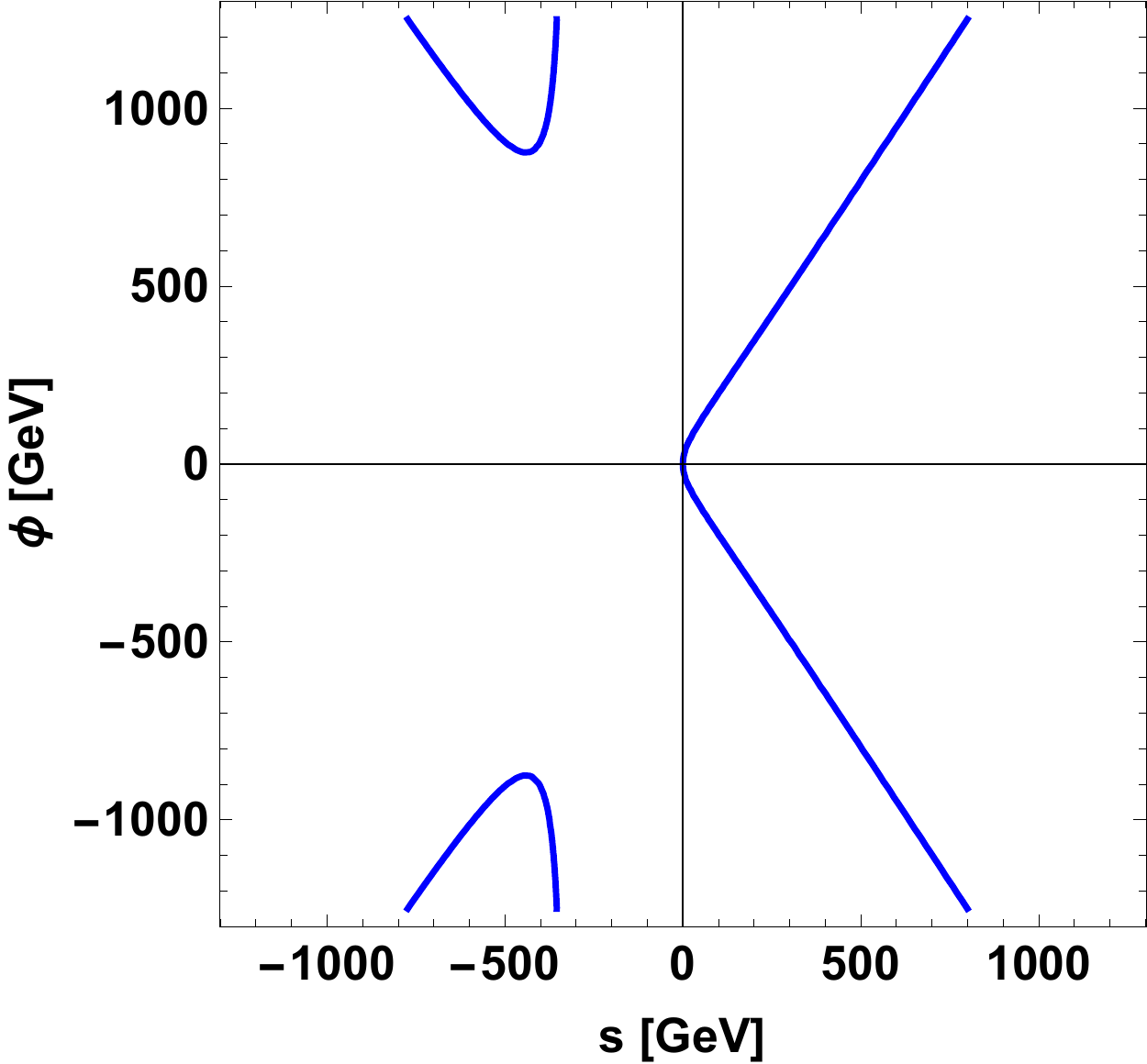}  } \\
  	  \shortstack{\textbf{C} \\ {\small $(+,+,-,+)$} \\ {\small $(+,-,+,+)$} \\ {\small $(+,-,-,+)$} } & \parbox[c]{1em}{
 		\includegraphics[width=1in]{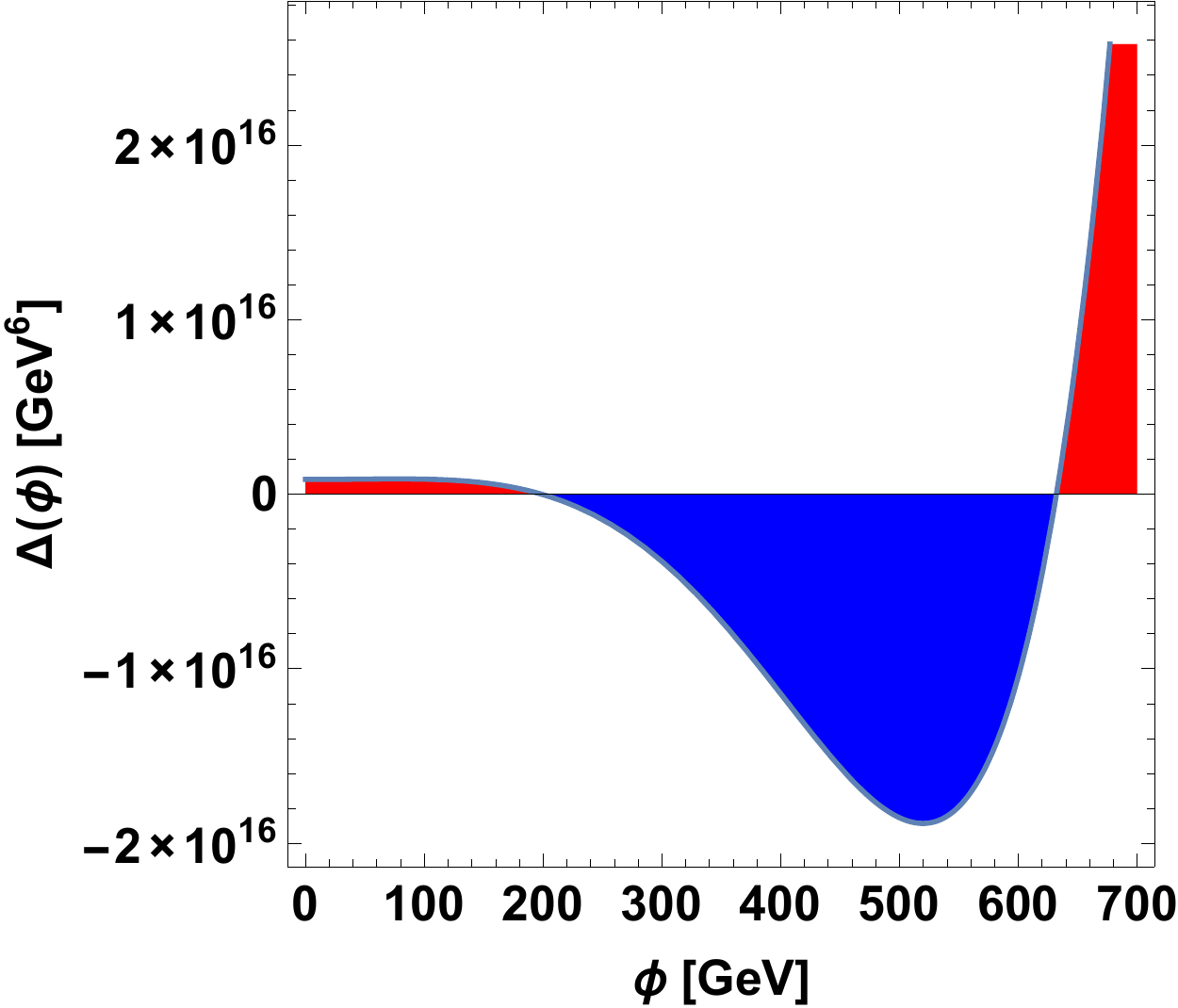}  } & \parbox[c]{1em}{
 		\includegraphics[width=1in]{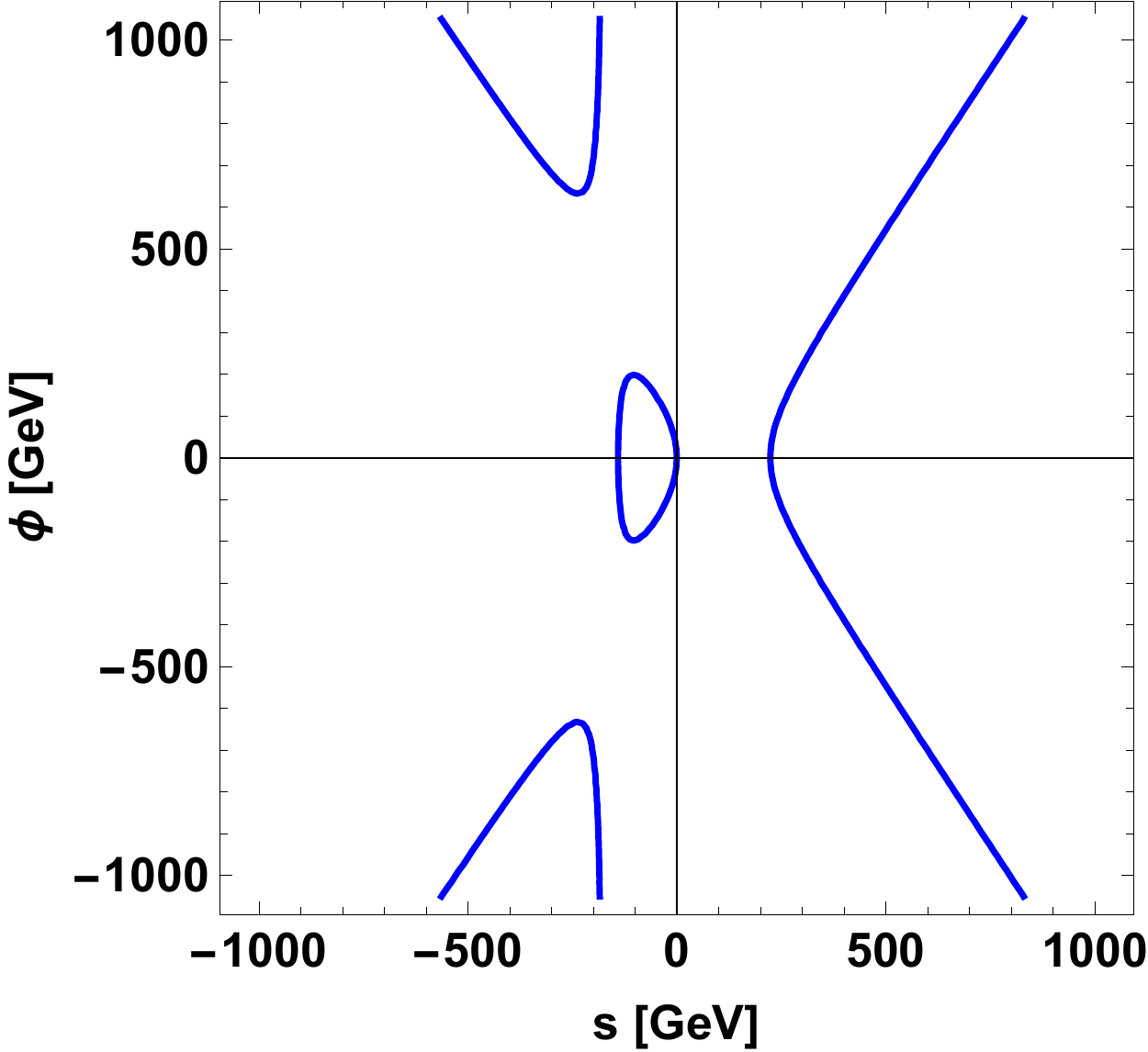}  } &
 	  \shortstack{\textbf{D} \\ {\small $(+,-,+,-)$}}  & \parbox[c]{1em}{
   		\includegraphics[width=1in]{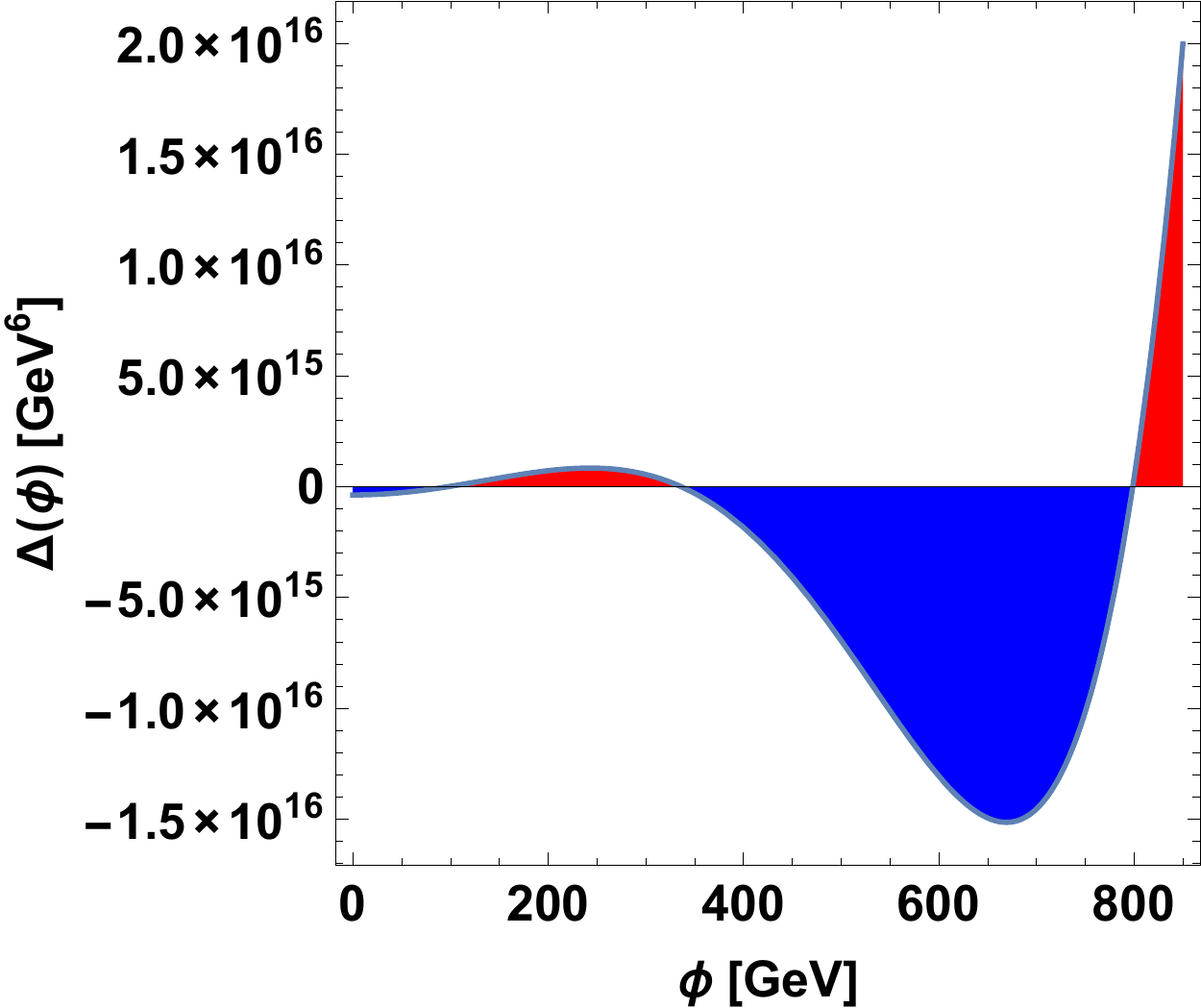}  } & \parbox[c]{1em}{
   		\includegraphics[width=1in]{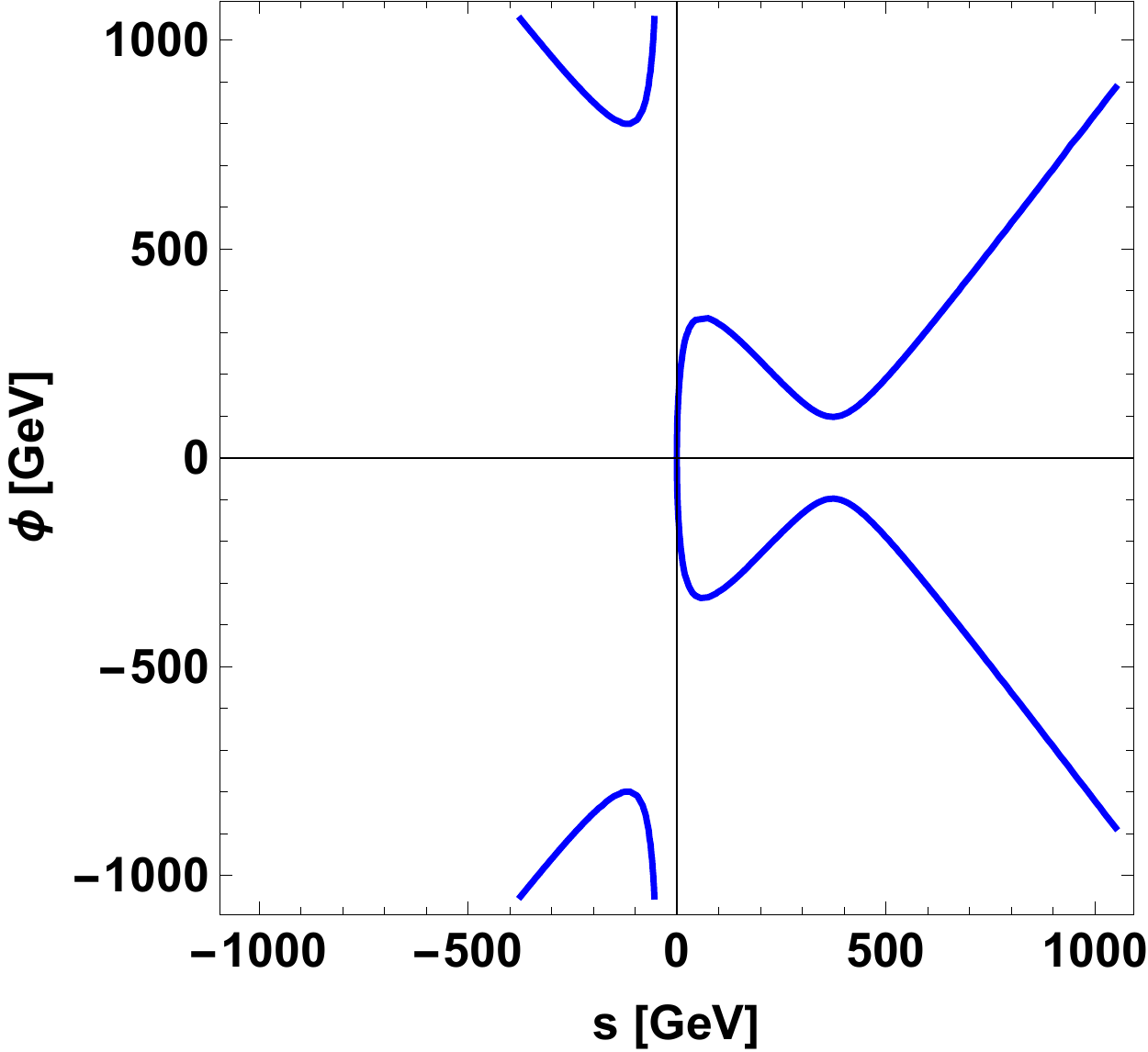}  } \\
   	  \shortstack{\textbf{E} \\ {\small $(-,+,+,+)$} \\ {\small $(-,-,+,+)$} \\ {\small $(-,-,-,+)$} } & \parbox[c]{1em}{
  		\includegraphics[width=1in]{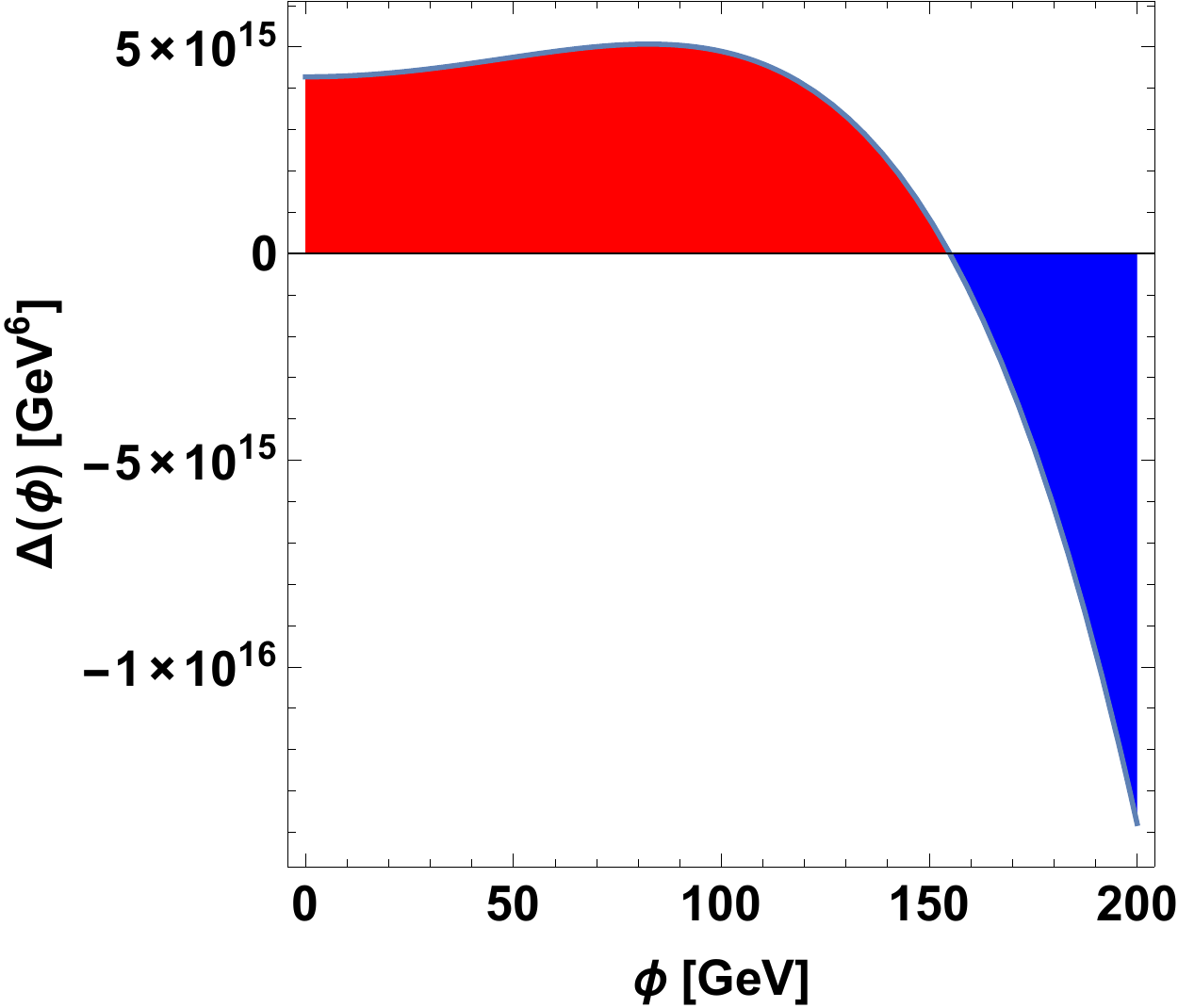}  } & \parbox[c]{1em}{
  		\includegraphics[width=1in]{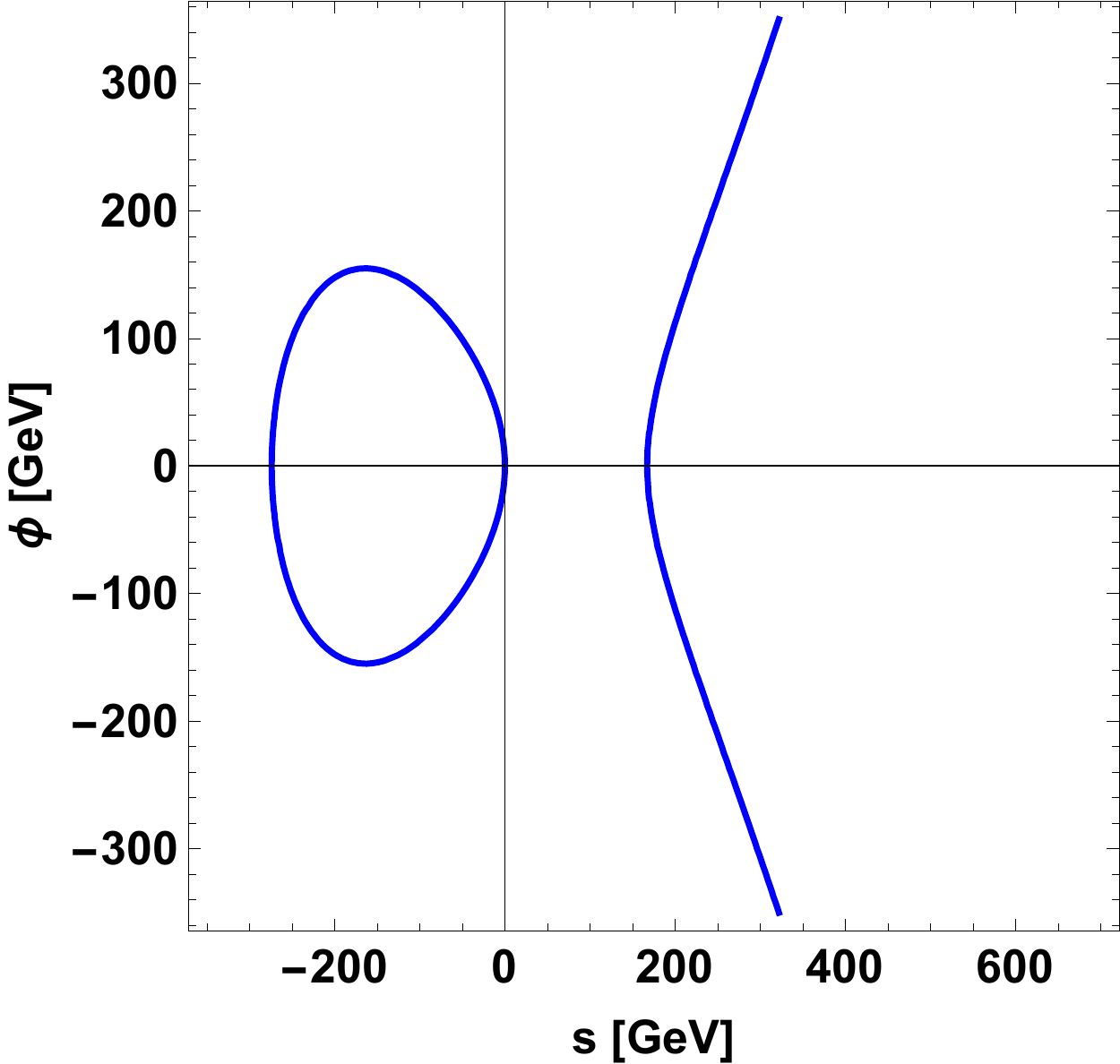}  } &
  	  \shortstack{\textbf{F} \\ {\small $(-,+,+,-)$} \\ {\small $(-,+,-,-)$} \\ {\small $(-,-,+,-)$}}  & \parbox[c]{1em}{
    		\includegraphics[width=1in]{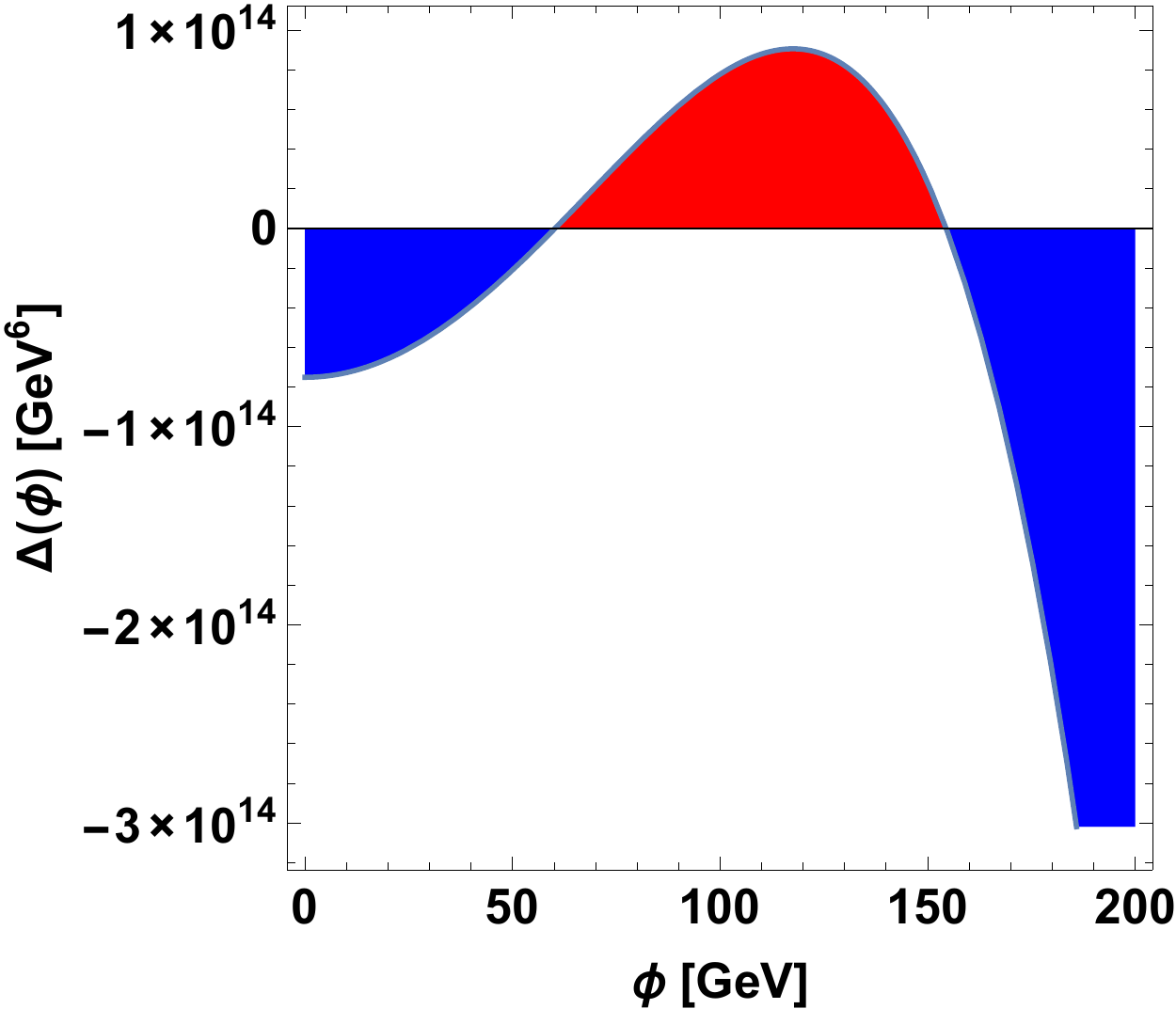}  } & \parbox[c]{1em}{
    		\includegraphics[width=1in]{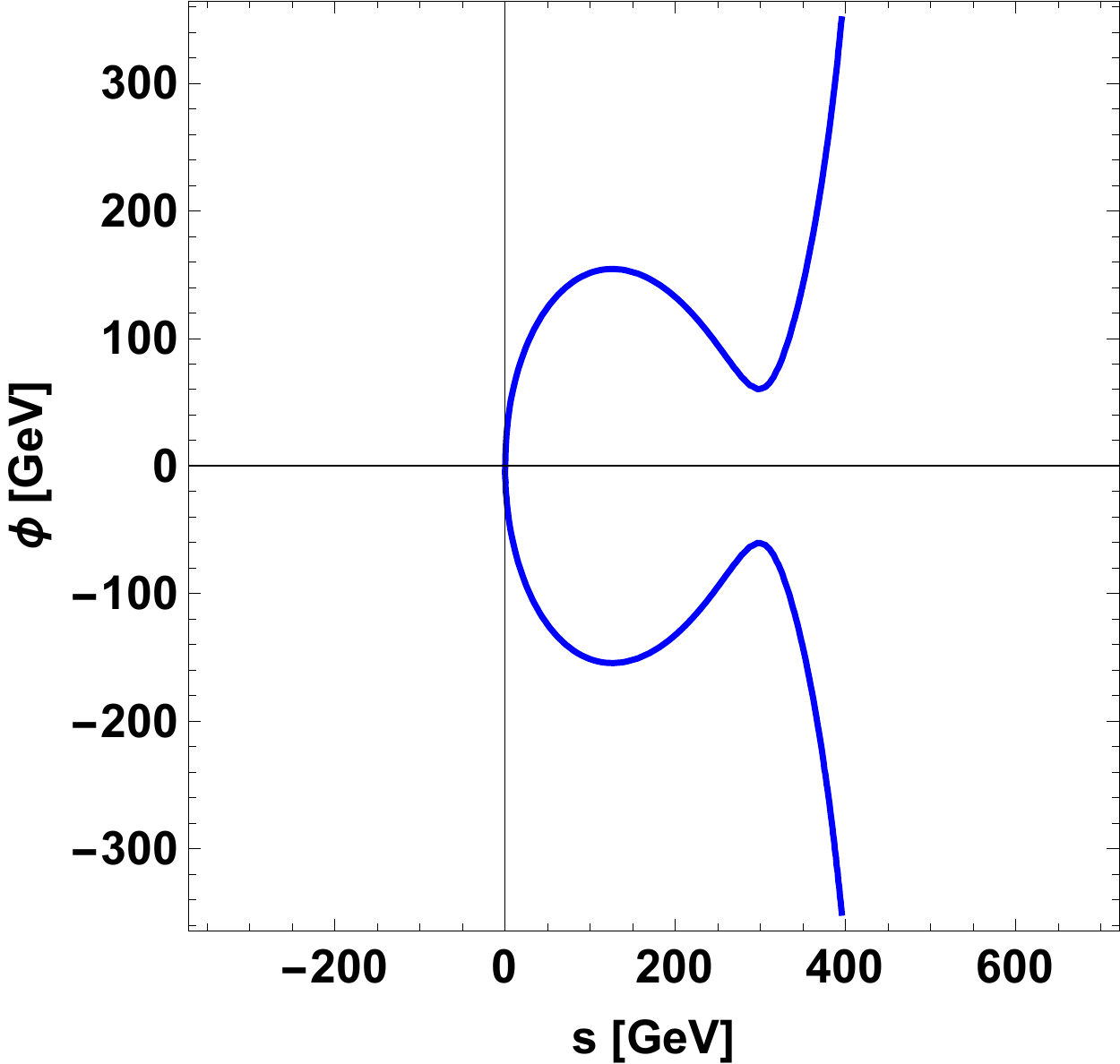}  } \\
	  \shortstack{\textbf{G} \\ {\small $(-,+,-,+)$}} & \parbox[c]{1em}{
	   	\includegraphics[width=1in]{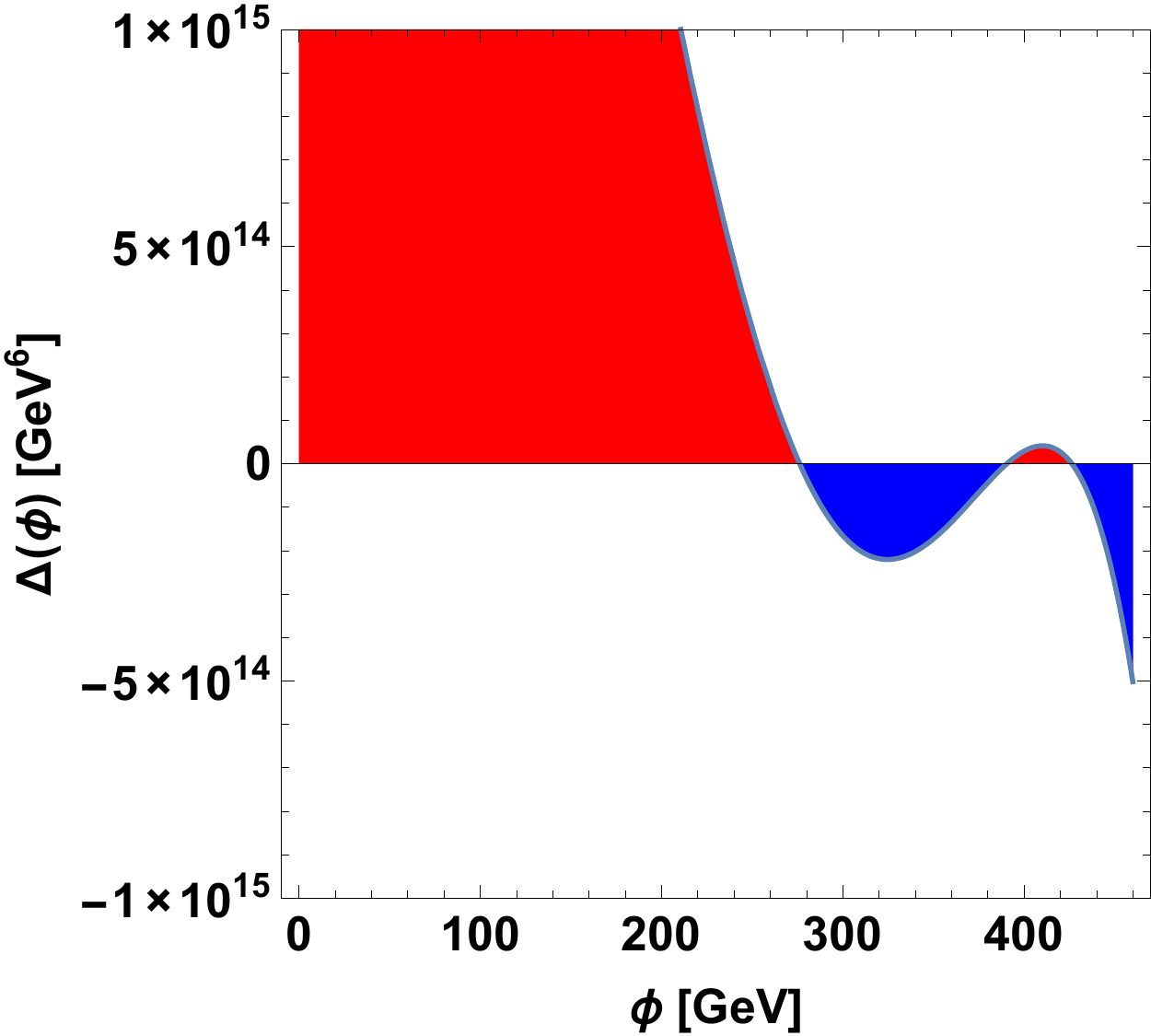}  } & \parbox[c]{1em}{
	   	\includegraphics[width=1in]{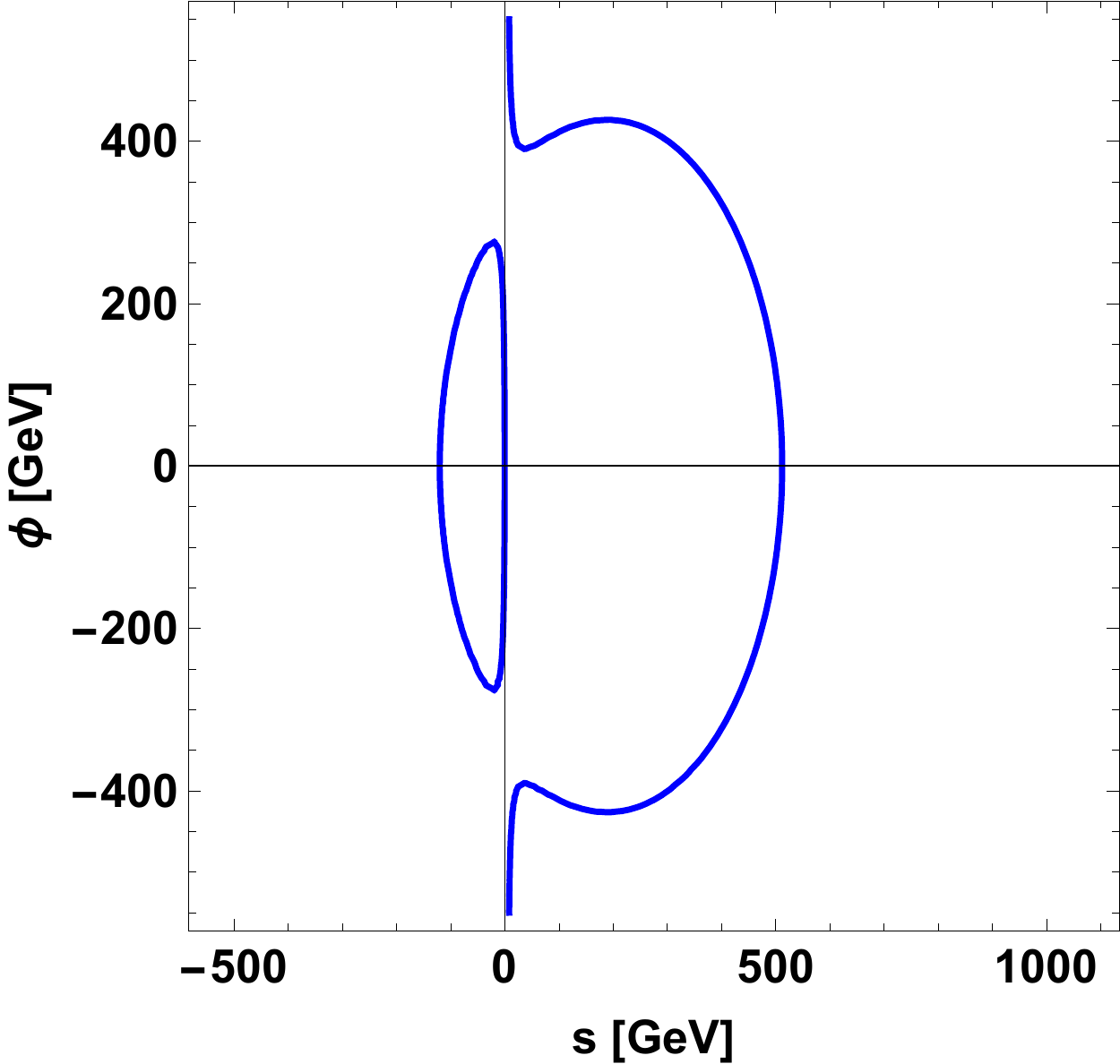}  } &
	  \shortstack{\textbf{H} \\ {\small $(-,-,-,-)$}}  & \parbox[c]{1em}{
	     \includegraphics[width=1in]{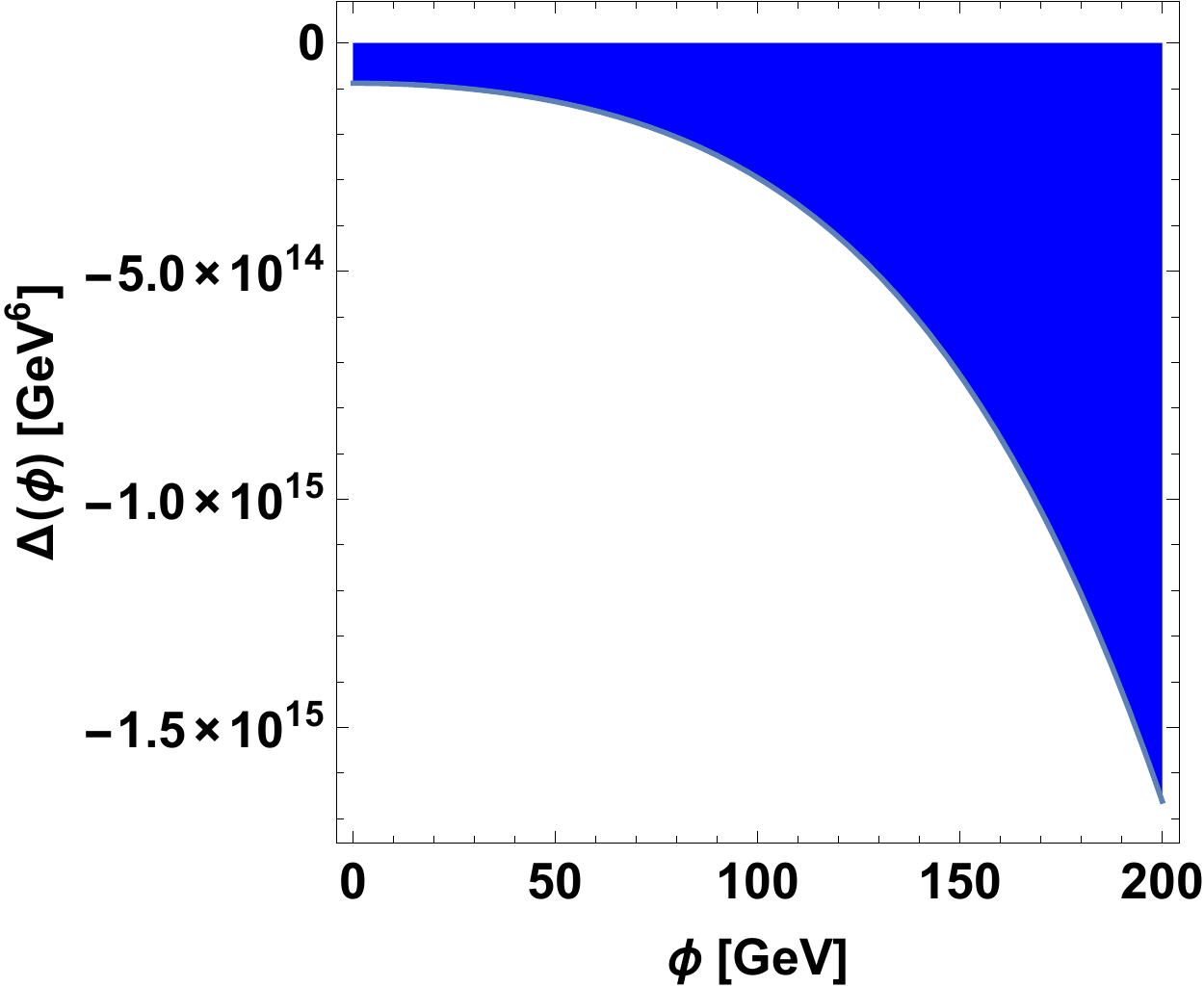}  } & \parbox[c]{1em}{
	     \includegraphics[width=1in]{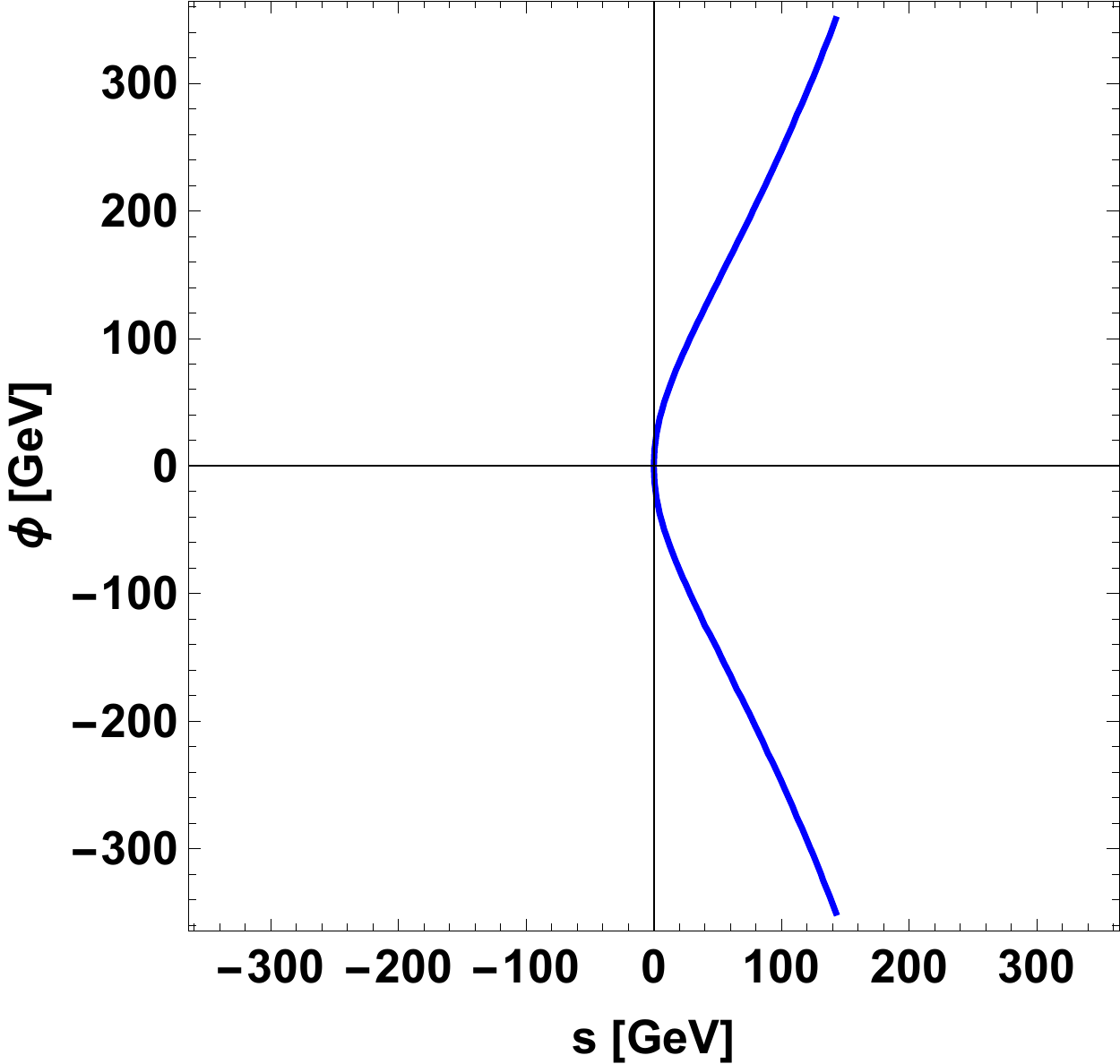}  } \\
      \hline \hline \hline
	  \shortstack{Type \\ {\small $(a, d)$} } & {\small discriminant $\Delta(\phi)$} & {\small $s$ curve {\tiny $\frac{\partial V}{\partial s} = 0$} }  & 
	  \shortstack{Type \\ {\small $(a, d)$} } & {\small discriminant $\Delta(\phi)$} & {\small $s$ curve {\tiny $\frac{\partial V}{\partial s} = 0$} }  \\
      \hline 
	  \shortstack{\textbf{A} \\ {\small $(+,+)$} \\ $\Delta'<0$ }  & \parbox[c]{1em}{
  		\includegraphics[width=1in]{fig/DelphiA.pdf}  } & \parbox[c]{1em}{
  		\includegraphics[width=1in]{fig/ScurveA.pdf}  } &
  	  \shortstack{\textbf{B} \\ {\small $(+,-)$}  \\ $\Delta'<0$ }  & \parbox[c]{1em}{
    		\includegraphics[width=1in]{fig/DelphiB.pdf}  } & \parbox[c]{1em}{
    		\includegraphics[width=1in]{fig/ScurveB.pdf}  } \\
  	  \shortstack{\textbf{E} \\ {\small $(-,+)$}  \\ $\Delta'<0$ }  & \parbox[c]{1em}{
  		\includegraphics[width=1in]{fig/DelphiE.pdf}  } & \parbox[c]{1em}{
  		\includegraphics[width=1in]{fig/ScurveE.pdf}  } &
  	  \shortstack{\textbf{H} \\ {\small $(-,-)$}  \\ $\Delta'<0$ }  & \parbox[c]{1em}{
	     \includegraphics[width=1in]{fig/DelphiH.pdf}  } & \parbox[c]{1em}{
	     \includegraphics[width=1in]{fig/ScurveH.pdf}  } \\
      \hline \hline
  \end{tabular}
  \end{center}
\end{table}

In the table, 
the middle column shows how the sign of $\Delta(\phi)$ varies with $\phi$ (red indicates a possitive discriminant while blue indicates a negative one). The typical shape of the corresponding $s$ curve is given in the last column. The first colume shows the condition for these types -- when the $\Delta'=b^2c^2-4ac^3-27a^2d^2+18abcd$ is positive, all the signs of $a,b,c,d$ should be specified; when $\Delta' < 0$, only $a,d$ need to be specified.

The upper half of the types (A,B,C,D) have positive coefficient $a$, and 3 asymptotic values at large $\phi$. For the lower half (E,F,G,H), where $a<0$, there is only one asymptotic value at large $\phi$. Also note that the left half of the types (A,C,E,G) have positive coefficient $d$, which leads to multiple branches (a branch is defined by a topologically connected part of the curve) in the small $\phi$ region. The right half, however, have negative $d$ and only one such branch. We name the branches present at small $\phi$ as "relevant branches". Usually, the "irrelevant branches", which only appear in large $\phi$ region like in type (B,C,D), are not important in our analysis. This reduces the set of types by identification $B\sim H$, $C\sim E$ and $D\sim F$. 

From the point of view of Lagrangian parameters, we find that the most important coefficients $a$ and $d$ are controlled by $\tilde{\lambda}_{s\phi}$ and $\tilde{\mu}_3^2 + 4\tilde{\lambda}_s\tilde{\mu}_s^2$. The latter changes with temperature mainly through Eq.~\ref{quadratic_T}. Moreover, the asymptotic value of the $s$ curve at large $\phi$ (the middle one when $a>0$) coincides with the center of $\phi$ curve $s_*$.

In the next section, we will show how the analysis of the $\phi$ curve and $s$ curve could help classify the phase transition patterns.
%


\section{Strong First Order Phase Transition}
\label{sec:phase}



Once we obtain the full effective potential, we could investigate how the vacuum state evolves with temperature.
At each temperature, we find the true vacuum by looking for the global minimum of the potential in the $(\phi, s)$ field space. 
It is known that at zero temperature, the global minimum of the scalar potential is at $(v_0, u_0)$, with $v_0=246$ GeV and $u_0$ as an input parameter.
%
After we turn on and increase the temperature, we track the position of the global minimum, seeking the sign of a phase transition.
In the context of effective potential, the fields are defined to be the first derivative of the free energy with respect to the corresponding particle source, and thus acts like an order parameter of the phase transition.
If the transition of the field values between the two phases is continuous, it is a second order phase transition.
Otherwise, if there is a discontinuous variation of the fields, it is a first order phase transition.
The first order phase transition proceeds by bubble nucleation of the broken phase at around the critical temperature.
The bubbles grow and coalesce, and finally turn the whole universe into the broken phase.

As discussed in the introduction, to have successful baryogenesis, it is essential to have a strong enough first order phase transition, so that $E_{\rm sph}(T_c)/T_c \ge 45$ inside the bubbles.
It has been shown in literature~\cite{Ahriche:2007jp} that for a singlet extended model the sphaleron energy is approximately proportional to the $\phi$ vev: $E_{\rm sph}(T_c) \sim v_c$.
Moreoever, the bubble expansion and wall velocity in the singlet extended model have been discussed in Ref.~\cite{Kozaczuk:2015owa}.
In our model there is a similar scalar sector. Thus the discussions about the bubble expansion and sphaleron process in literatures are also applied to our model. 
Therefore, similar to singlet extended model, we add the following criterion to our scan to pick out the events of successful baryogenesis:
\bea
	\xi = \frac{v_c}{T_c} \ge 1.
\eea
We also perform a consistent check by calculating the sphaleron profiles and the sphaleron energy at the critical temperature numerically. 

\begin{figure}[!htb]
\begin{center}
\includegraphics[width=0.4\textwidth]{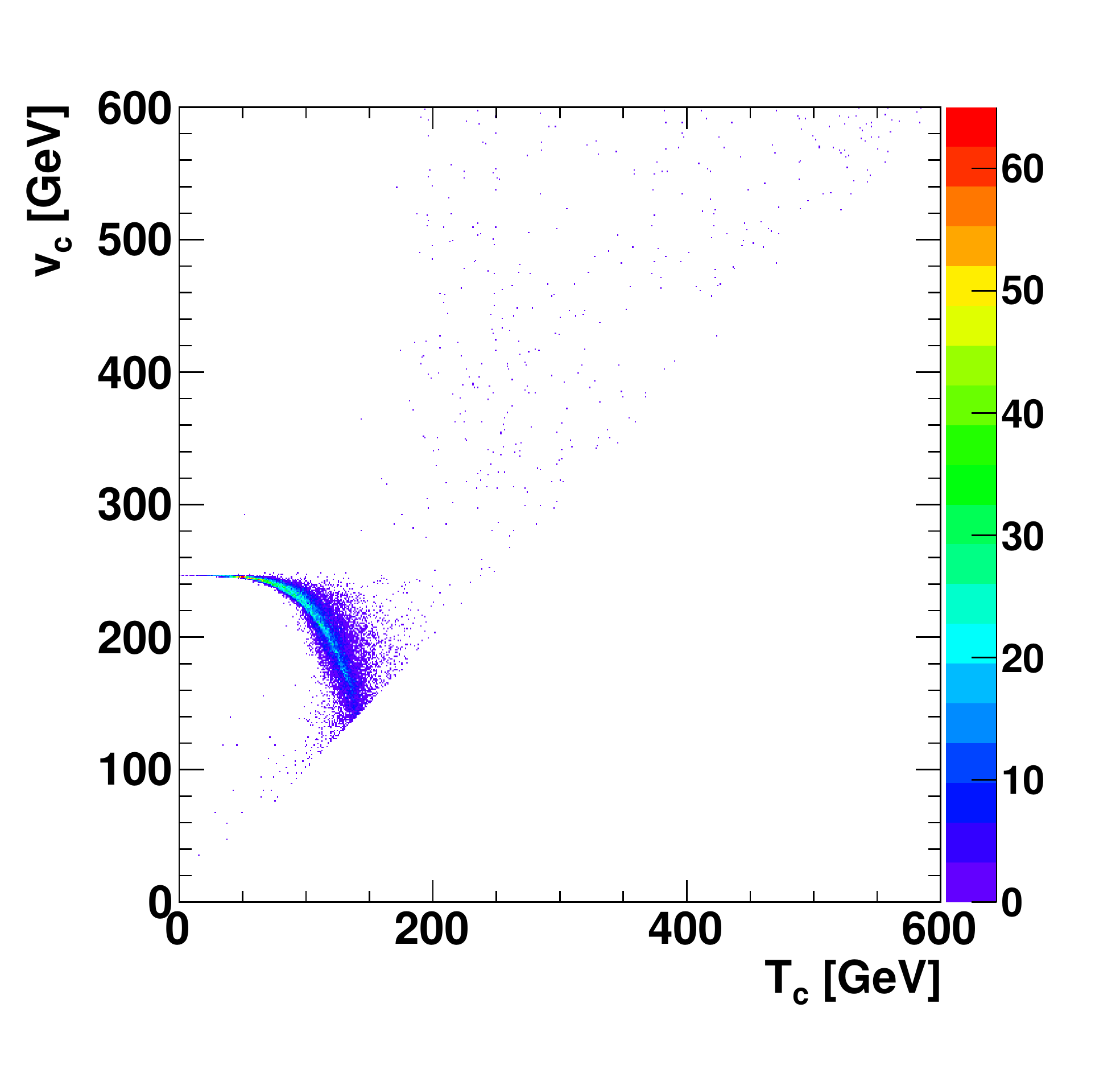}
\caption{\small The allowed value of the critical temperature $T_c$ versus the $\phi$ vev $v_c$ at the critical temperature from a random scan over the parameter space. The scatter points are selected to satisfy the SFOPT in our random numerical scan. The color palette on the right shows the density of the scatter points in one GeV interval. } 
\label{fig:TcvcFOPT}
\end{center}
\end{figure}

To determine the parameter region in which the strong first order phase transition could happen, we perform a random scan over the parameter space. 
The procedure is the following.
We have quite a few independent parameters 
\bea
	 \lambda_s, \quad \lambda_{s\phi}, \quad \mu_3, \quad  \mu_{s\phi}, \quad M, \quad  y', \quad y_s,
\eea
in addition to the singlet vev $u_0$ in the zero temperature. 
The parameters $\lambda_\phi$ and $y_t$ are determined by the Higgs mass and the top quark mass that are already known. 
We choose the input parameters from the ranges
\footnote{
When the singlet vev $u_0$ was chosen to be larger than 600GeV, the data with critical temperature larger than 200GeV, and a $v_c$ even larger, accumulates. They are mostly the case IIIB as introduced later, and are not our focus in this paper. That's why we chose a smaller range for $u_0$. }
\bea
	 &|\lambda_{s\phi}| \le 1.5, \,\, 0 < \lambda_{s}  \le 2,\,\,  |y'| \le 1.5, \,\,|y_s| \le 1.5, \\
	 &|u_0| \le 600, \,\, |\mu_3| \le 800,   \,\, | \mu_{s\phi}| \le 1000,  \,\, 0 \le M \le 1200. 
\eea
Given the input parameters, the full effective potential in the field space is calculated.
Then for each temperature, we utilize the MINUIT subroutine~\cite{James:1975dr} to find the global minimum of the effective potential.
As the temperature increases, we track the change of the global minimum at each step in our numerical scan.
Additional care should be taken: if the minimum moves to very large field values, it may indicates the vacuum instability.
If the global minimum becomes the symmetric one $(0, \langle s\rangle )$ at certain temperature, 
we perform a fine scan near the temperature until we find the critical temperature $T_c$ and the corresponding vevs $(v_c, u_b)$ in the broken phase.
After obtaining the $T_c$ and $v_c$, we use the washout condition to pick out the strong first order phase transition, eliminating the data points that has $\xi < 1$.
We randomly scan $10^6$ parameter points, among which 25818 parameter points pass all the requirements.
%
%
Figure~\ref{fig:TcvcFOPT} shows the distribution of the successive data points in the $T_c-v_c$ plane. 
From the figure we notice that for the parameter region we scanned, the critical temperature is typically less than 200 GeV, while $v_c$ is smaller than its zero-temperature value $v_0=246$ GeV. 
As expected, the higher the critical temperature is, the smaller the $\phi$ vev gets to before the transition, and this correlation is clearly shown in the figure.
Furthermore, this range of critical temperature is quite safe from the bound of the CP violation strength that we discussed in Sec.\ref{sec:cpv}.

\subsection{Phase Transition Patterns}

\begin{figure}[!htb]
\begin{center}
\includegraphics[width=0.4\textwidth]{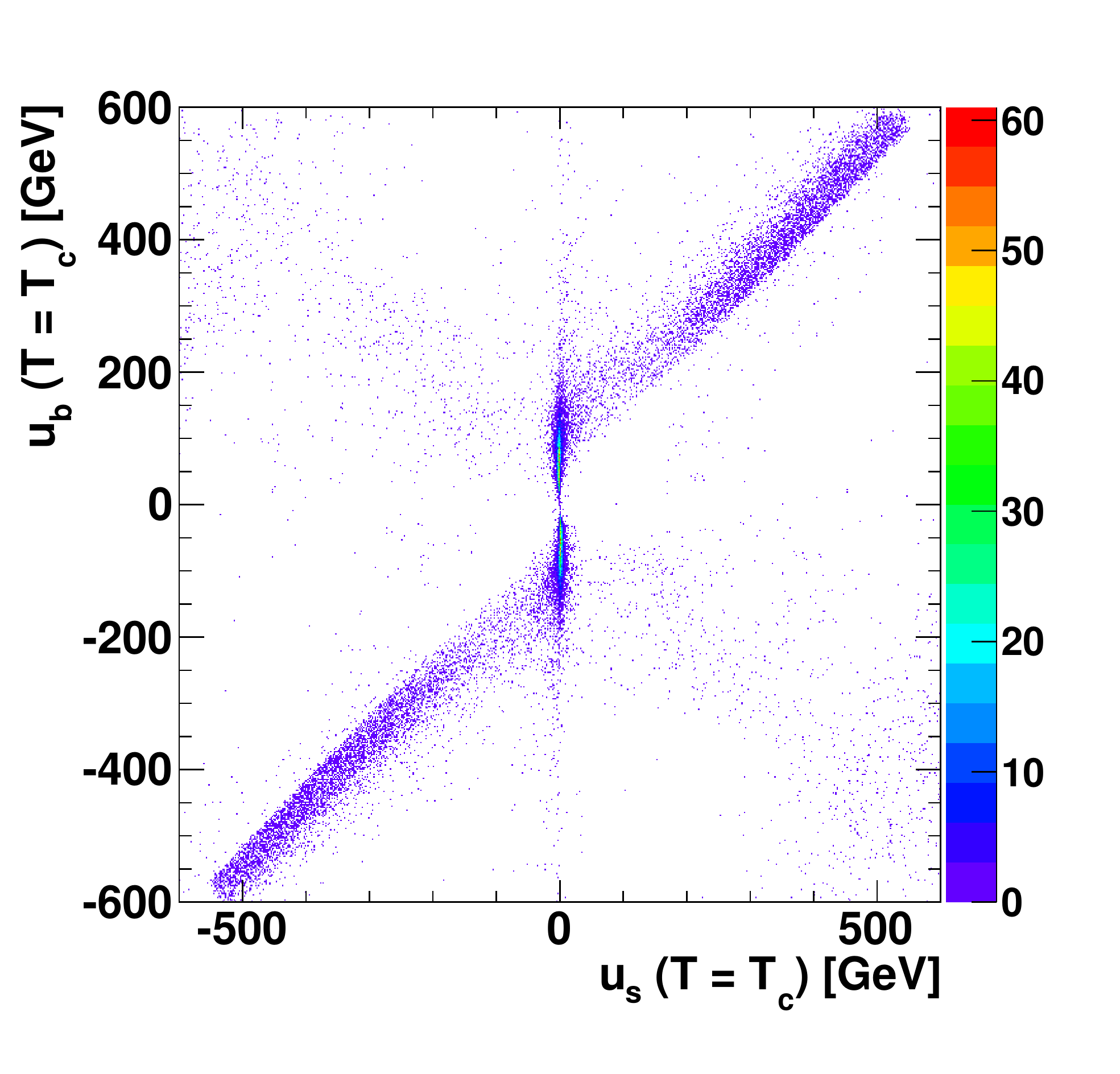} 
\includegraphics[width=0.4\textwidth]{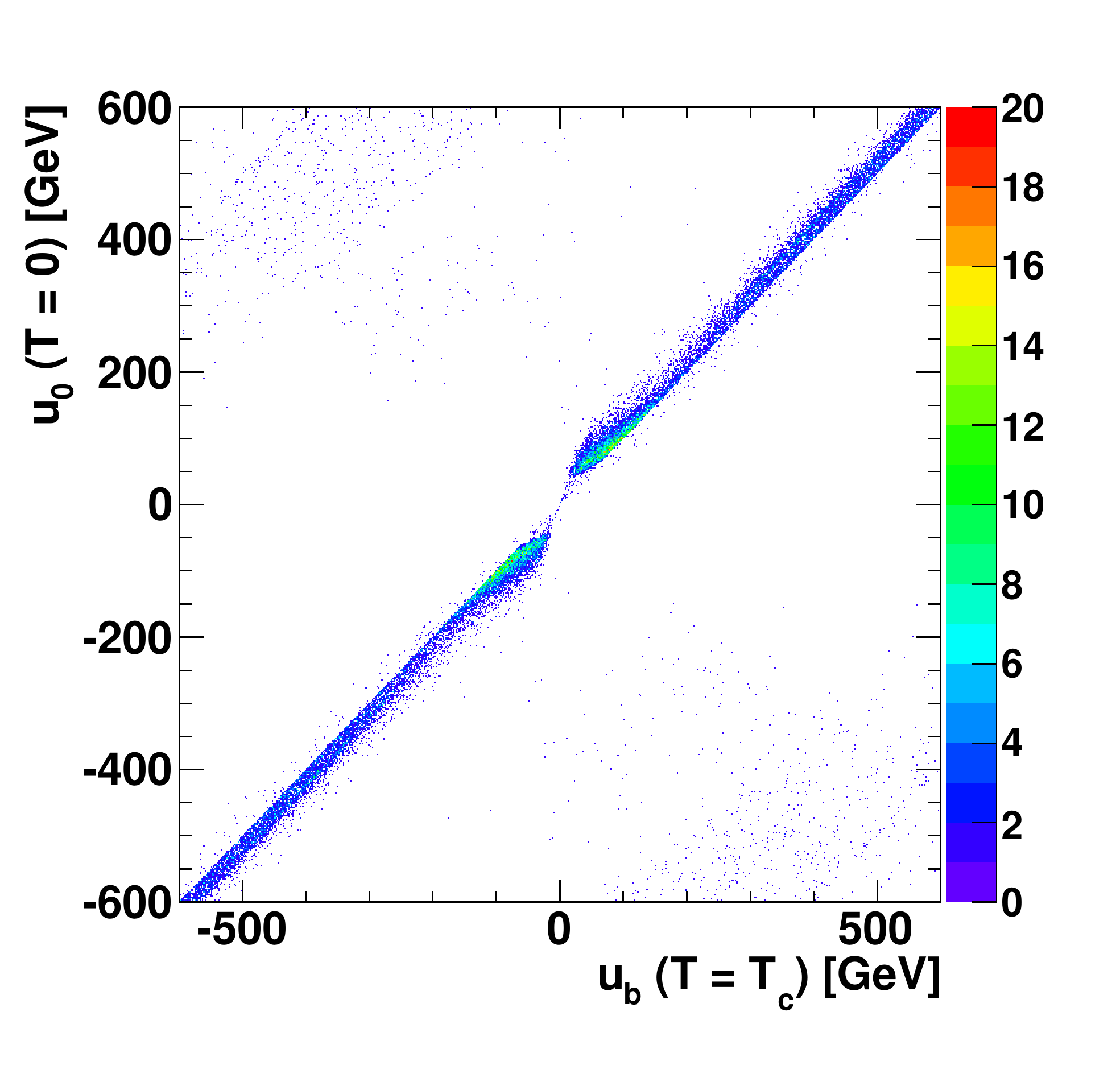} 
\caption{\small The allowed value of the contour $(u_s, u_b)$ (left) and $(u_b, u_0)$ (right) from a random scan over the parameter space.  The color palette on the right shows the density of the scatter points in one GeV interval. 
}
\label{fig:scansssb}
\end{center}
\end{figure}

The patterns of phase transition are described by the critical values of the scalar fields during the transition, defined as $(0,u_s)$ and $(v_c,u_b)$ in the previous section.
In the left panel of Figure~\ref{fig:scansssb}, we show the correlation between $u_b$ and $u_s$, indicating the $s$ value jump during the phase transition.
We notice that there are several distinct allowed regions: 
Region I: the vev $u_s$ is very close to zero while $u_b$ is non-zero; Region II: the vev $u_s$ is non-zero linearly correlated with the $u_b$; Region III: both the vev $u_s$ and $u_b$ are scattered over the second and fourth quadrants. 
The Region III can be further classified by the correlation between $u_s$ and $u_0$, which characterizes the evolution of $s$ value before the transition. 
This is shown in the right panel of the Figure~\ref{fig:scansssb}. 
Typically, the vev $u_0$ and $u_b$ are strongly correlated, because there is not much time for $s$ to move very far without jumping. 
However, there are also cases where $u_0$ and $u_b$ are quite uncorrelated. 
This indicates that between the zero temperature and the critical temperature, there should be another phase transition, though the $Z_2$ symmetry is not restored yet. 
Further analysis shows that the region that the $u_0$ and $u_b$ are not linearly correlated only appear in Region III.
So we separate the Region III into two subregions:
one-step phase transition (IIIa) and multi-step phase transition (IIIb). 
Hopefully, these four regions of parameters correspond to four different phase transition patterns.

\begin{figure}[!htb]
\begin{center}
\includegraphics[width=0.3\textwidth]{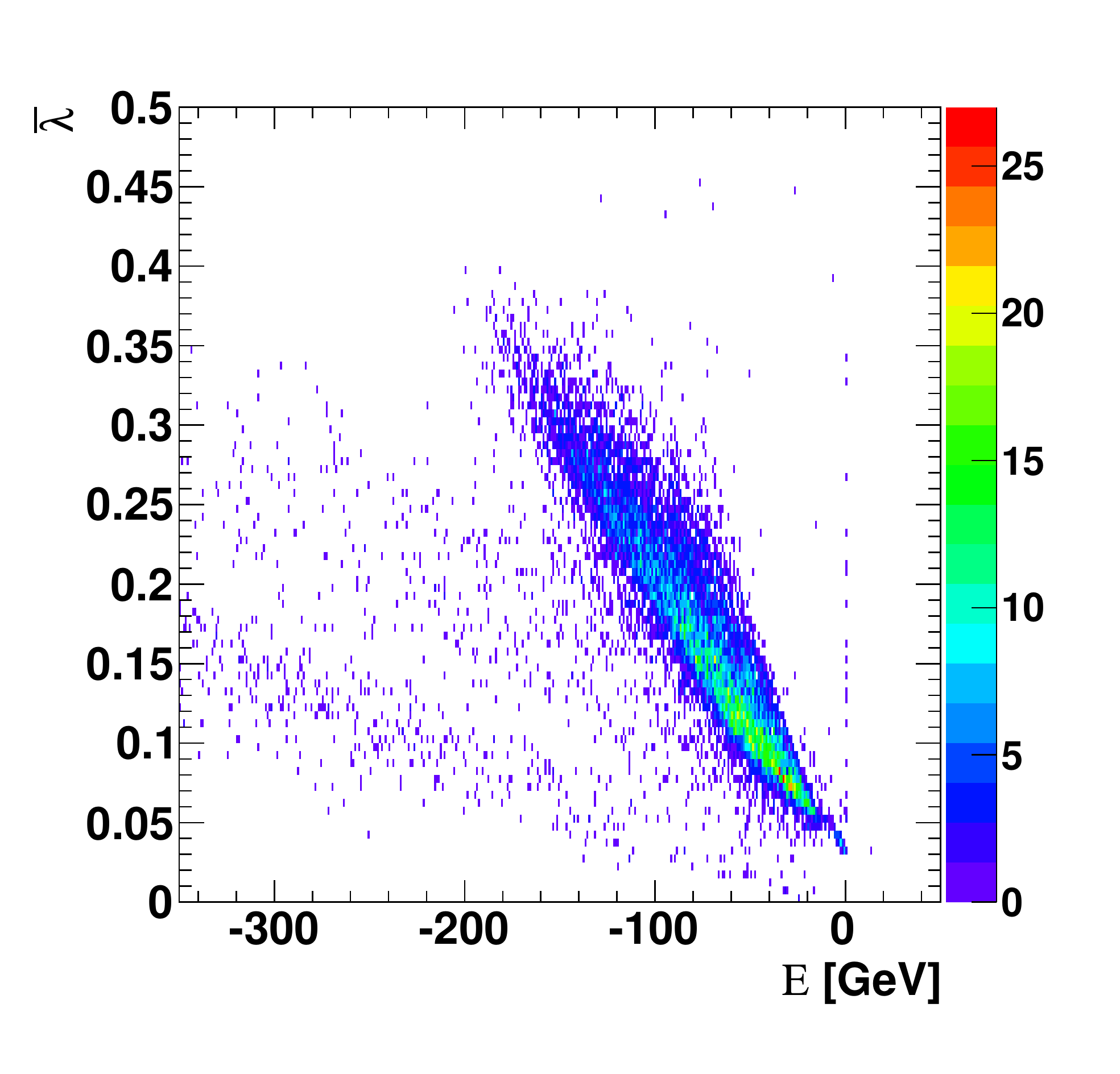} 
\includegraphics[width=0.3\textwidth]{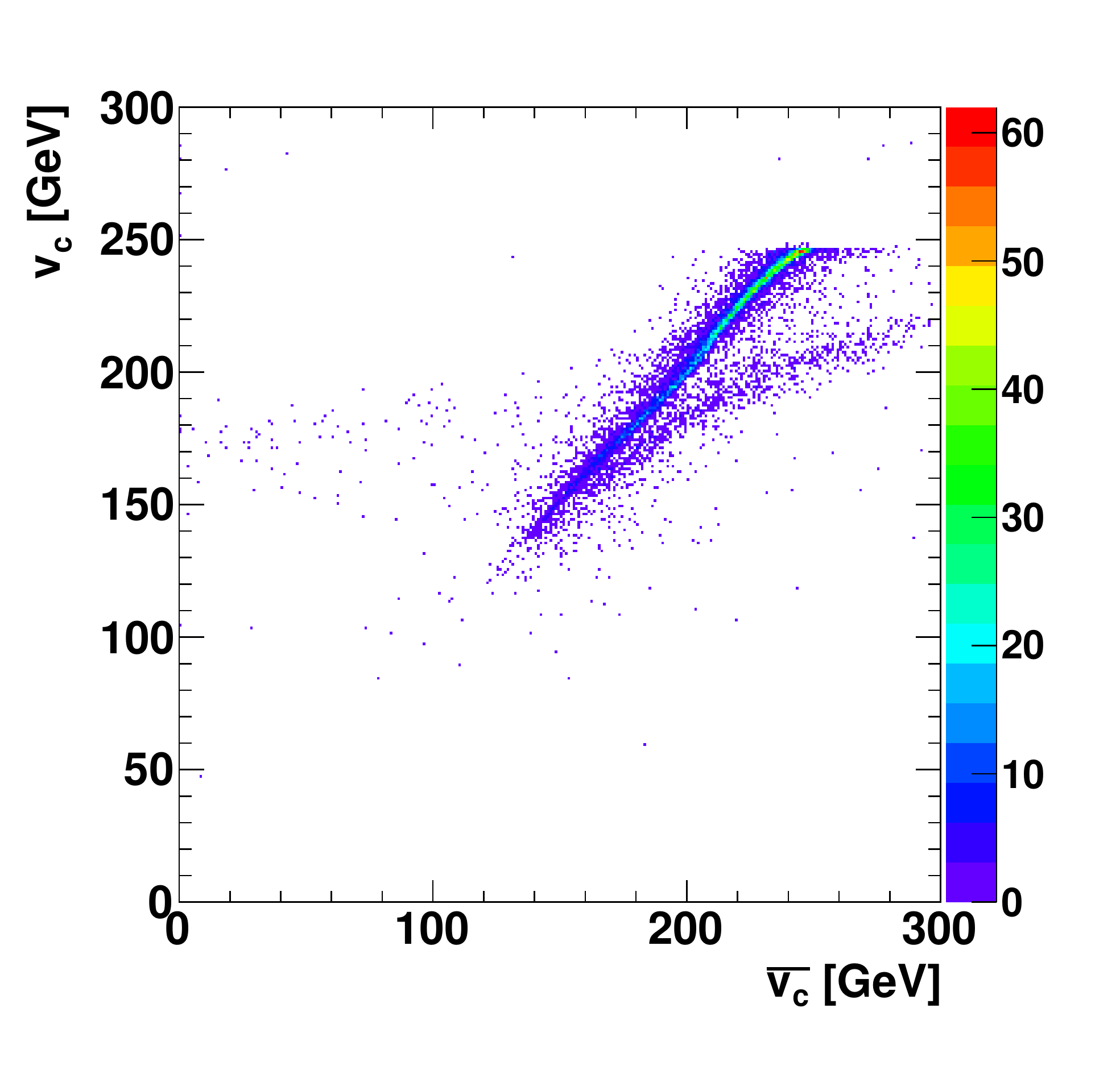} 
\includegraphics[width=0.3\textwidth]{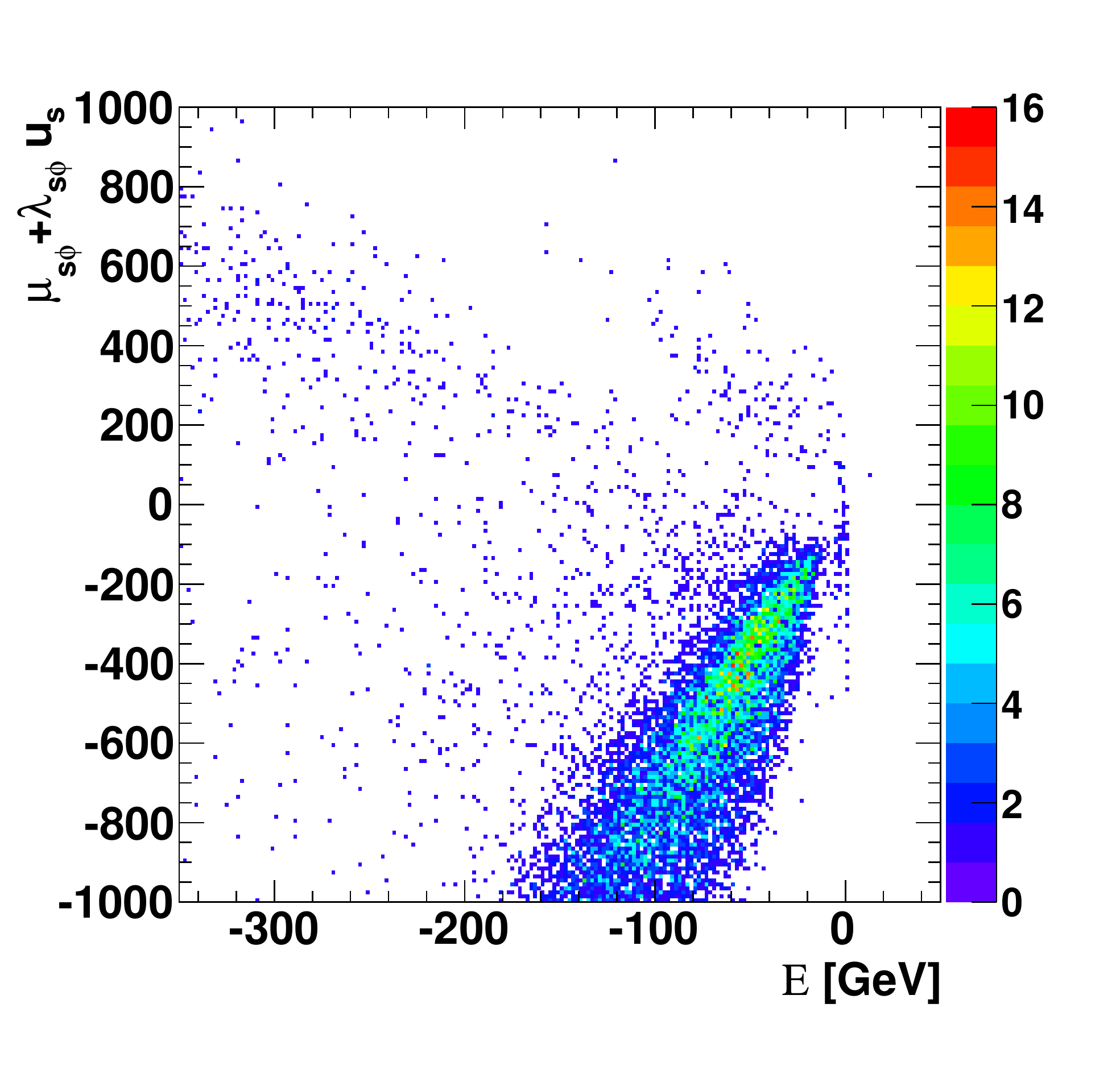} 
\caption{\small The allowed value of the contour $({\mathcal E}, \bar{\lambda})$ (left), $(\bar{v}_c, v_c)$ (middle) and $({\mathcal E}, \mu_{s\phi}+\lambda_{s\phi}u_s)$ (right) from a random scan over the parameter space. The color palette on the right shows the density of the scatter points in one GeV interval. 
}
\label{fig:scanElamda}
\end{center}
\end{figure}

In order to help estimating the parameter preferences, we also utilize the polar coordinates of the scalar fields as in Eq.~\ref{eq:Veffradial}.
In terms of the non-zero vacuum value $\bar{\rho}$, we obtain the $\phi$ value for the approximate potential:
\bea
\bar{v}_c = \bar{\rho}c_{\alpha} = -\frac{2{\cal E}}{\bar{\lambda}}c_{\alpha}.
\label{eq:vcbar}
\eea
A necessary condition of having strong first order phase transition is that $\bar{v}_c$ is positive and large. 
However, the expressions of the $\cal E$ and $\bar{\lambda}$ cannot be expressed as functions of only the model input parameters. They also involve the information of the features of the phase transition, such as the angle $\alpha$ and the symmetric minimum $u_s$. Therefore, we would like to further utilize the classification of the transition patterns to get more information on these features.

Several approximations can be made here, according to the correlations between the parameters that we found. First, in the Region II and perhaps also Region I, where $u_s$ and $u_b$ differ only a little, we thus assume that $\alpha \ll 1$. For Region I, $u_s$ is very small in comparison with all the other massive parameters. After these approximations, the relations between $\bar{v}_c$ and the couplings will be more manifest.
In Figure~\ref{fig:scanElamda} we plot the comparison between the $\bar{v}_c$ and the true $v_c$ that we obtain from the scan (left), and the correlation between $\cal E$ and $\mu_{s\phi}+\lambda_{s\phi}u_s$ (right). The left panel tells us that in most cases $\bar{v}_c$ is a good approximation to the true value. The right panel shows the correlation between $\cal E$ and its main part under the assumption of $\alpha \ll 1$.

In order to analyze the four regions in detail, we also have to rely on the shape of the scalar potential. The $\phi$ curve and $s$ curve at the critical temperature help us understand the origins of the different patterns, as they determine the distribution of the potential minima.   
The $s$ curve could have one, two or three relevant branches, depending on the types as shown in Table~\ref{tab:scurve}. Therefore, it could have one or three intersections with the line $\phi=0$. We call the minima along the line $\phi=0$ "symmetric minima", while others are called "broken minima". 
Next, we notice that the broken minimum must be the intersection between the quadratic branch of the $\phi$ curve and one of the $s$ curve branches. This $s$ branch must have an intersection with the $s$ axis, which may be a symmetric minimum. We would like to define the barrier between the broken minimum and this symmetric minimum as a "single-branch barrier", while those between minima on different branches are called "inter-branch barrier". We discovered that the single-branch barrier usually has much smaller width along $s$ direction than the inter-branch barriers, due to the limited stretch of the $s$ curve along the $s$ direction. It implies that a transition through the single-branch barrier should have closely related $u_s$ and $u_b$. Finally, in terms of the above features of the shape of potential, let us discuss the four phase transition patterns in detail. 

\subsubsection{Pattern I: Single-branch barrier transition, with $u_s\sim 0$}

According to the discussion in Sec.\ref{sec:potential}, there are three cases for the small $\phi$ behavior of the $s$ curve, one of which is usually negligible. In the other two significant cases, the one with three roots will be discussed in the next part. Let's focus on the other case, which has only one relevant branch that intersects $\phi=0$ at around $(0,0)$.
It indicates that this case mainly corresponds to the Region I of the parameter scan. 
%
%

%
%
%
%
%
%
%

\begin{figure}[!htb]
\begin{center}
\includegraphics[width=0.4\textwidth]{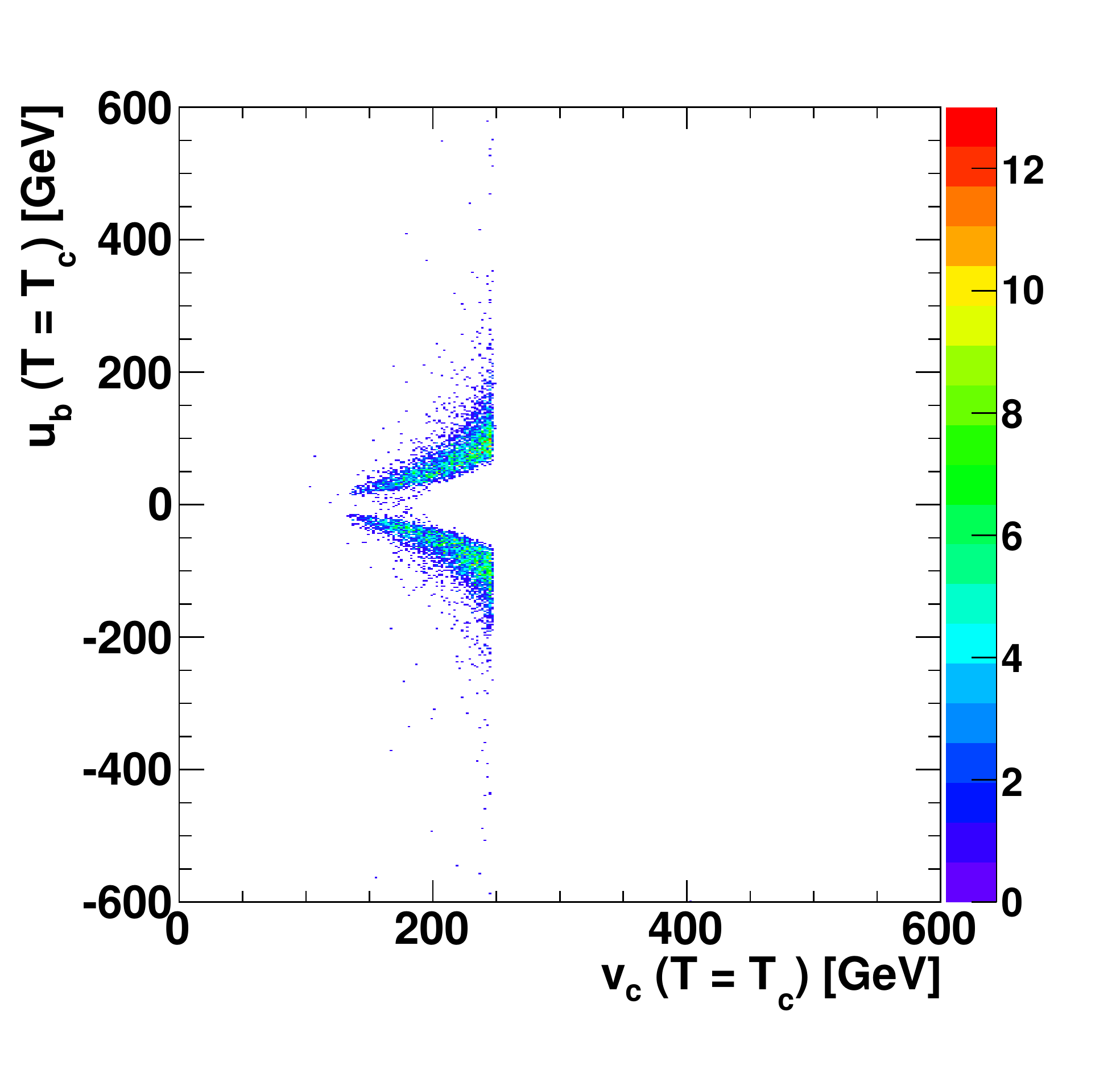} 
\caption{\small The allowed values of the broken minimum $(v_c, u_b)$ in pattern I. Here the phase transition happens between the symmetric minimum $(0, 0)$ and the broken minimum. The color palette on the right shows the density of the scatter points in one GeV interval. 
}
\label{fig:scan1vcsb}
\end{center}
\end{figure}

In Figure~\ref{fig:scan1vcsb}, we show the allowed values of the broken minimum $(v_c, u_b)$ in this case, while the symmetric minimum is always at $(0,u_s\sim 0)$. The value of $v_c$ is upper-bounded by its zero-temperature value $v_0=246$ GeV, implying a decrease of $\phi$ before the transition, and is also lower-bounded at about $100$ GeV by the condition $\xi>1$. The distribution of the broken minimum clearly sketches the shape of the $s$ curve.

\begin{figure}[!htb]
\begin{center}
\includegraphics[width=0.3\textwidth]{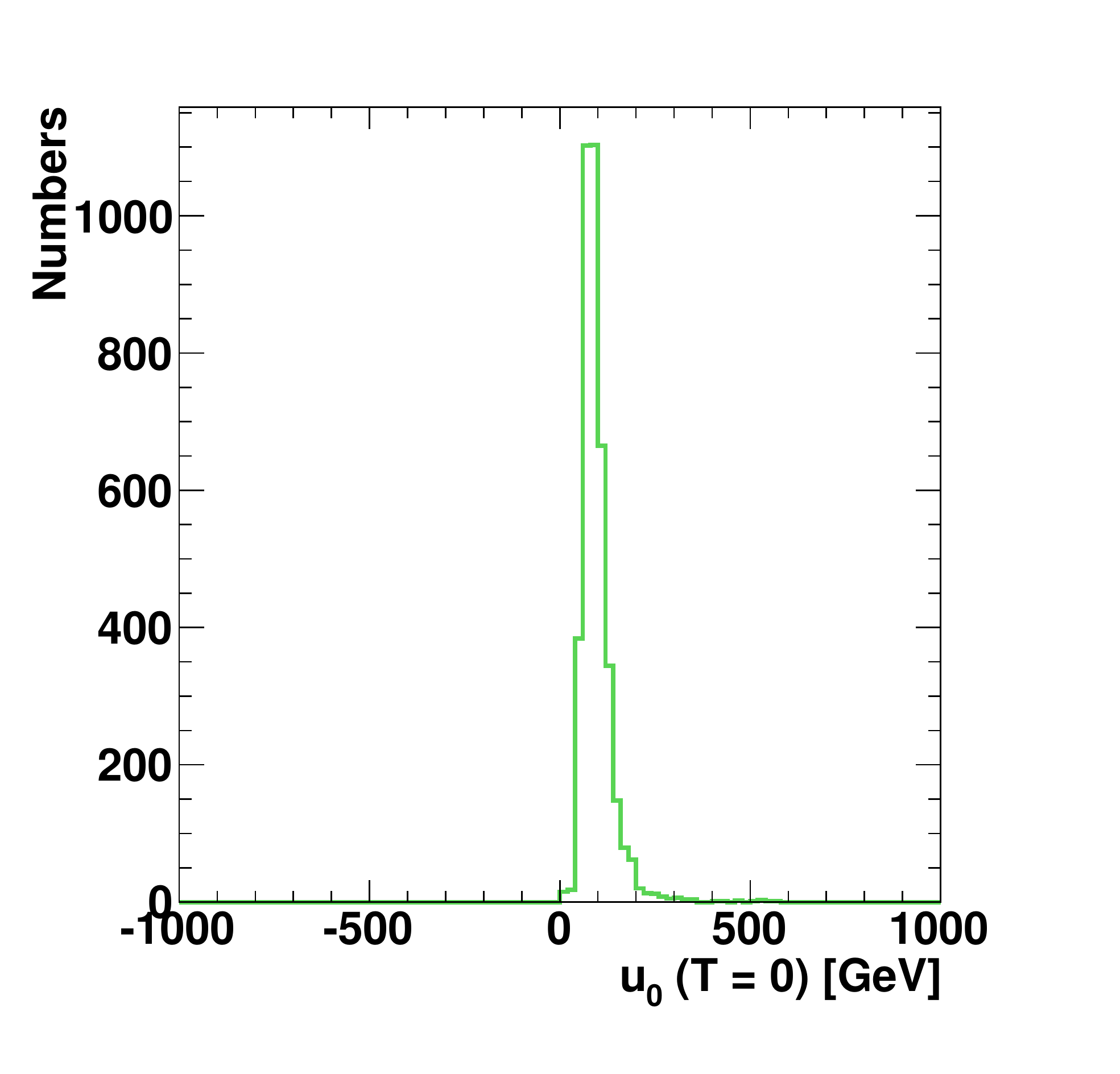} 
\includegraphics[width=0.3\textwidth]{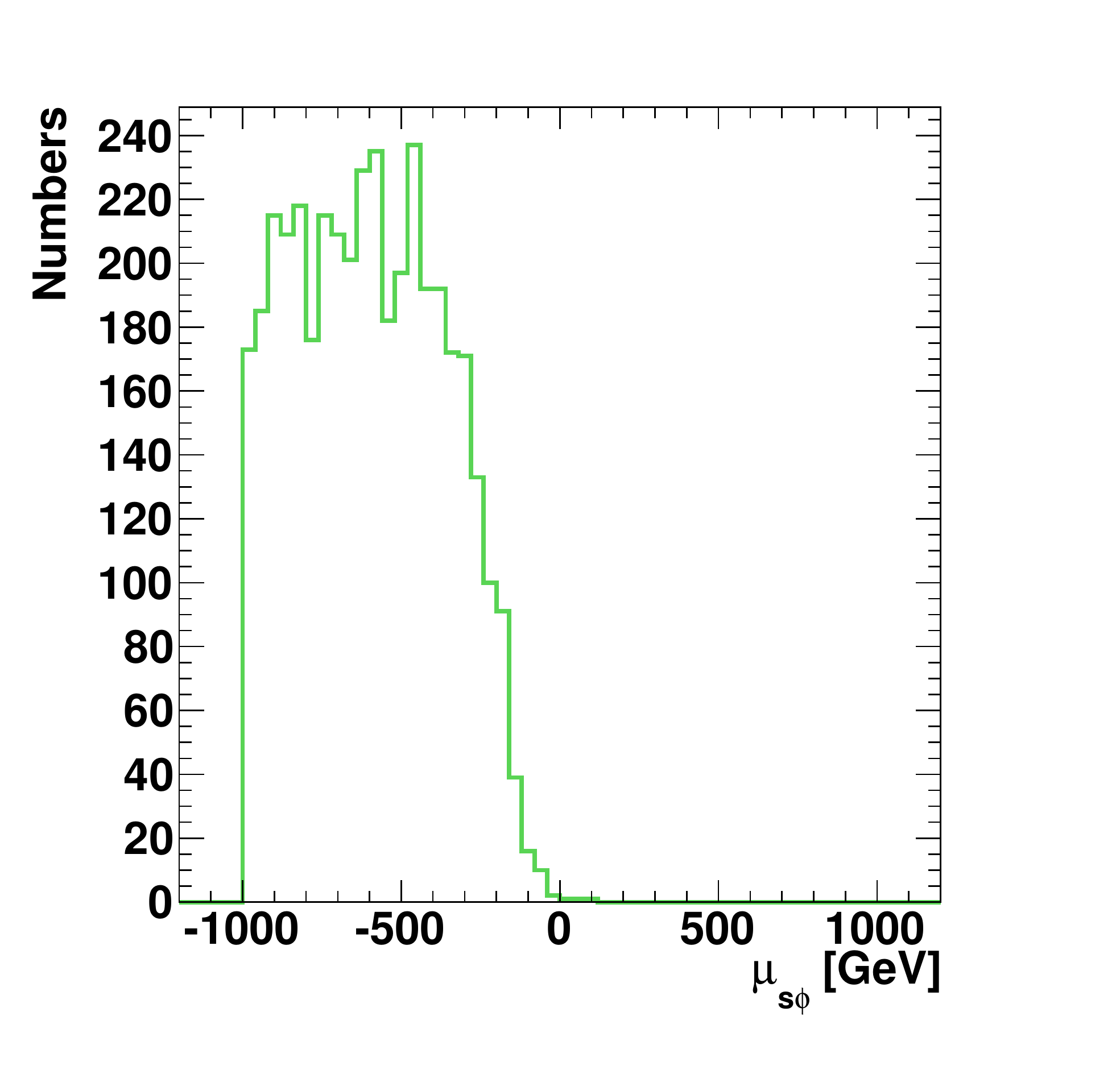} \\
\includegraphics[width=0.3\textwidth]{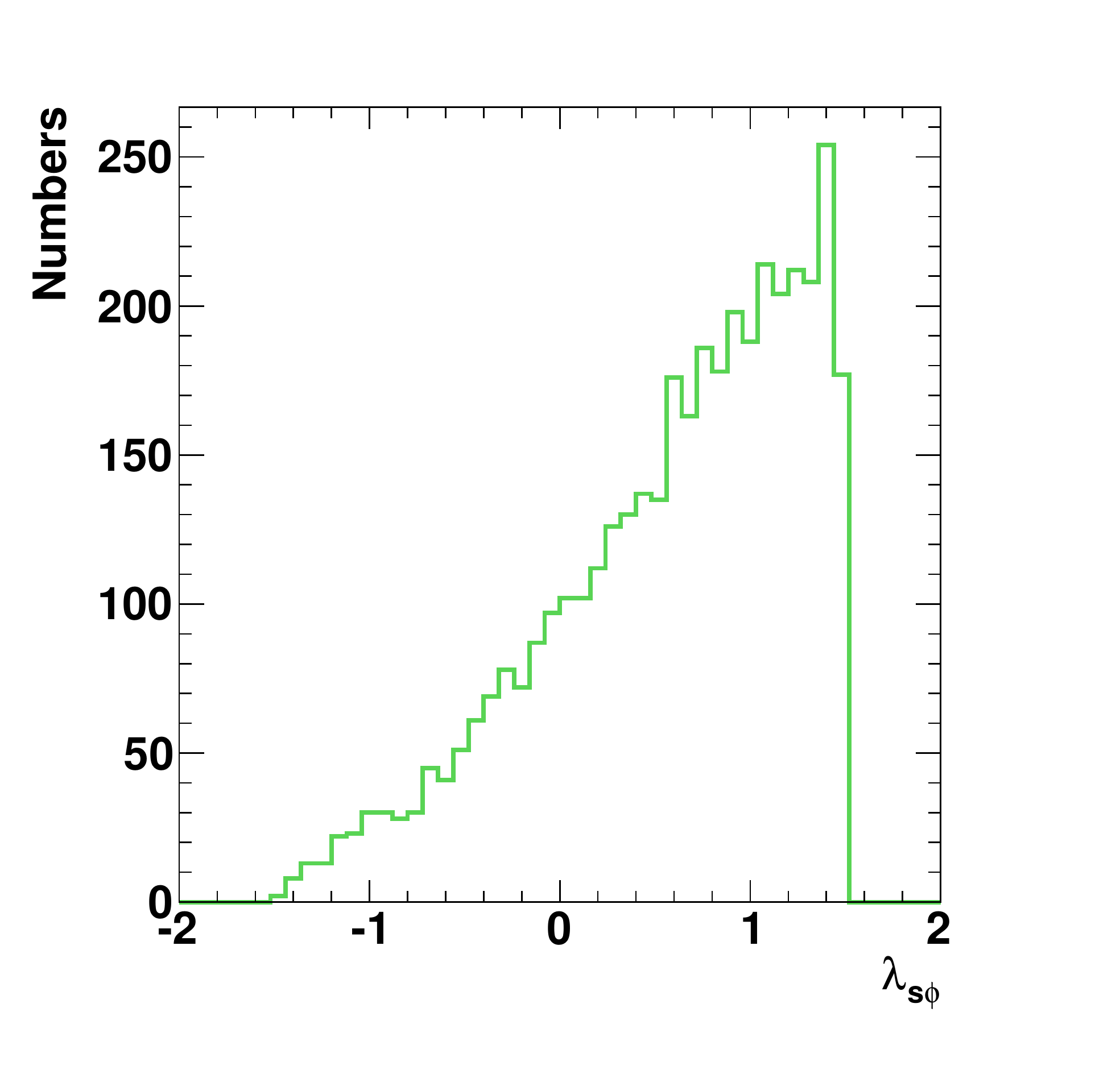} 
\includegraphics[width=0.3\textwidth]{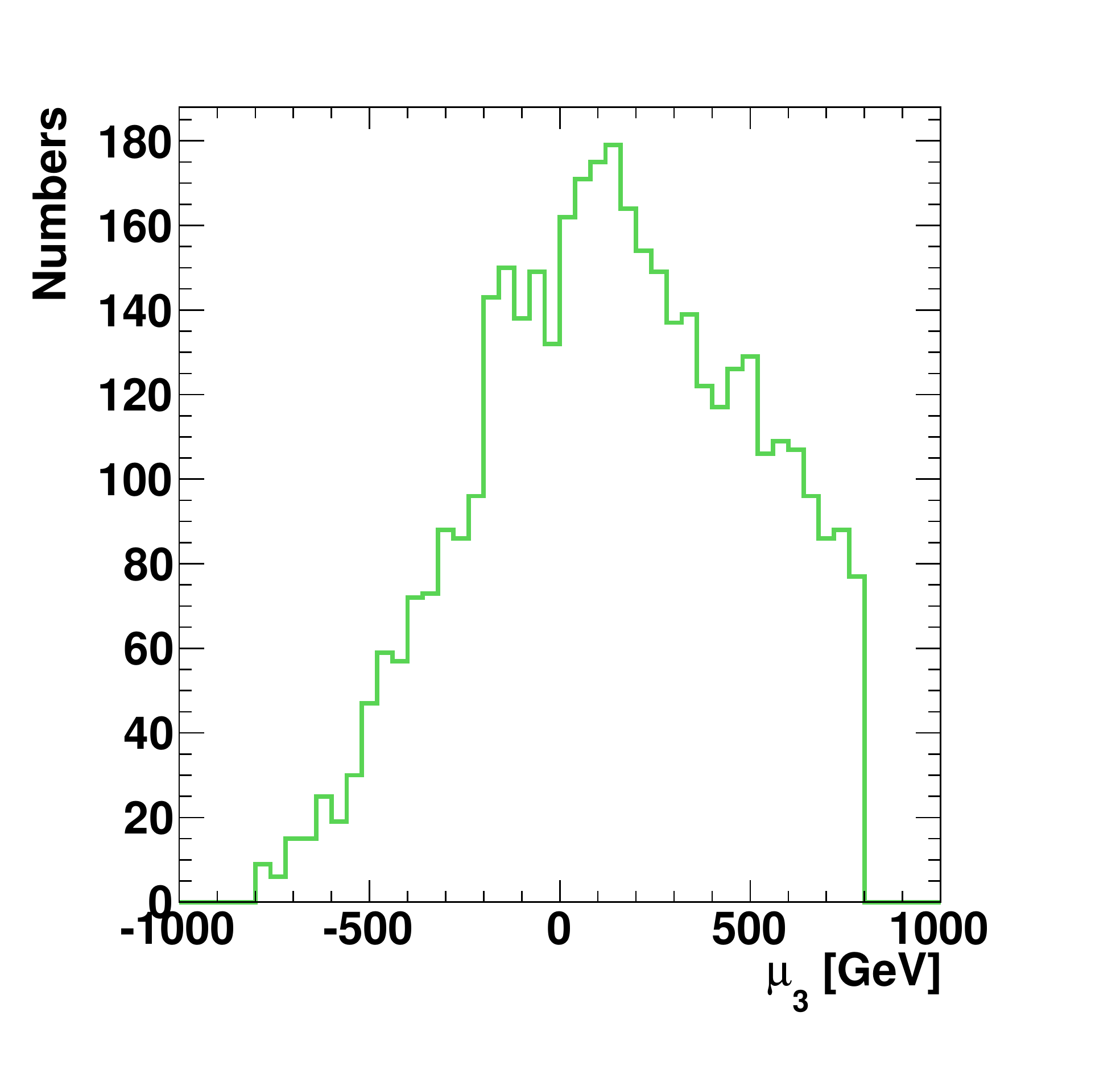} 
\caption{\small In the pattern I with $u_0 > 0$,  the allowed values of the model parameters $u_0$, $\mu_{s\phi}$,  $\lambda_{s\phi}$ and $\mu_3$ are shown. The color palette on the right shows the density of the scatter points in one GeV interval. 
}
\label{fig:scan1musph3}
\end{center}
\end{figure}

Let's see what can be inferred for the model parameters. Without loss of generality, we choose the benchmark points with only positive $u_0$, to investigate the sign preferences of other parameters. Here are the observations:
\begin{itemize}
	\item As shown above, the $s$ curve is a single branch curve across the point $(0,0)$, which indicates that the zero-temperature vev $u_0$, which is on the same branch, should also be small, but not zero due to the bend of the curve. 
	
	\item In the light of the previous discussions, there are several approximations we can employ for $\cal E$: $u_s \approx 0$ and $\alpha \ll 1$. As a result, the only important term in $\cal E$ is the $\mu_{s\phi}$ term:
	\bea
	{\cal E} \approx \frac{1}{2}\mu_{s\phi}s_{\alpha}c_{\alpha}^2.
	\eea
	 In order to get a large and negative $\cal E$, large and negative $\mu_{s\phi}$ should be favored.

	\item We would like to argue that positive $\tilde{\lambda}_{s\phi}$ leads to shapes of the curves that are much more favored by the strong first order phase transition. One may notice that for negative $\tilde{\lambda}_{s\phi}$, both $\phi$ curve and $s$ curve are hyperbola-like. Two hyperbolas could not make the twisting intersection needed for the existence of degenerate minima. Although the parameter $\tilde{\mu}_{s\phi}$ causes a deviation from perfect hyperbola of the $s$ curve, it's still harder for such case to have first order phase transition. Therefore, positive $\lambda_{s\phi}$ should be strongly preferred.
	
	\item The one-branch condition for $s$ curve requires that the coefficient $d$ in the polynomial $\Delta(\phi)$ is negative, thus $\tilde{\mu}_s^2 < -\frac{\tilde{\mu}_3^2}{4\tilde{\lambda}_s}$. Large $\tilde{\mu}_3^2$ would compress the parameter region that satisfies this condition. Thus in this pattern, we expect that small $\tilde{\mu}_3^2$ is favored.
\end{itemize}
Fortunately, our numerical results from the scan do exhibit the above features, as shown in Figure~\ref{fig:scan1musph3}. 

\begin{figure}[!htb]
\begin{center}
\includegraphics[width=0.3\textwidth]{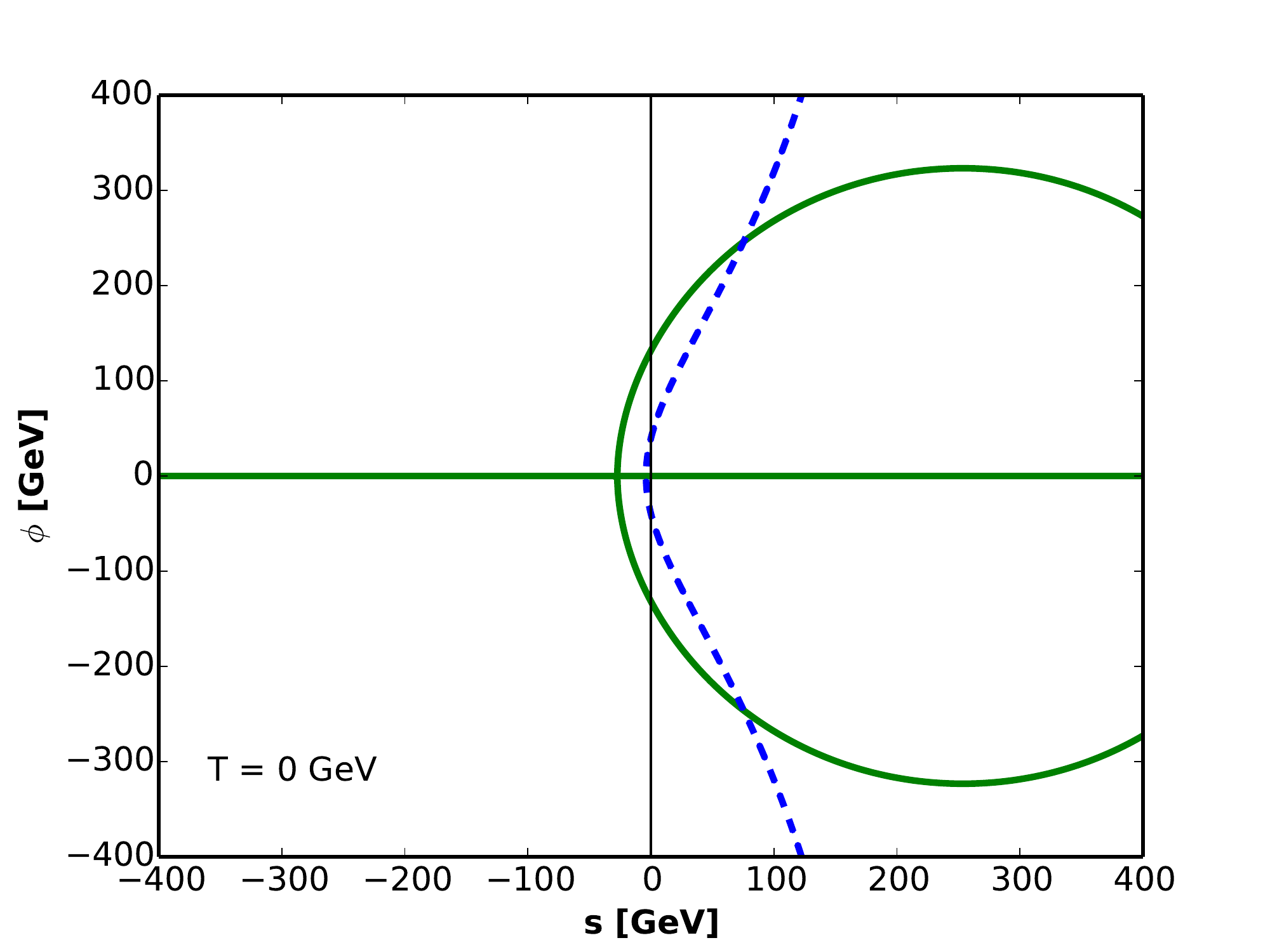} 
\includegraphics[width=0.3\textwidth]{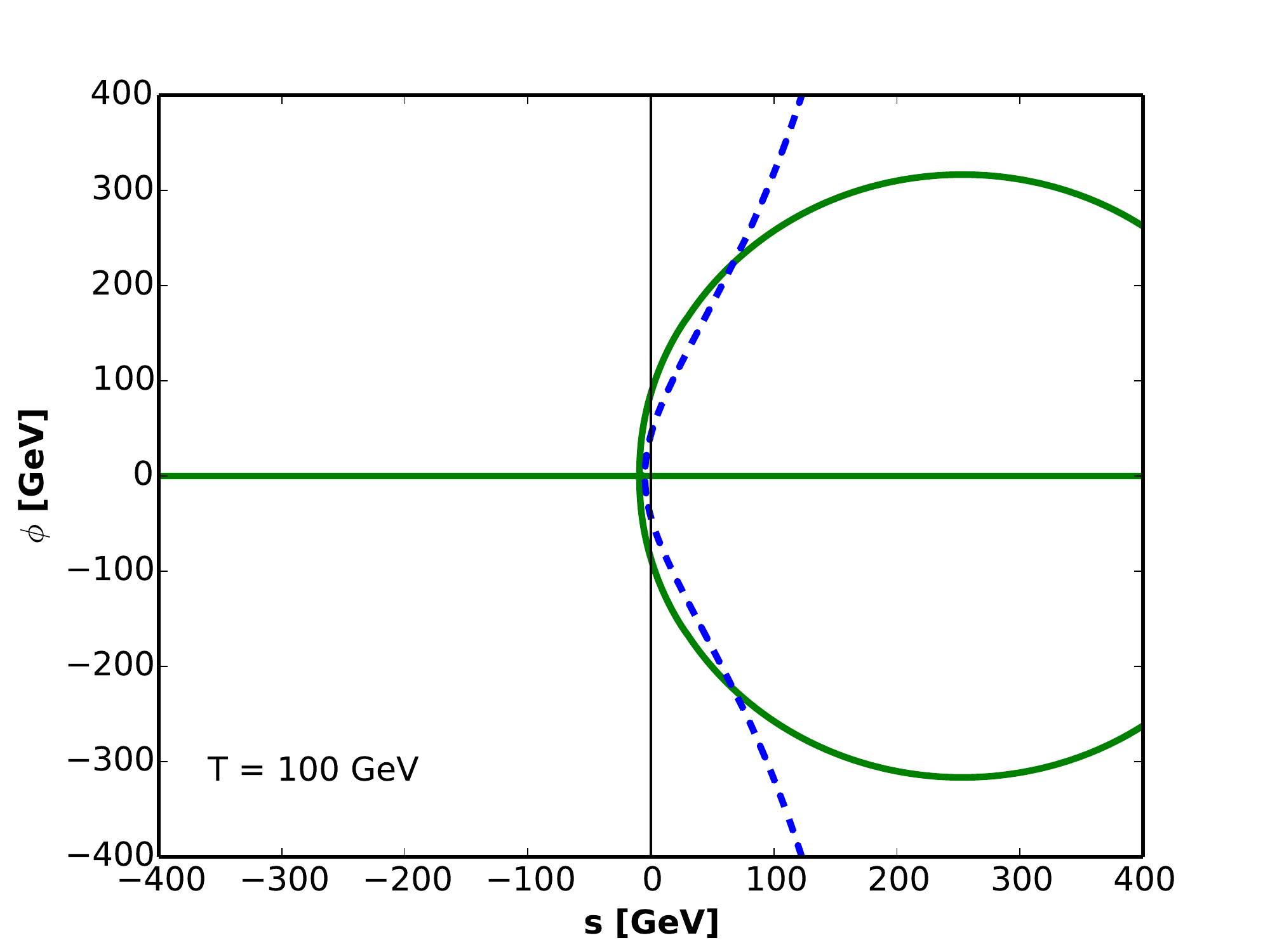} \\
\includegraphics[width=0.3\textwidth]{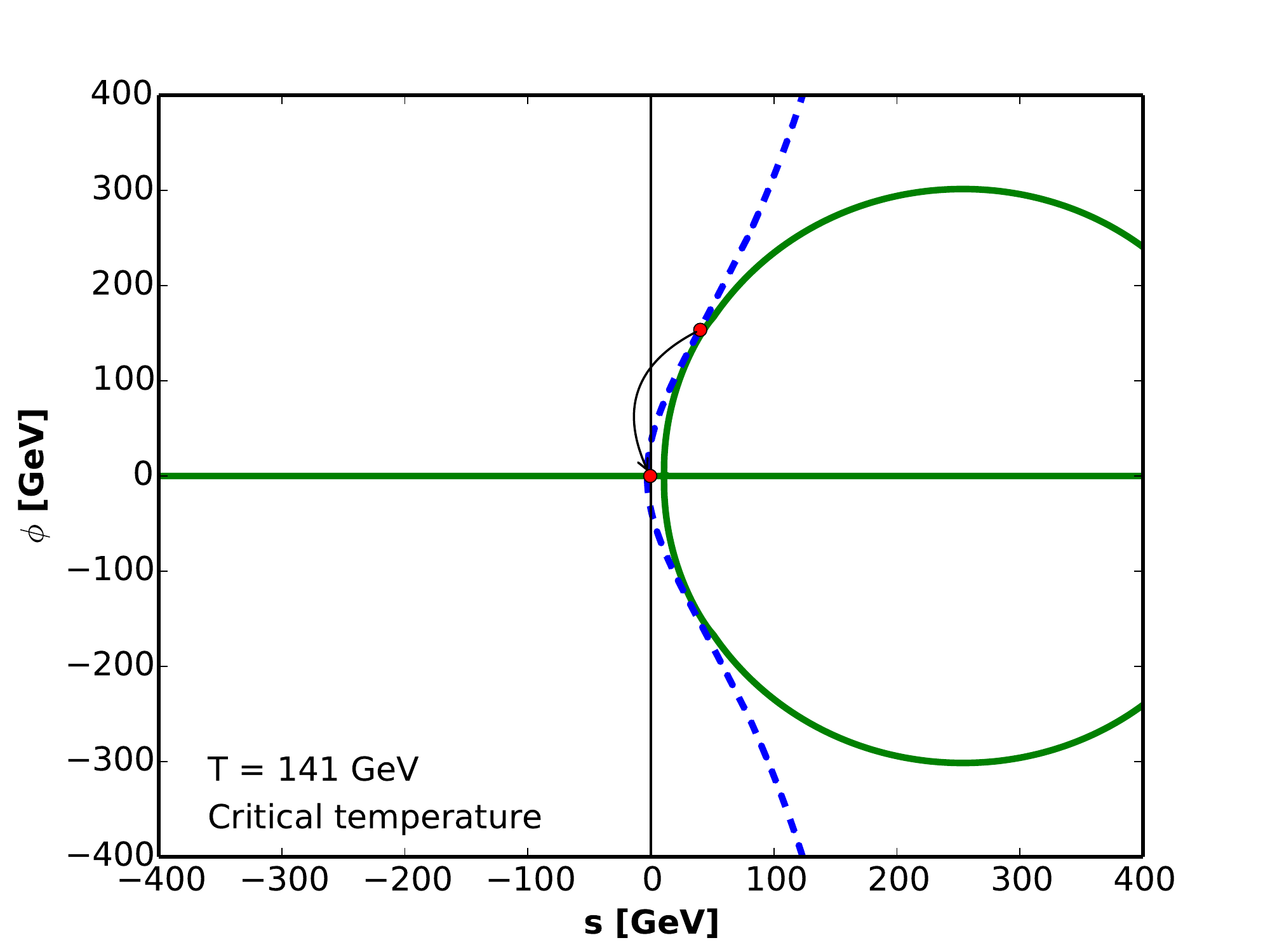} 
\includegraphics[width=0.3\textwidth]{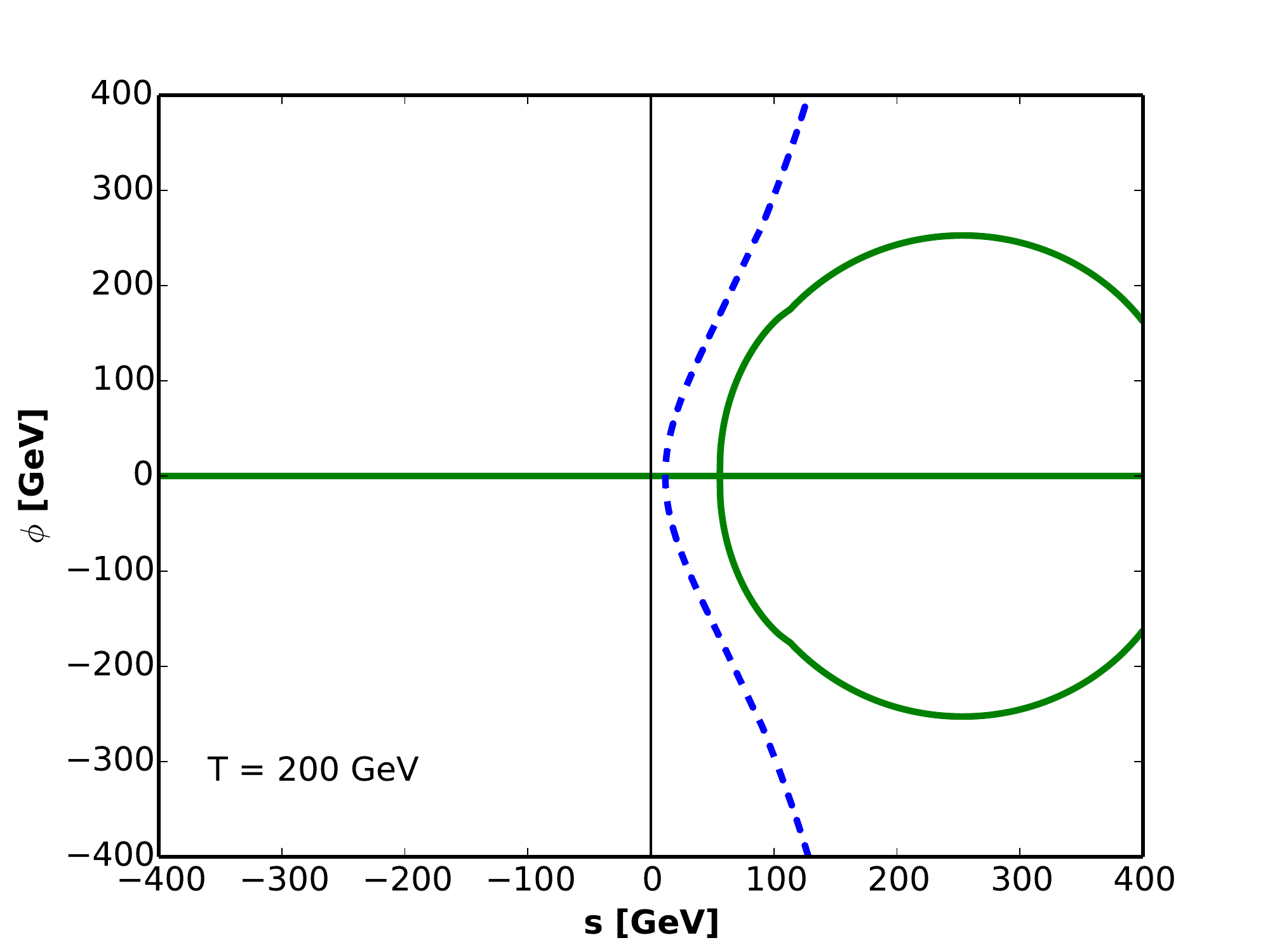}
\caption{\small In the phase transition pattern I, the $s$ curve (dashed blue) and $\phi$ curve (solid green) in the different temperatures: 
zero temperature, below critical temperature, at the critical temperature, and 
above critical temperature. A thermal barrier is developed during phase transition. The color palette on the right shows the density of the scatter points in one GeV interval. }
\label{fig:phase1}
\end{center}
\end{figure}

\begin{figure}[!htb]
\begin{center}
\includegraphics[width=0.3\textwidth]{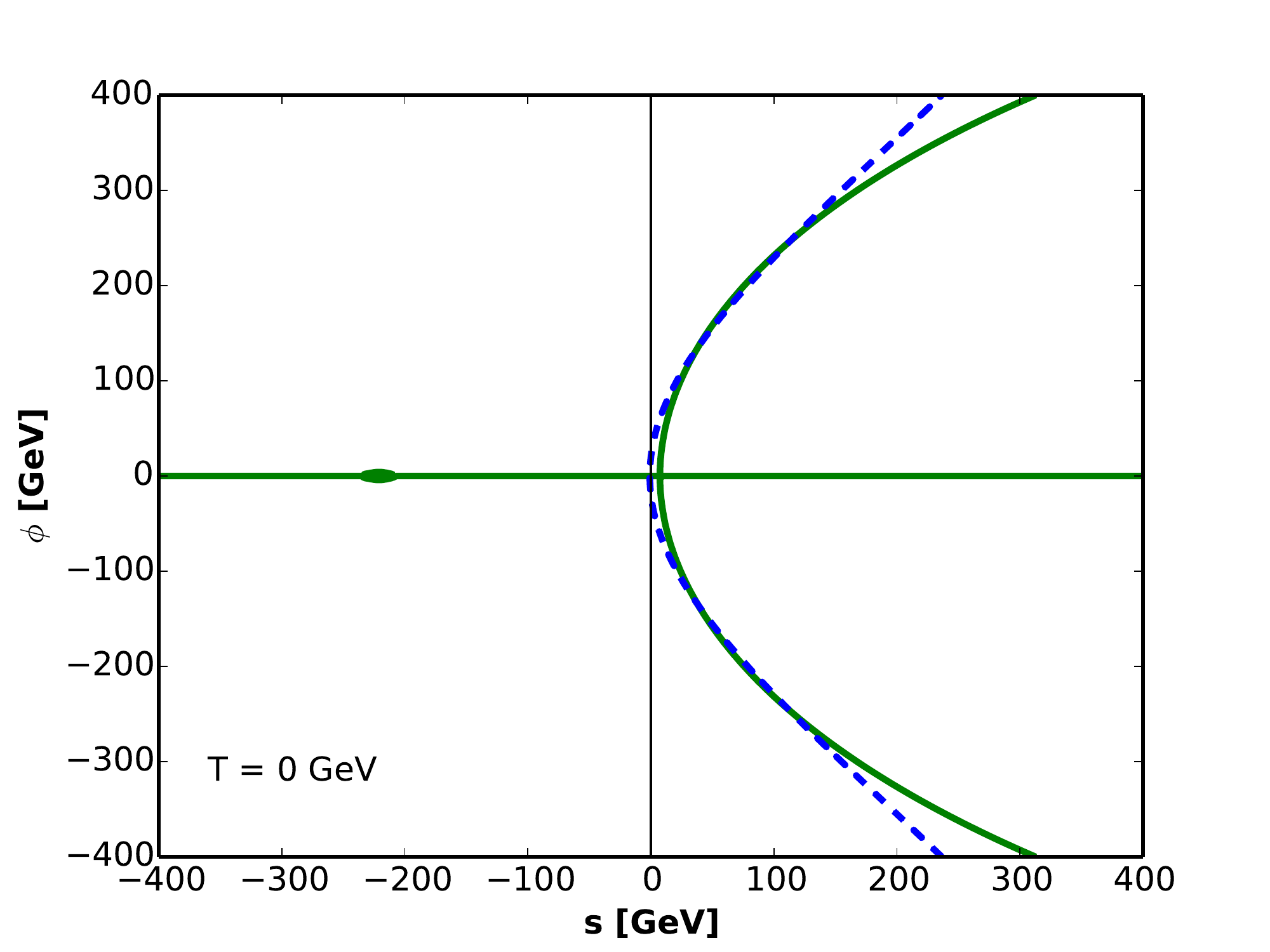} 
\includegraphics[width=0.3\textwidth]{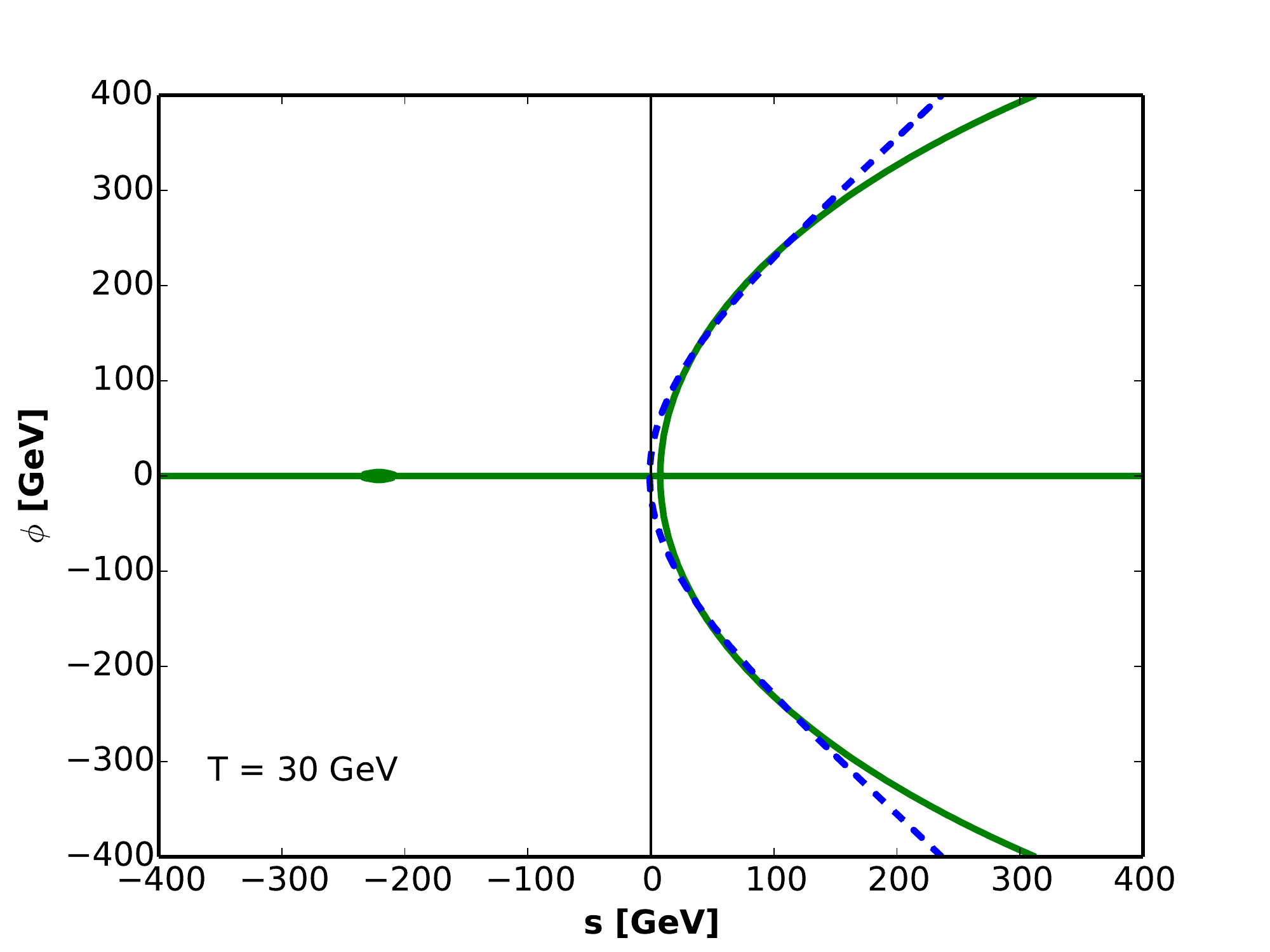} \\
\includegraphics[width=0.3\textwidth]{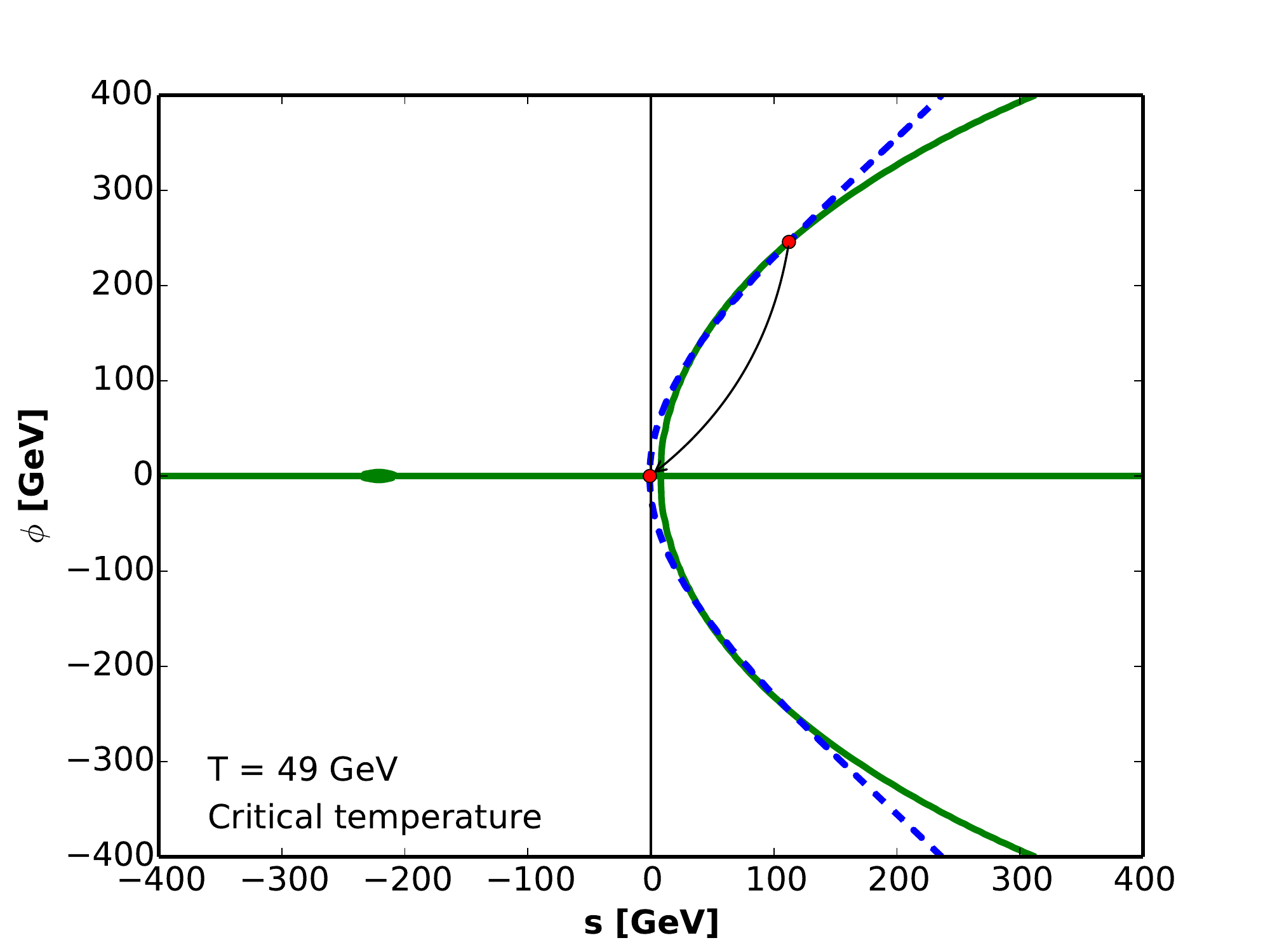} 
\includegraphics[width=0.3\textwidth]{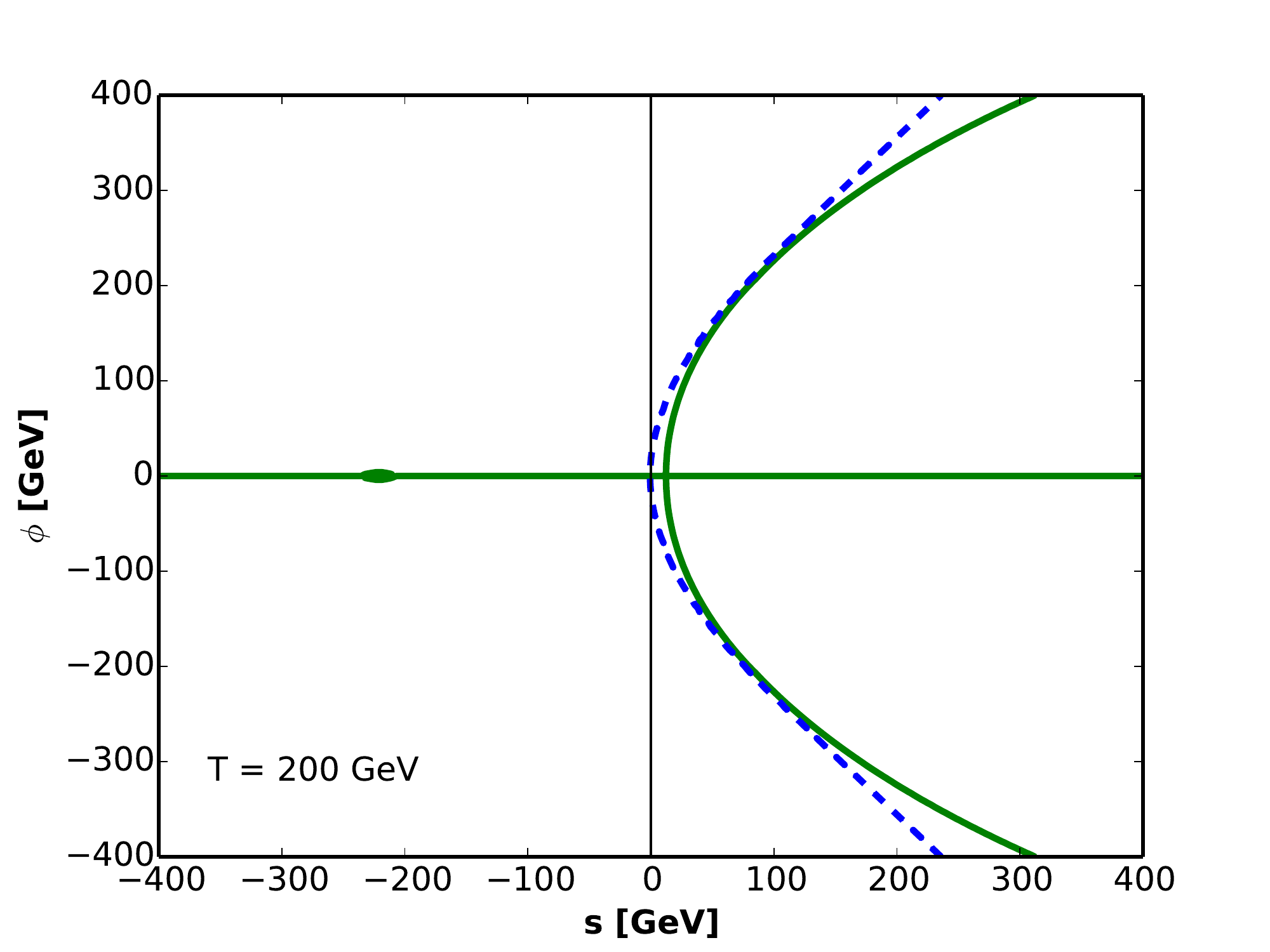}
\caption{\small In the phase transition pattern I, same as the Figure~\ref{fig:phase1}, but a tree-level barrier is developed. }
\label{fig:phase1.1}
\end{center}
\end{figure}

Figure~\ref{fig:phase1} and Figure~\ref{fig:phase1.1} show the variations of the $\phi$ curve and $s$ curve with temperature. In both figure, the first diagram represents the configuration at zero temperature. The second, third and fourth diagrams represent the configuration below, at and above the critical temperature. In the diagram at critical temperature, an arrow was drawn to show how the phase transition happened, in the point of view of increasing temperature. Similar figures will be given for other patterns of phase transition later.

The difference between the two figures is whether the barrier exists or not at zero temperature. In Figure~\ref{fig:phase1}, only broken minimum exists at zero temperature, then the symmetric one is developed during the heating. In Figure~\ref{fig:phase1.1}, the symmetric minimum already exists at zero temperature, but with higher potential than the broken one. The existence of the barrier at the zero temperature is a new feature for the singlet-assisted electroweak phase transition, which does not happen in the traditional electroweak phase transition where the barrier must be thermally induced. This could be attributed to the existence of the term $\mu_{s\phi}$, as shown in the expression of $\mathcal{E}$ where only the $\mu_{s\phi}$ is important in pattern I. In other patterns, other terms in $\mathcal{E}$ could also contribute to the barrier. Unlike the SM, these contributions to $\cal E$ don't require a non-zero temperature, hence zero-temperature barrier can exist.

%
%


%

\subsubsection{Pattern II: Single-branch barrier transition, with $u_s$ separated from 0}

Suppose the $s$ curve has at least two relevant branches. Assuming $\mu_3$ is small, we only consider the case when $\mu_s^2>0$. As mentioned earlier, there are two symmetric minima, one positive and the other negative, while the stationary point around $(0,0)$ must be a saddle point. 
Let's focus on the single-branch transition for now, which means that the symmetric minimum $(0,u_s^{\rm inter})$ on a different branch from the broken minimum has higher potential than $(0,u_s^{\rm single})$ on the same branch as the broken minimum. The other case will be discussed later.

\begin{figure}[!htb]
\begin{center}
\includegraphics[width=0.4\textwidth]{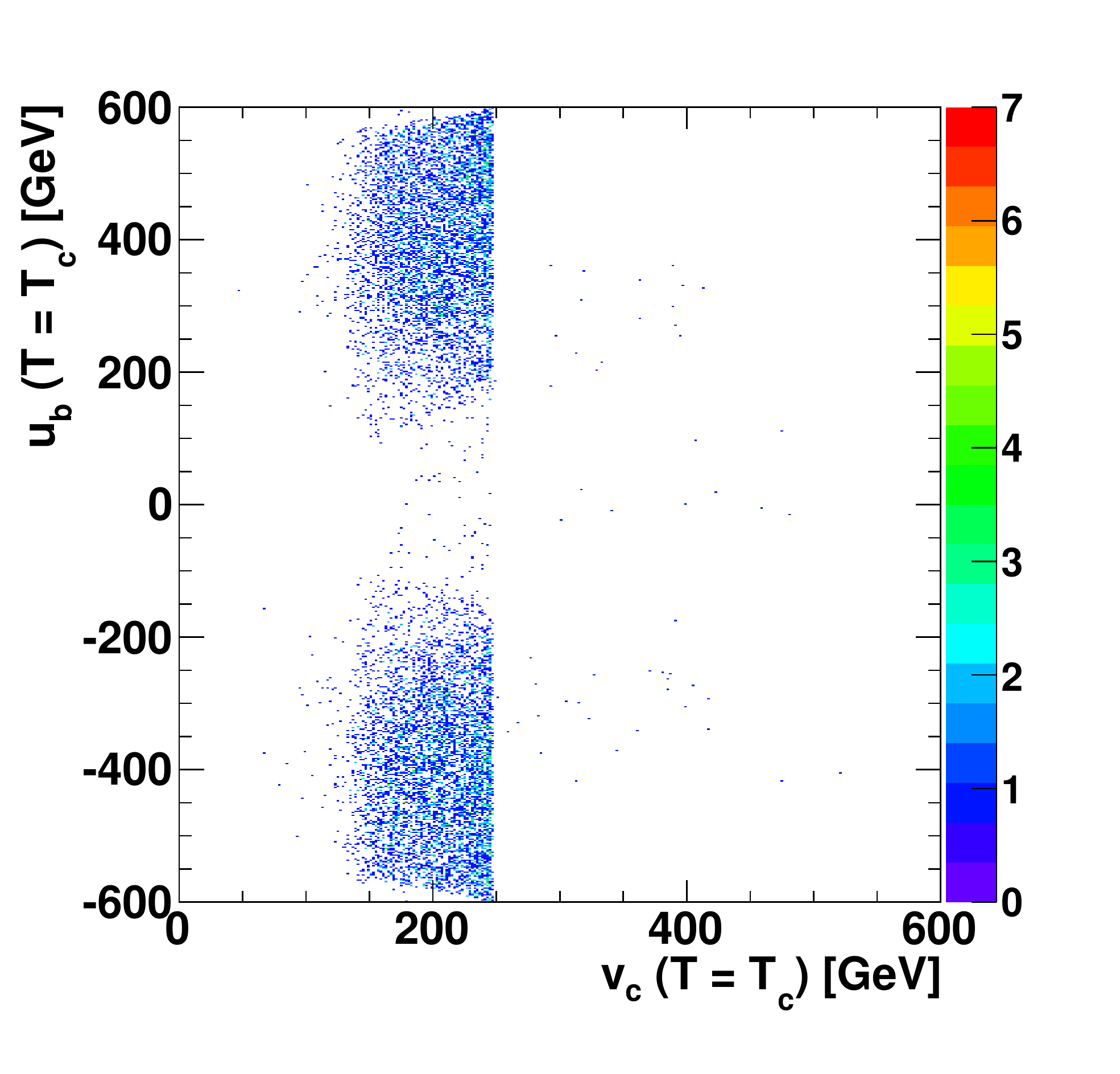} 
\caption{\small The allowed values of the broken minimum $(v_c, u_b)$ in pattern II. The color palette on the right shows the density of the scatter points in one GeV interval. 
}
\label{fig:scan2vcsb}
\end{center}
\end{figure}

Now that the symmetric minimum chosen by the phase transition is other than $(0,0)$, we are convinced that $u_s = u_s^{\rm single}$ can't be close to 0. That being said, the gap in Figure~\ref{fig:scan2vcsb} is well understood. Meanwhile, $u_s$ is still strongly correlated with $u_b$ because they are on the same branch. It is precisely the characteristic of the transition pattern II.

\begin{figure}[!htb]
\begin{center}
\includegraphics[width=0.3\textwidth]{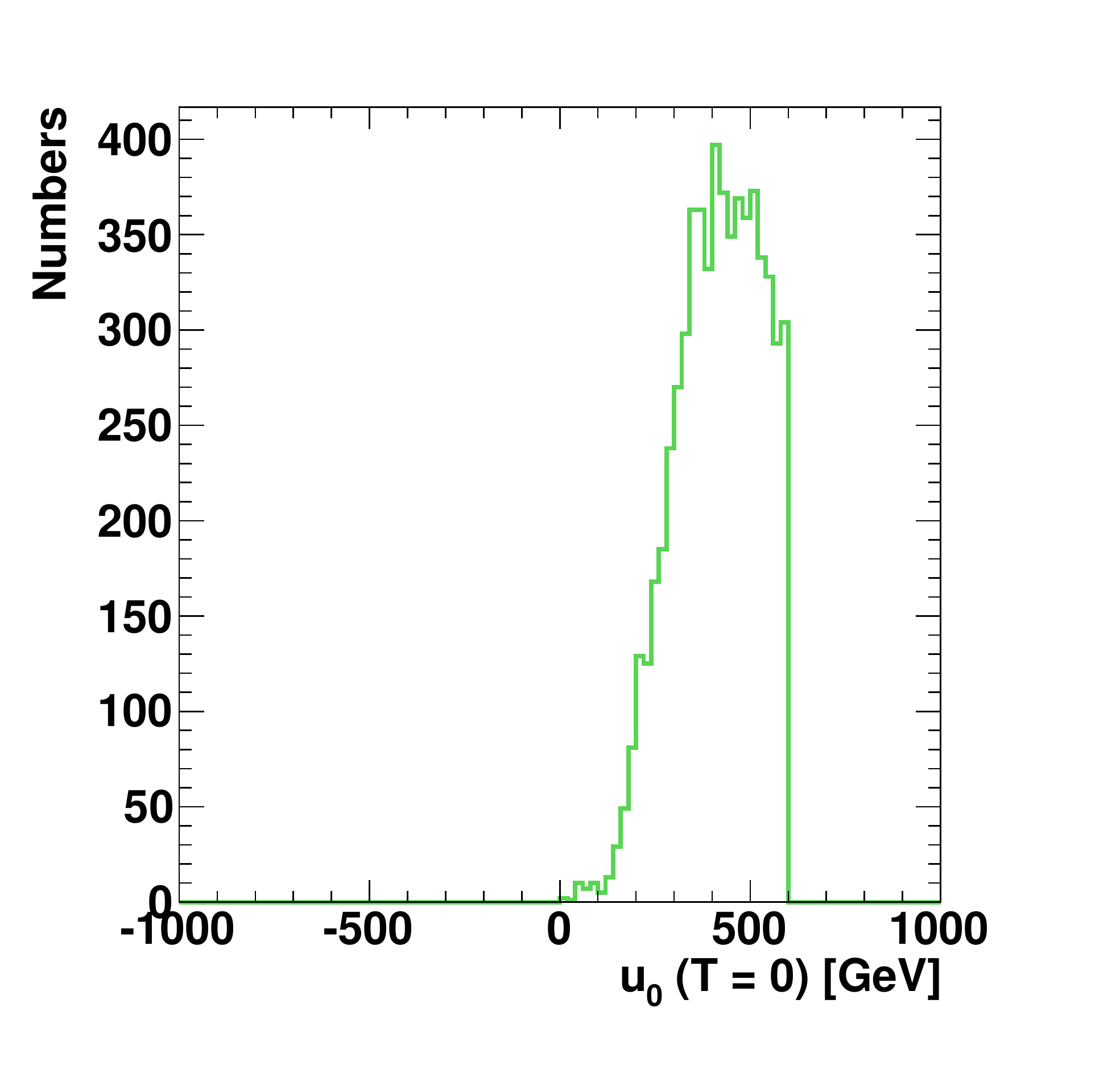} 
\includegraphics[width=0.3\textwidth]{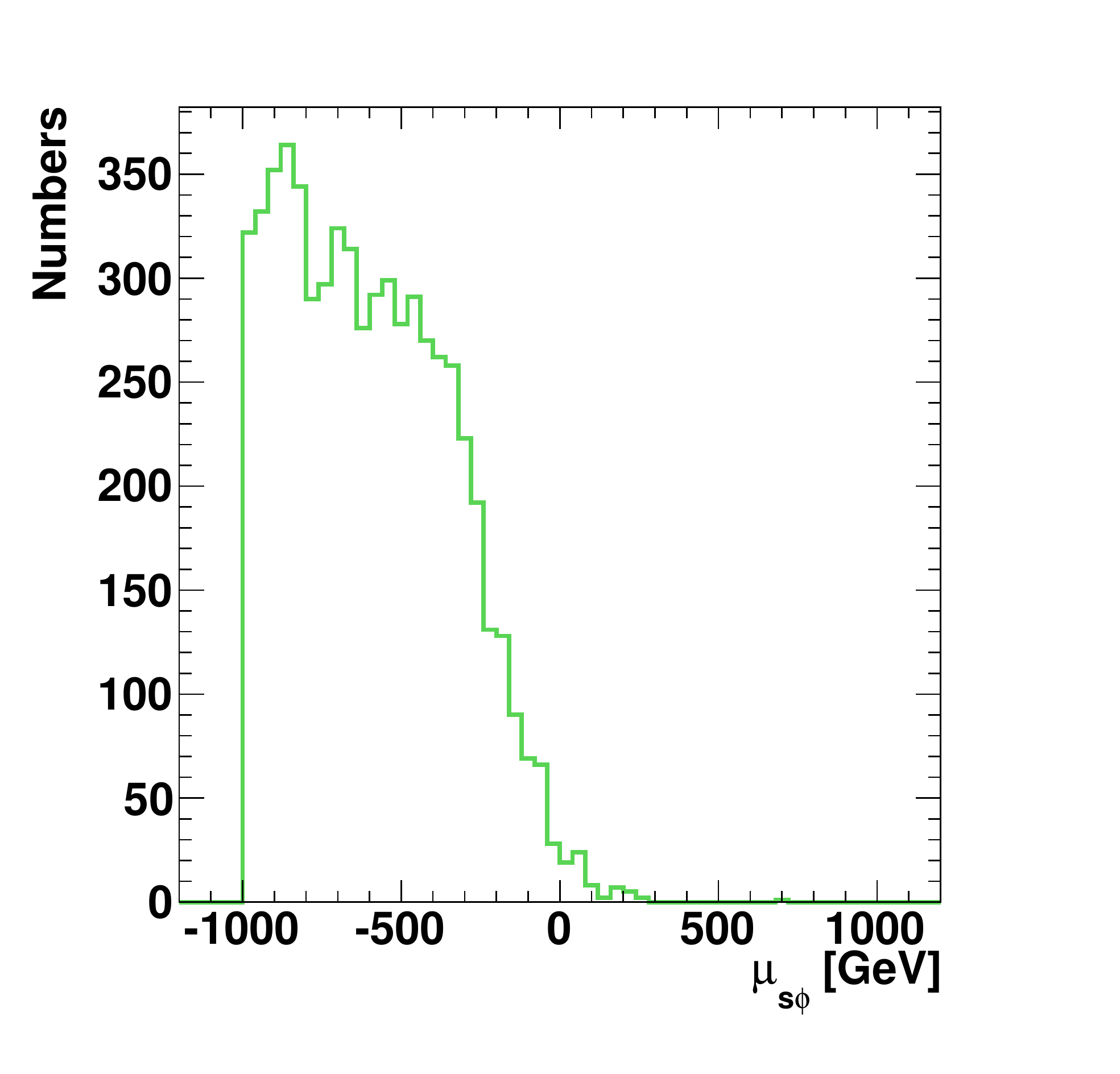} \\
\includegraphics[width=0.3\textwidth]{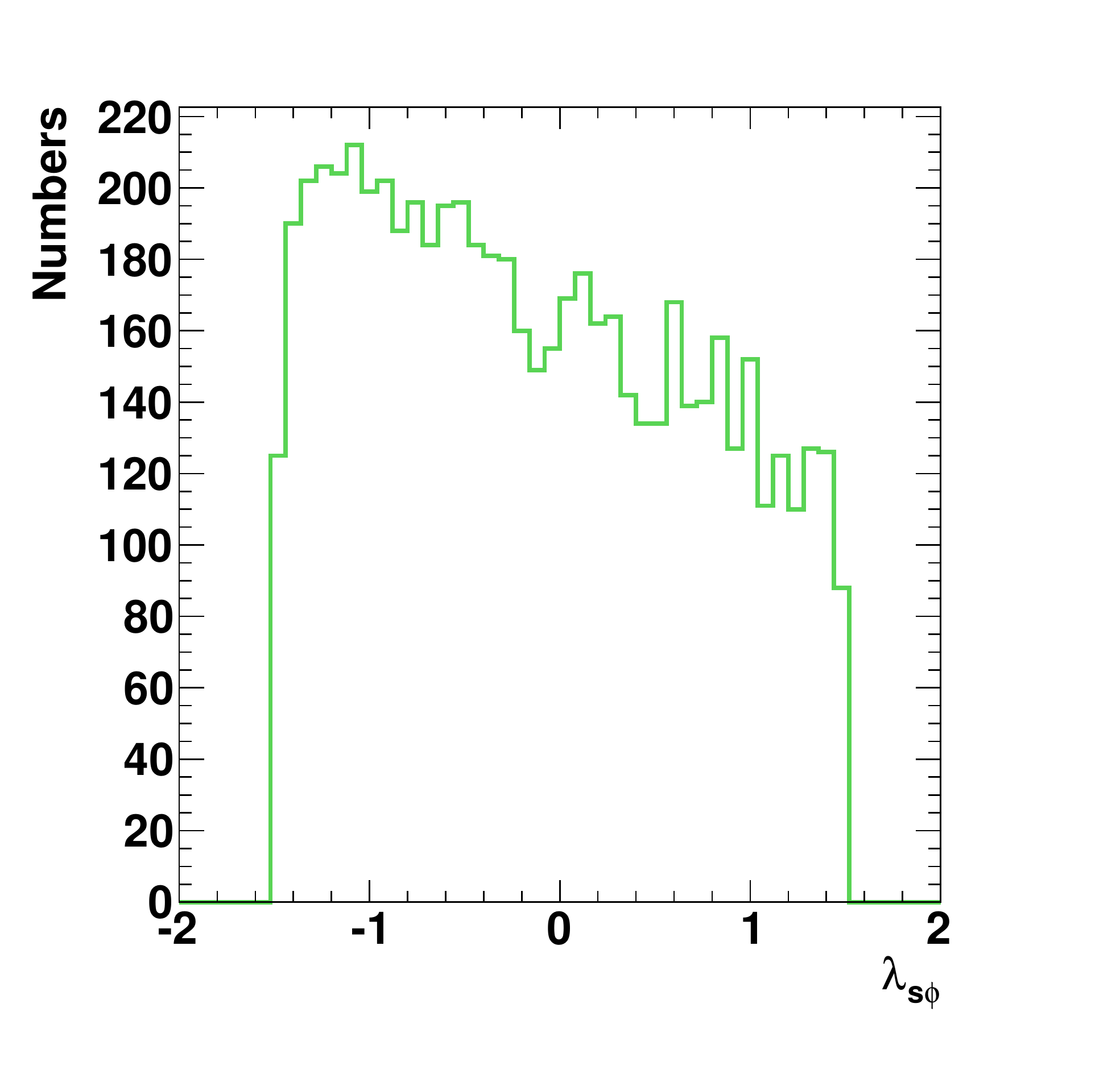} 
\includegraphics[width=0.3\textwidth]{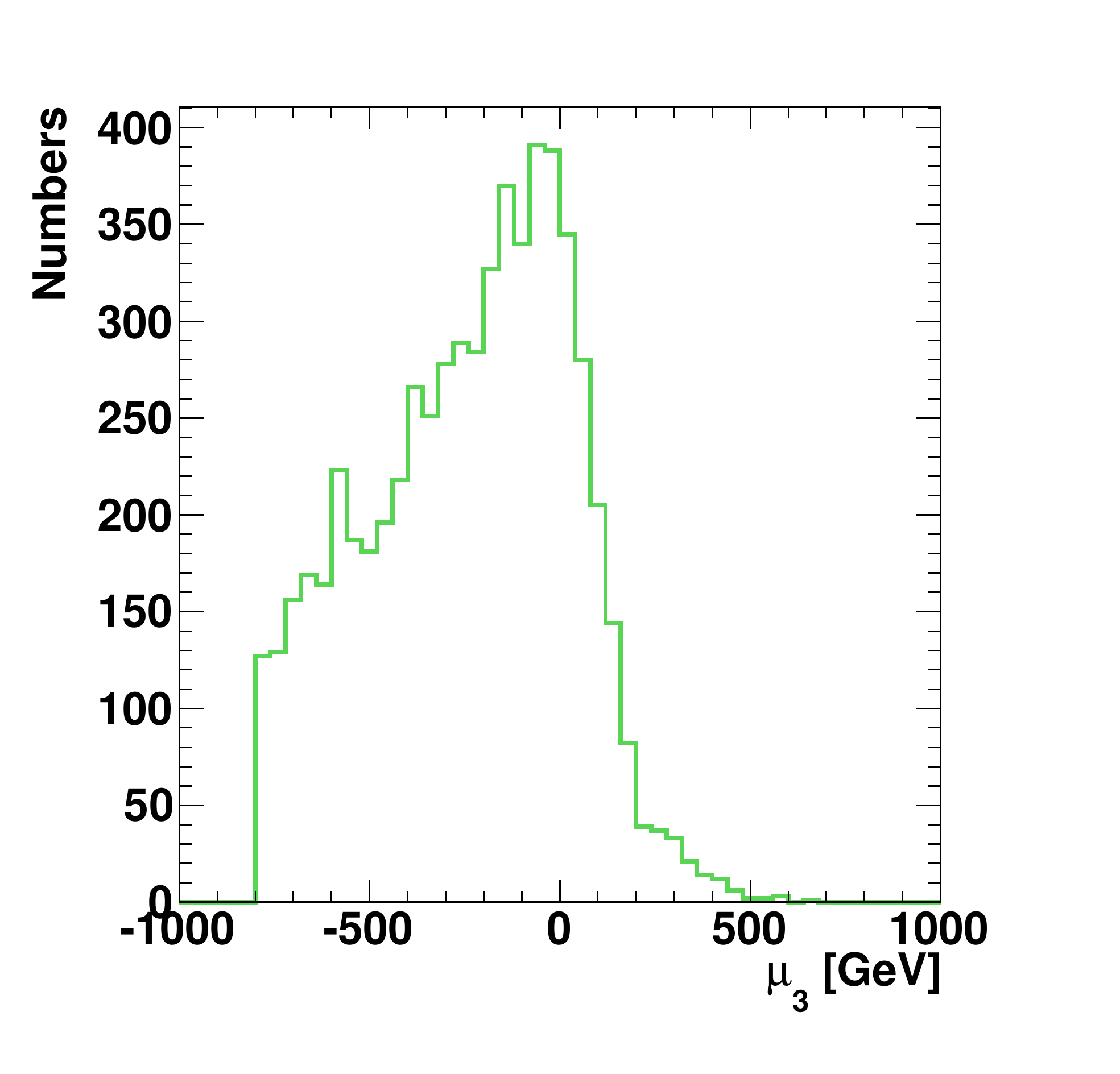} 
\caption{\small In the pattern II with $u_0 > 0$,  the allowed values of the model parameters $u_0$, $\mu_{s\phi}$,  $\lambda_{s\phi}$ and $\mu_3$ are shown. The color palette on the right shows the density of the scatter points in one GeV interval. 
}
\label{fig:scan2musph3}
\end{center}
\end{figure}

Let's investigate the parameter preferences in this pattern. 
As the whole branch of $s$ curve under consideration is apart from $s=0$, it is natural to have $u_0$ also apart from 0. Actually, all of the $u_0$, $u_b$ and $u_s$ are similar, which means that $s$ field is seemingly irrelevant with the phase transition, although it contributes a crucial coupling $\mu_{s\phi}$ that plays important role.
The preference of the $\mu_{s\phi}$ is similar as pattern I. 
A big difference from pattern I is that positive $\lambda_{s\phi}$ is strongly suppressed. One can understand it by looking at $\cal E$, which in this case has two significant terms: $\mu_{s\phi}$ and $\lambda_{s\phi}u_s$, the latter becoming significant due to the non-zero $u_s$. As we need a large negative $\mathcal{E}$, negative $\lambda_{s\phi}$ is preferred. Therefore, the advantage of positive $\lambda_{s\phi}$ is much weakened in pattern II.
Another interesting feature is that $\mu_3$ is mostly negative. This is essential to guarantee that among the two symmetric minima, the one with positive $s$ value has a lower potential, and tends to be the one chosen as the vacuum state at high temperature. Otherwise, inter-branch transition should happen.

\begin{figure}[!htb]
\begin{center}
\includegraphics[width=0.3\textwidth]{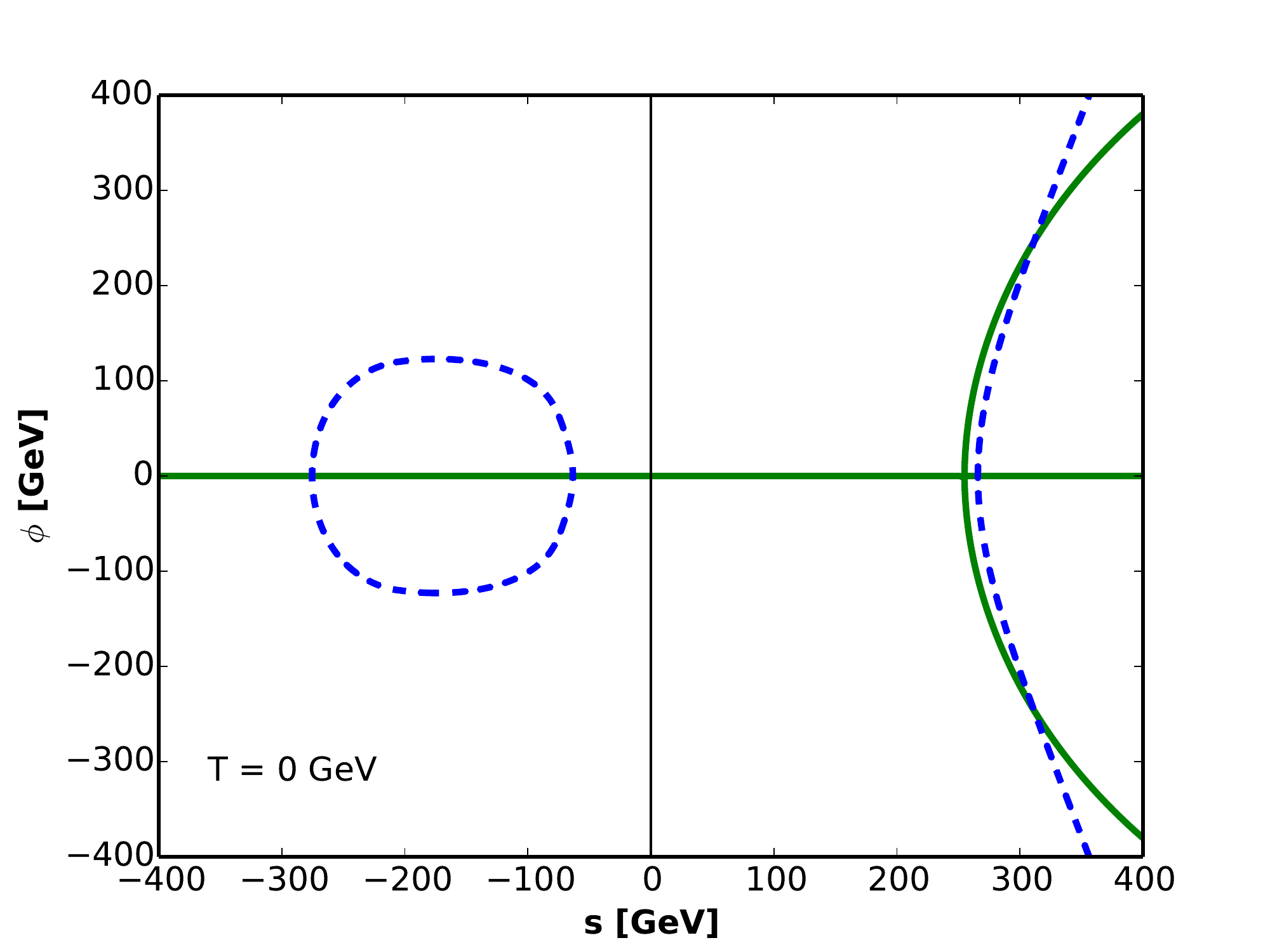} 
\includegraphics[width=0.3\textwidth]{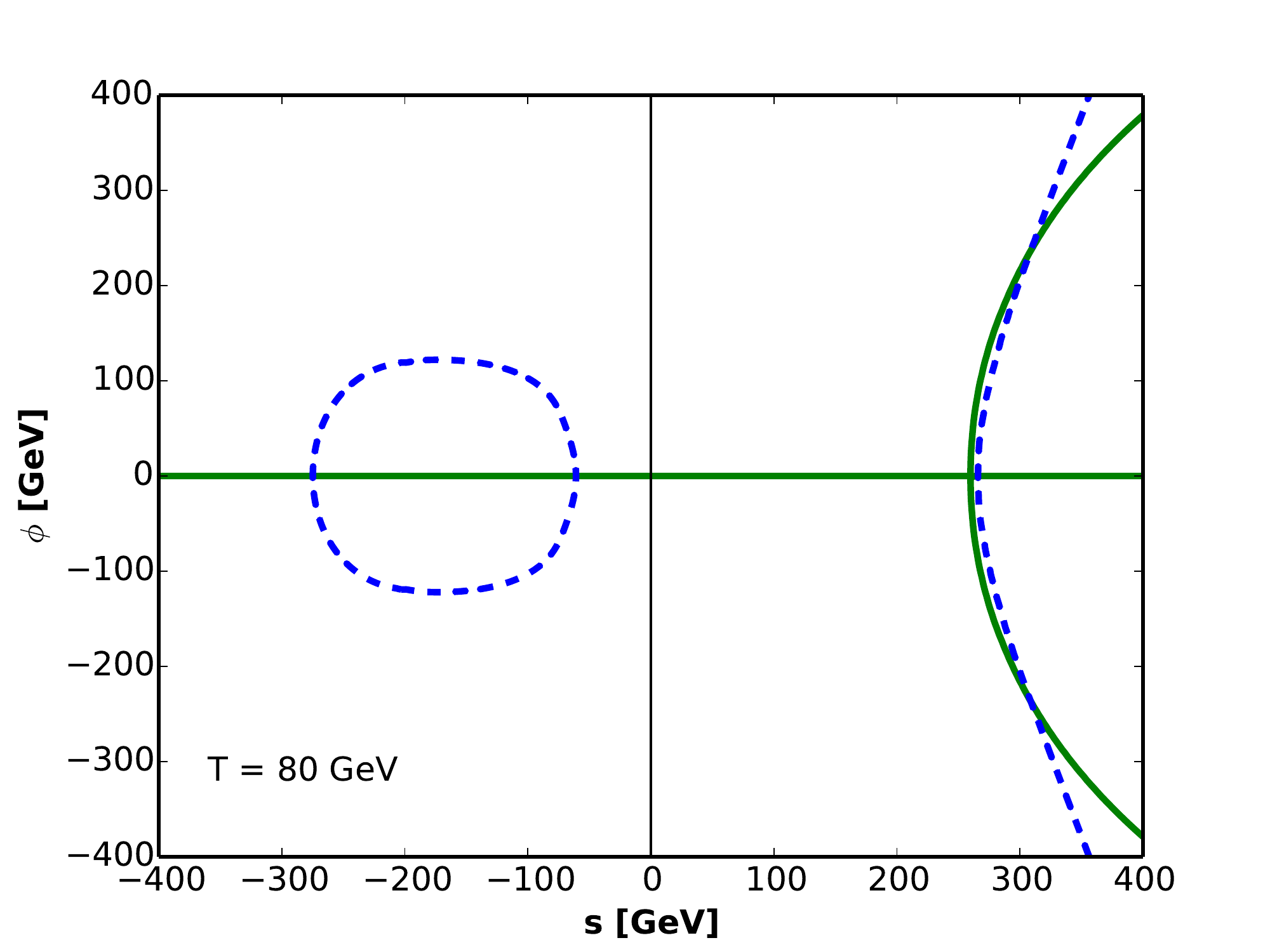}\\
\includegraphics[width=0.3\textwidth]{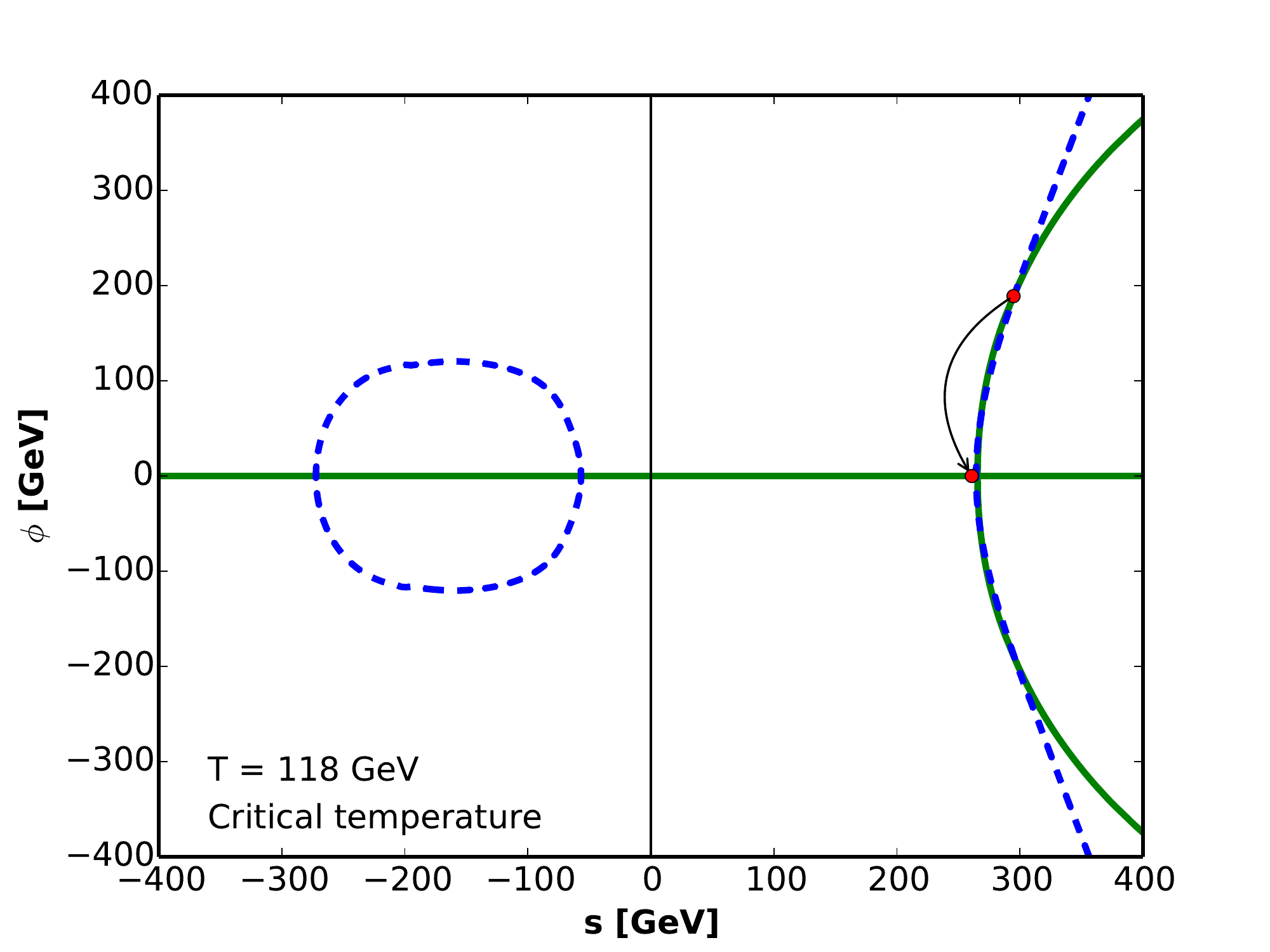} 
\includegraphics[width=0.3\textwidth]{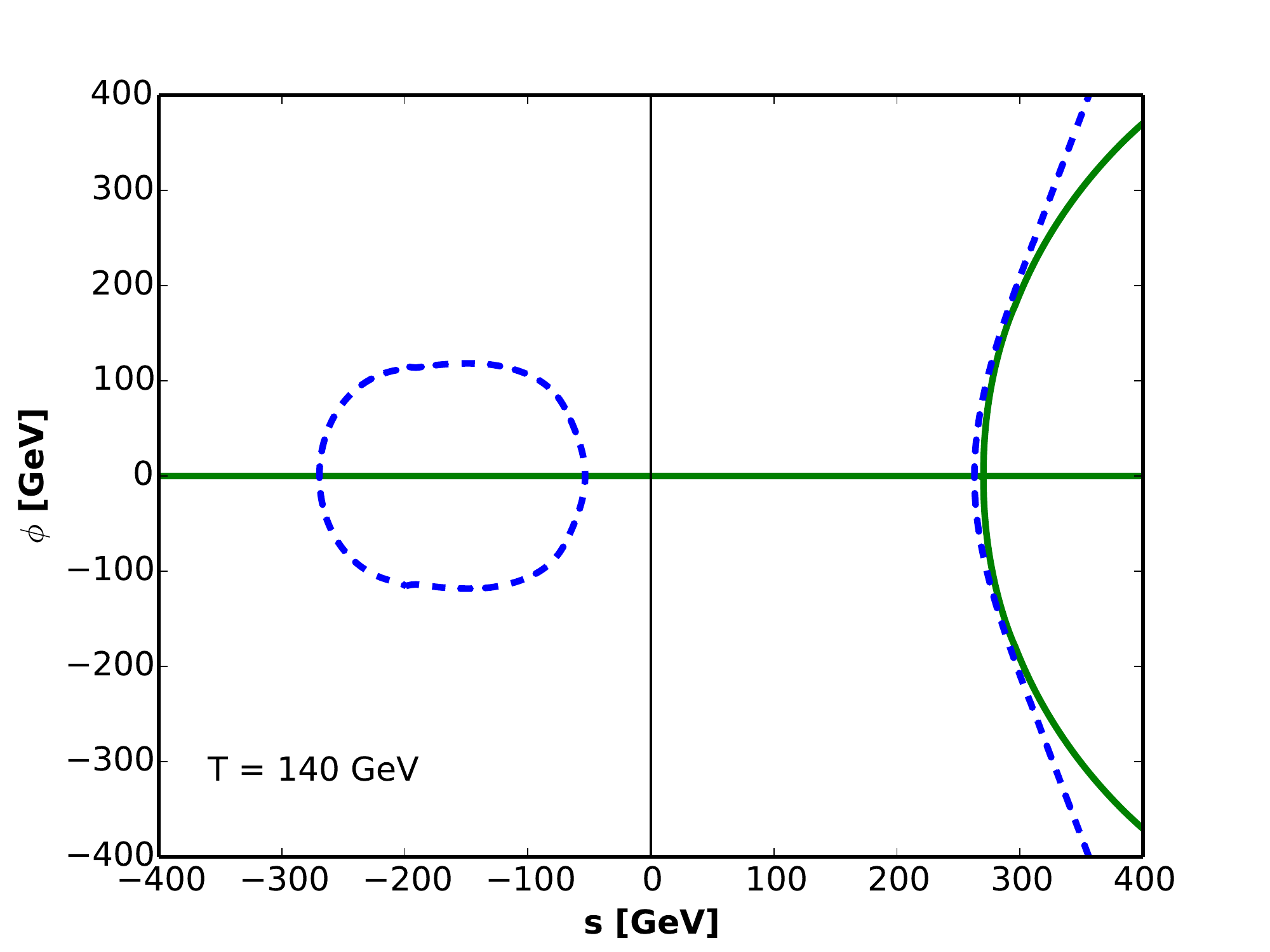}
\caption{\small In the phase transition pattern II, the curves are same as the Figure~\ref{fig:phase1}. }
\label{fig:phase2}
\end{center}
\end{figure}

The procedure of the phase transition is outlined in Figure~\ref{fig:phase2}. The only difference from pattern I is the existence of the symmetric minimum on the other branch. This minimum must already exist at zero temperature in this pattern, which induces a barrier that is impossible in traditional electroweak phase transition. 
%
%
As for the single-branch barrier, similar to the pattern I, it can be thermally induced, or already present at zero temperature.

%
%
%
%
%

\subsubsection{Pattern IIIa: Inter-branch barrier transition}

The setup in this case is similar to the pattern II, except that the phase transition occurs across the inter-branch barrier. 

\begin{figure}[!htb]
\begin{center}
\includegraphics[width=0.3\textwidth]{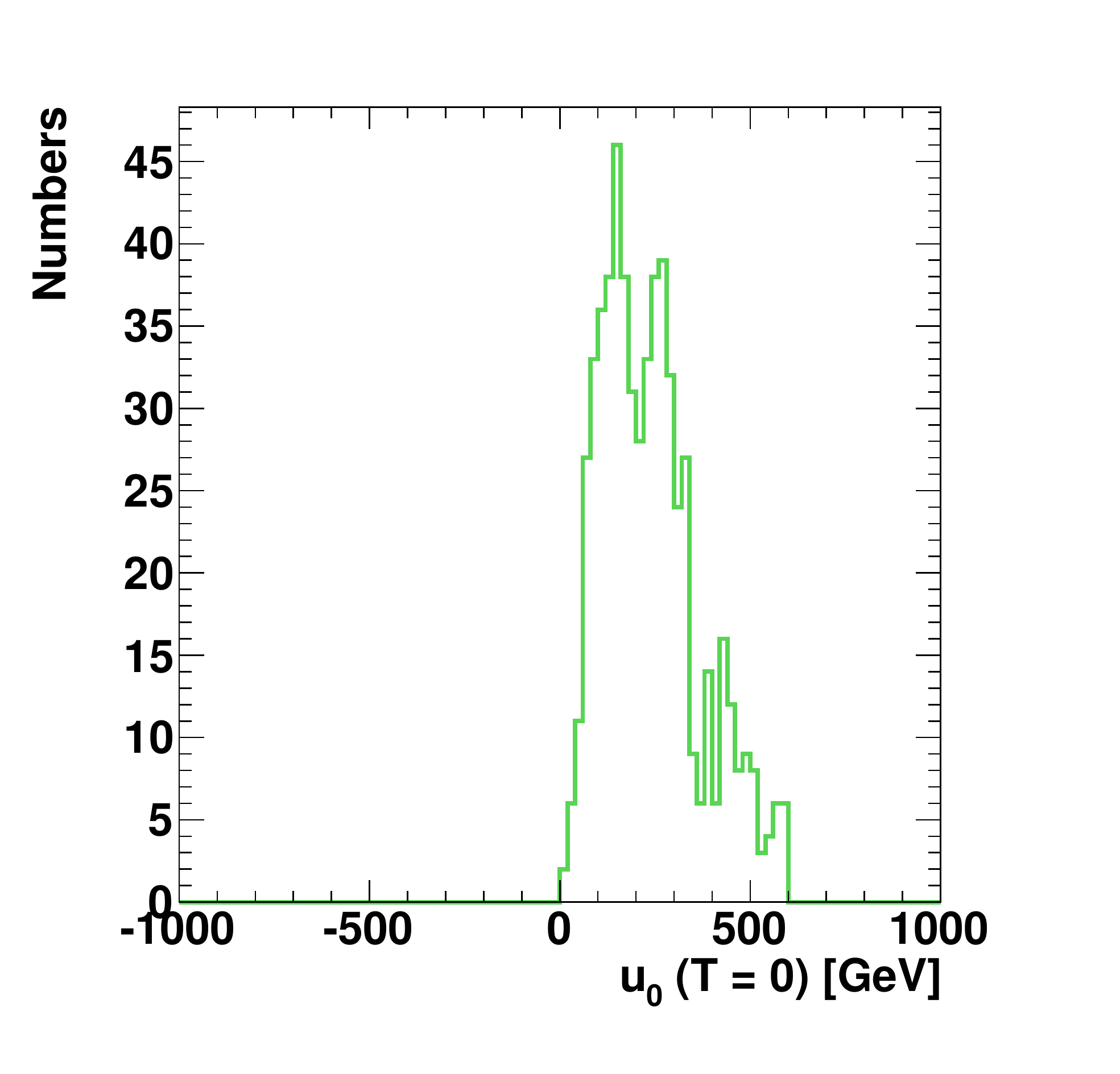} 
\includegraphics[width=0.3\textwidth]{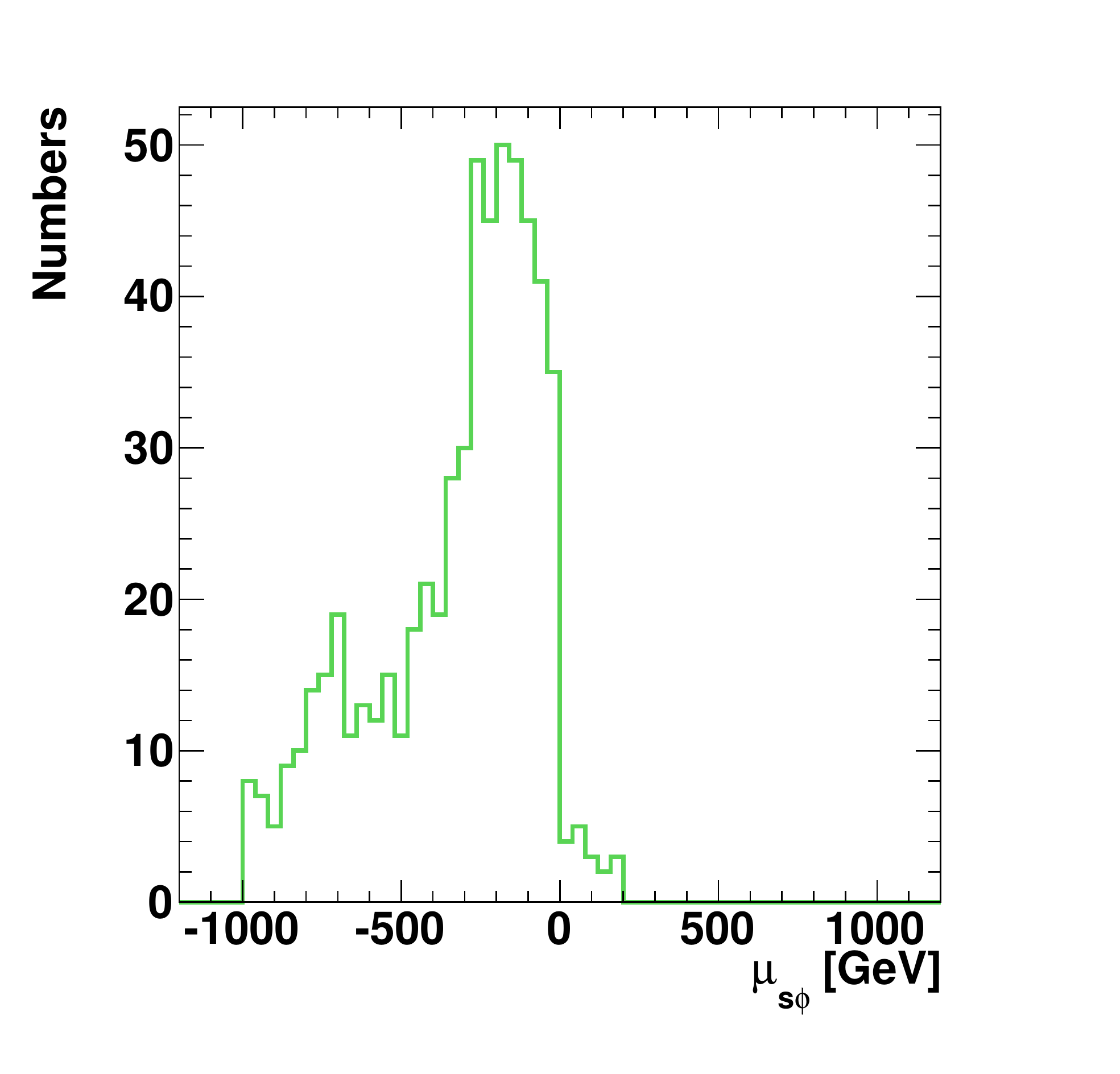}\\
\includegraphics[width=0.3\textwidth]{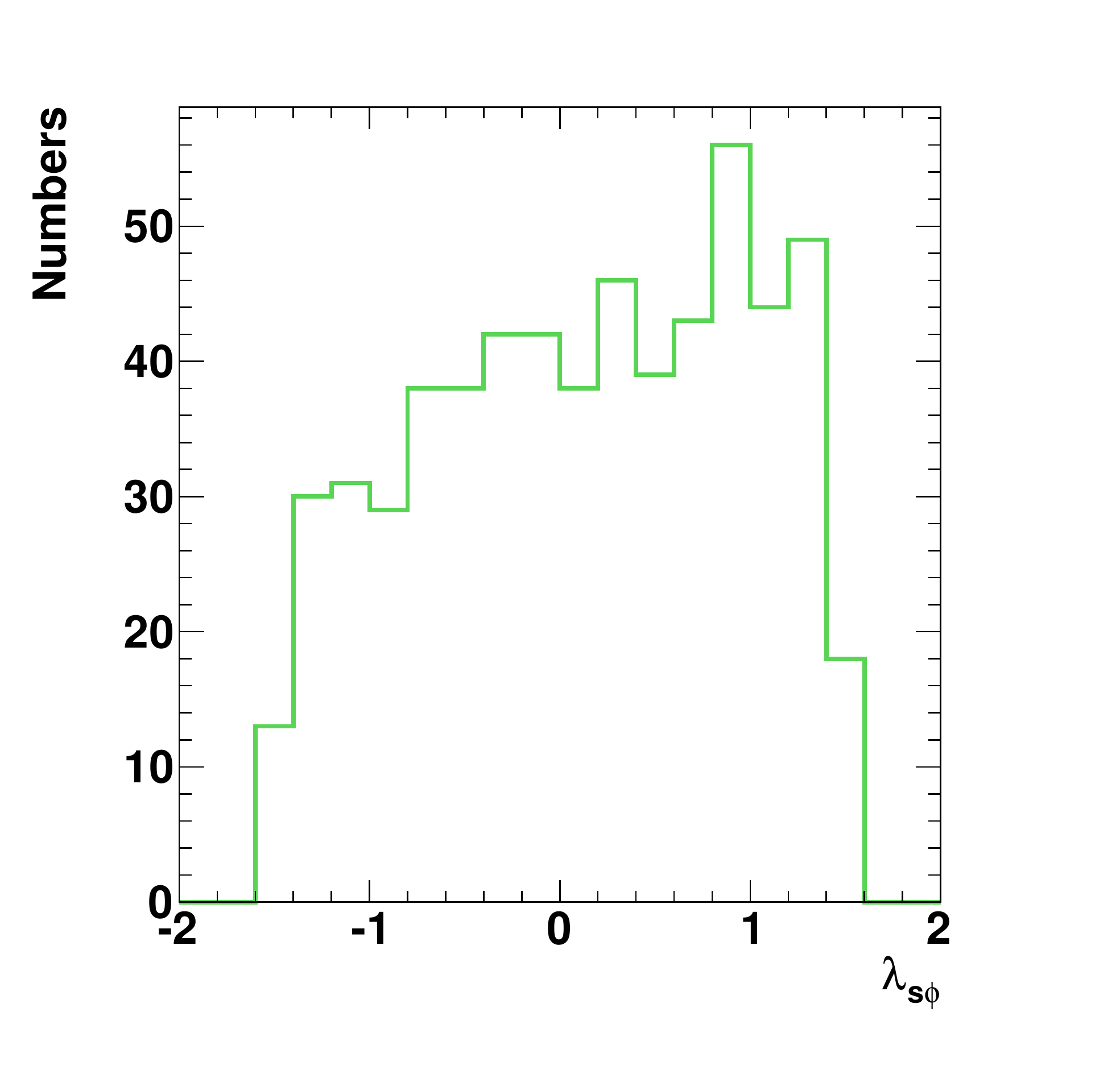} 
\includegraphics[width=0.3\textwidth]{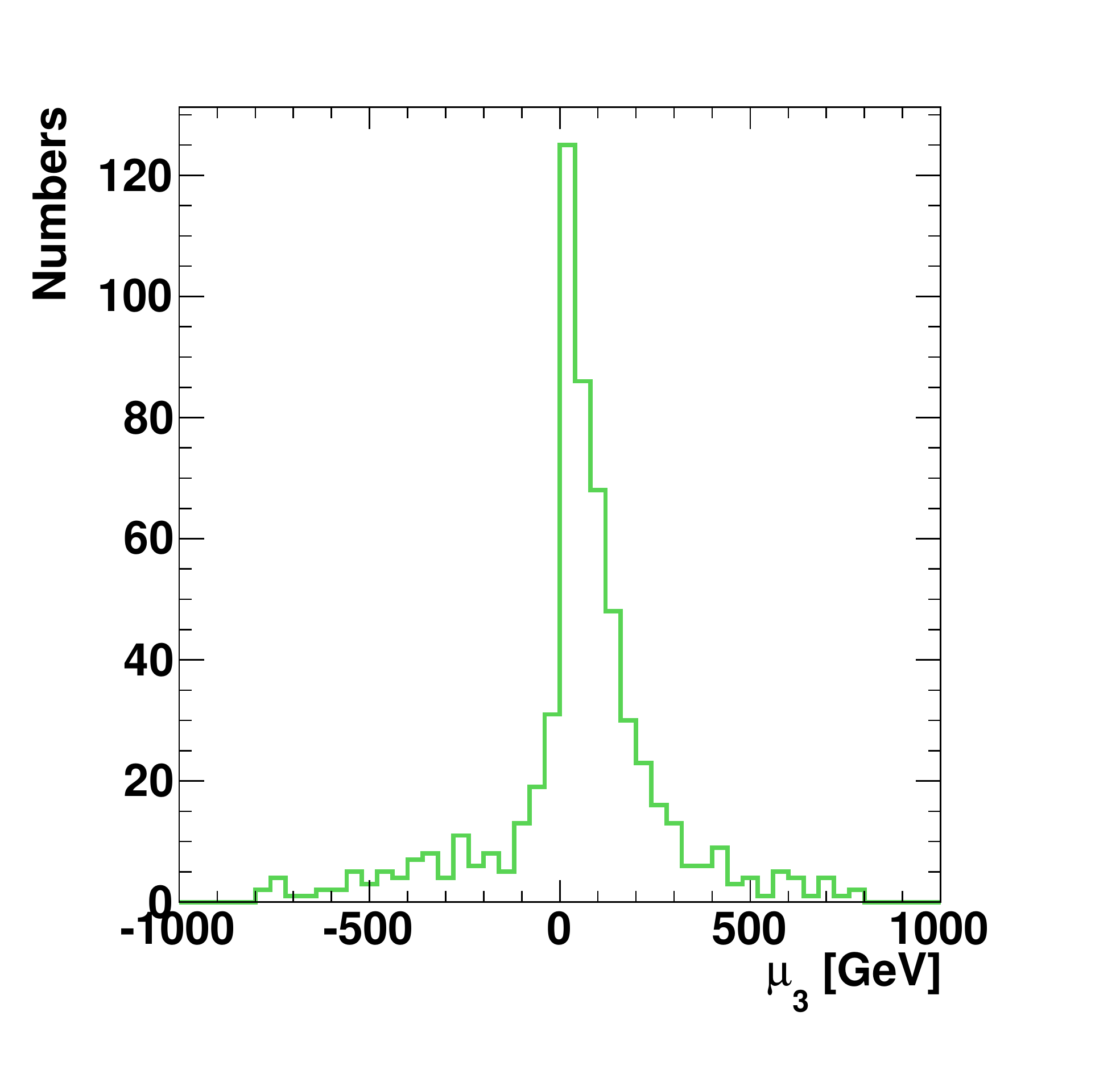} 
\caption{\small In the case IIIa with $u_0 > 0$,  the allowed values of the model parameters $u_0$, $\mu_{s\phi}$,  $\lambda_{s\phi}$ and $\mu_3$  are shown. The color palette on the right shows the density of the scatter points in one GeV interval. 
}
\label{fig:scan3musph3}
\end{center}
\end{figure}

As discussed in pattern II, $\mu_3$ determines the relative height between the potential at the two symmetric stationary points $(0,u_s^{\rm single})$ and $(0,u_s^{\rm inter})$, where $u_s^{\rm single}$ and $u_s^{\rm inter}$ are the scalar $s$ vev in single branch and inter-branch. It tends to be negative when we want a single-branch transition towards $(0,u_s^{\rm single})$, and for the same reason, it prefers positive values when we require an inter-branch transition towards $(0,u_s^{\rm inter})$. However, if $\mu_3$ is too large, it would be harder for us to get a large and negative $\cal E$, as in this case $s_{\alpha}$ is not small, and hence the $\mu_3s_{\alpha}^3$ term in $\cal E$ is not suppressed any more. In sum, we should have a small and positive $\mu_3$ in pattern IIIa, which is justified by the Figure~\ref{fig:scan3musph3}.

\begin{figure}[!htb]
\begin{center}
\includegraphics[width=0.3\textwidth]{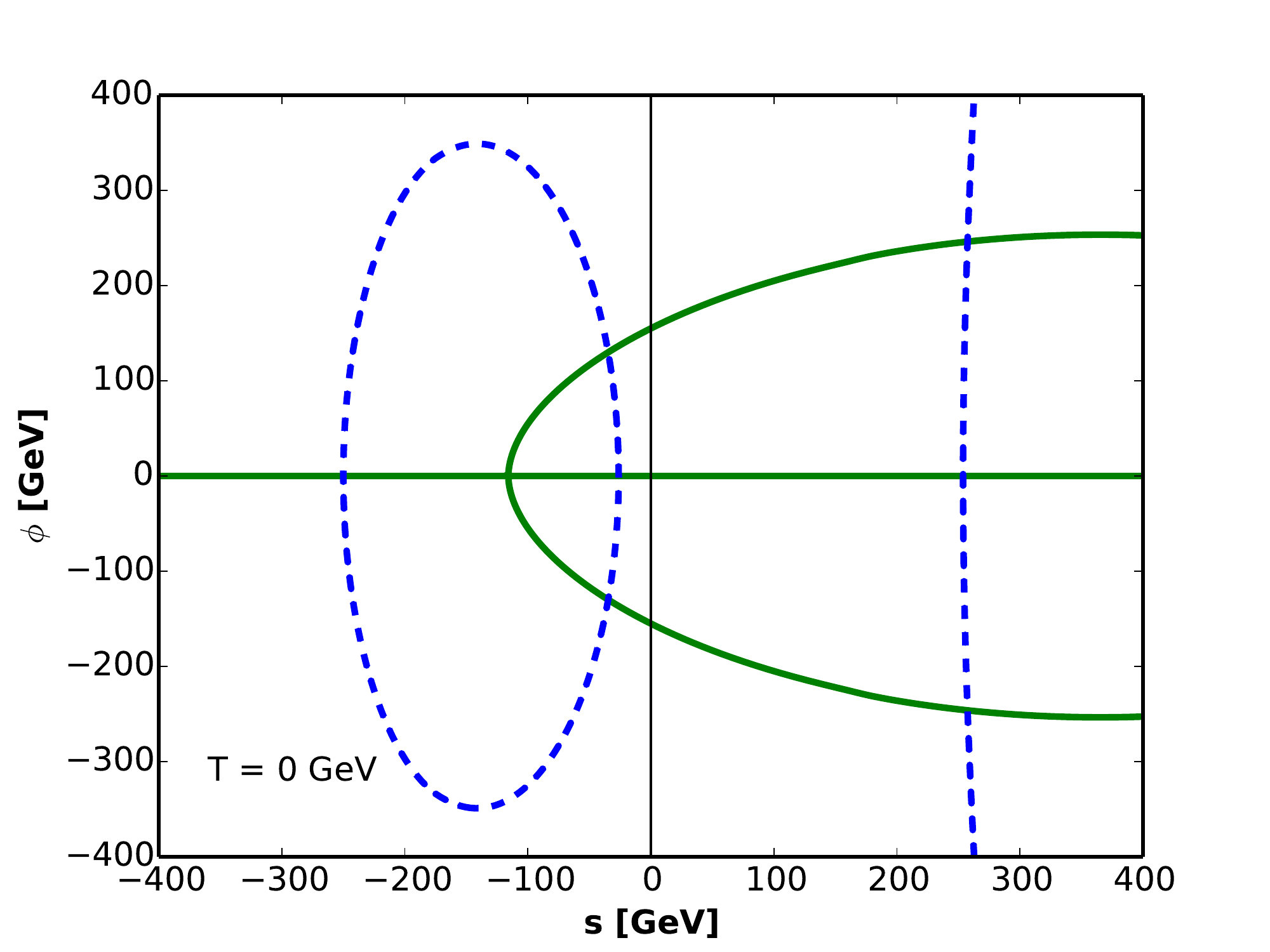} 
\includegraphics[width=0.3\textwidth]{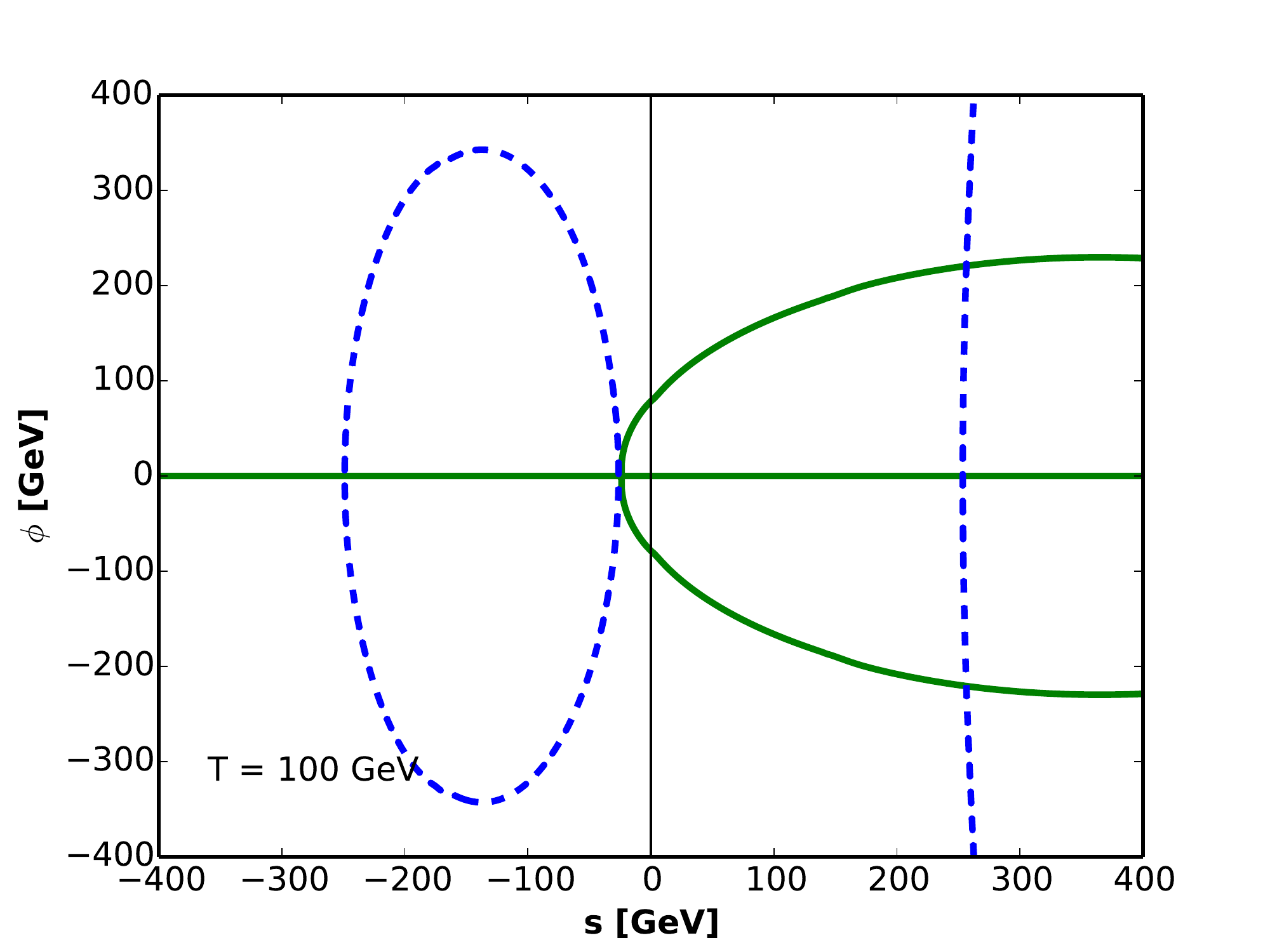} \\
\includegraphics[width=0.3\textwidth]{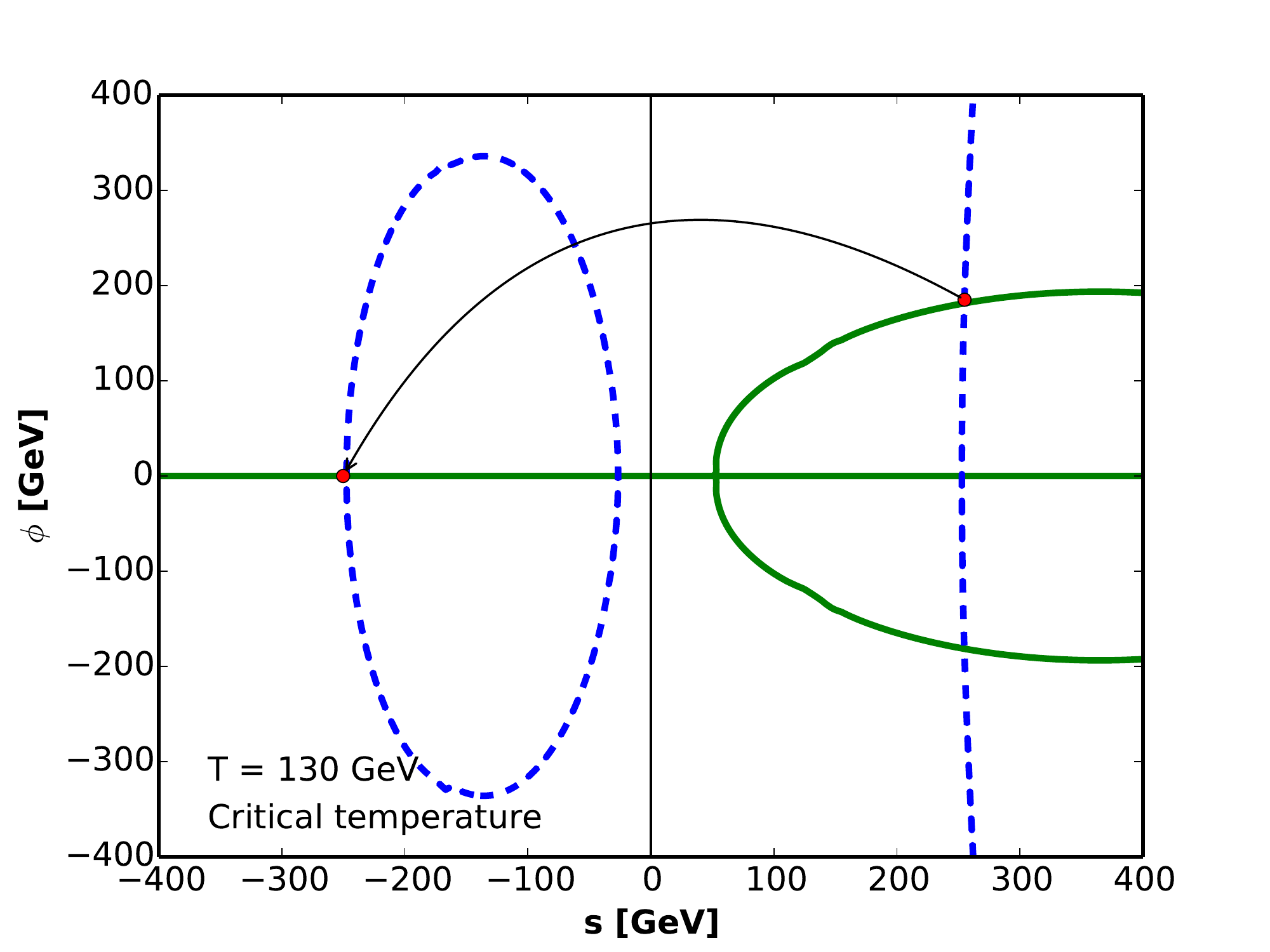} 
\includegraphics[width=0.3\textwidth]{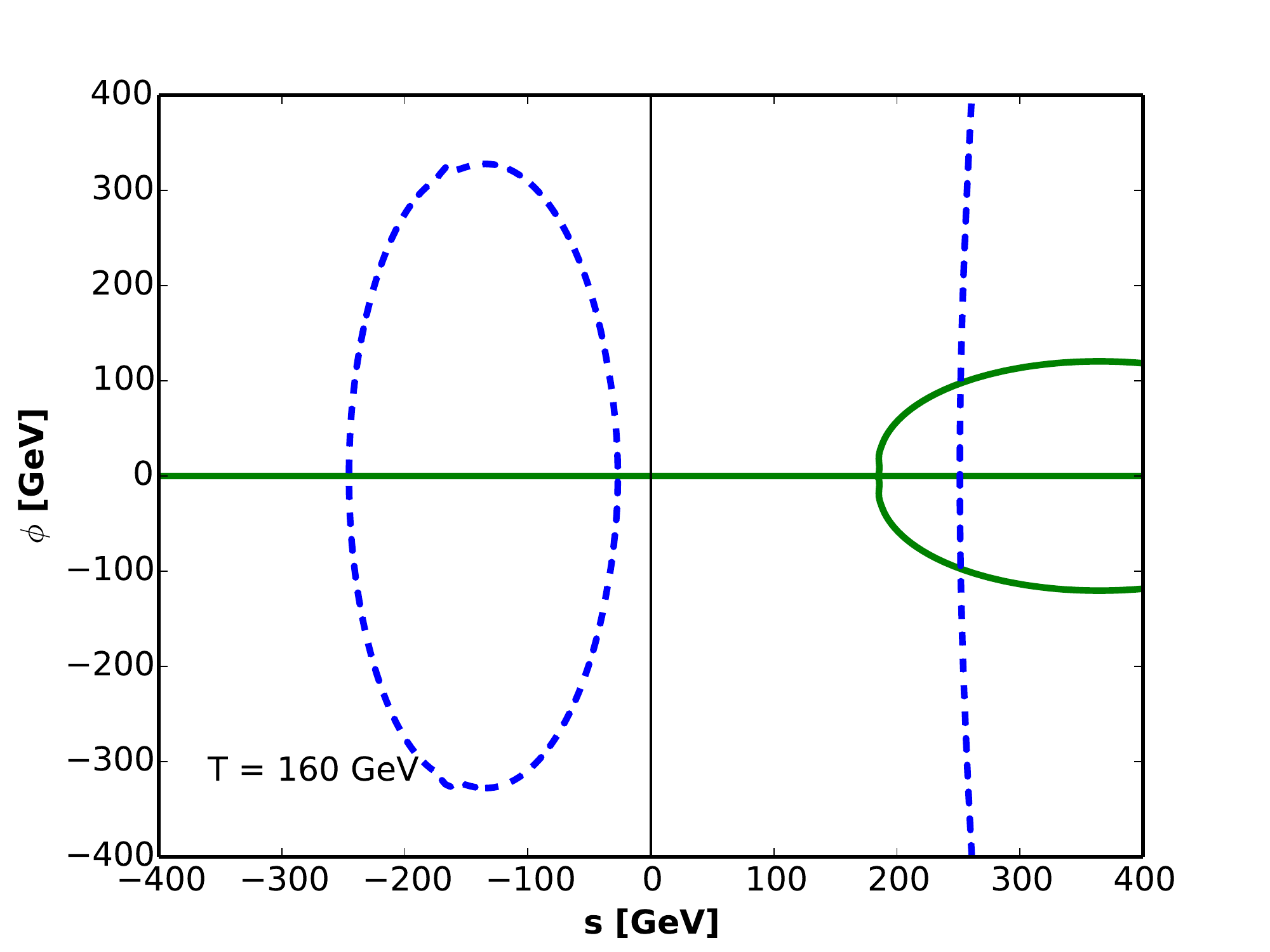}
\caption{\small In the phase transition pattern IIIa, the curves are same as the Figure~\ref{fig:phase1}. } 
\label{fig:phase3}
\end{center}
\end{figure}

The phase transition is shown in Figure~\ref{fig:phase3}, by exhibiting the variations of the $\phi$ curve and $s$ curve with temperature. The only difference from pattern II is that the phase transition happens across the inter-branche barrier. Nevertheless, as the $(0,u_s^{\rm single})$ need not be a minimum, no twisting intersection is required for the two curves, and hence there is no strong preference for $\lambda_{s\phi}$.

	
	

\subsubsection{Pattern IIIb: multi-step transitions }

\begin{figure}[!htb]
\begin{center}
\includegraphics[width=0.3\textwidth]{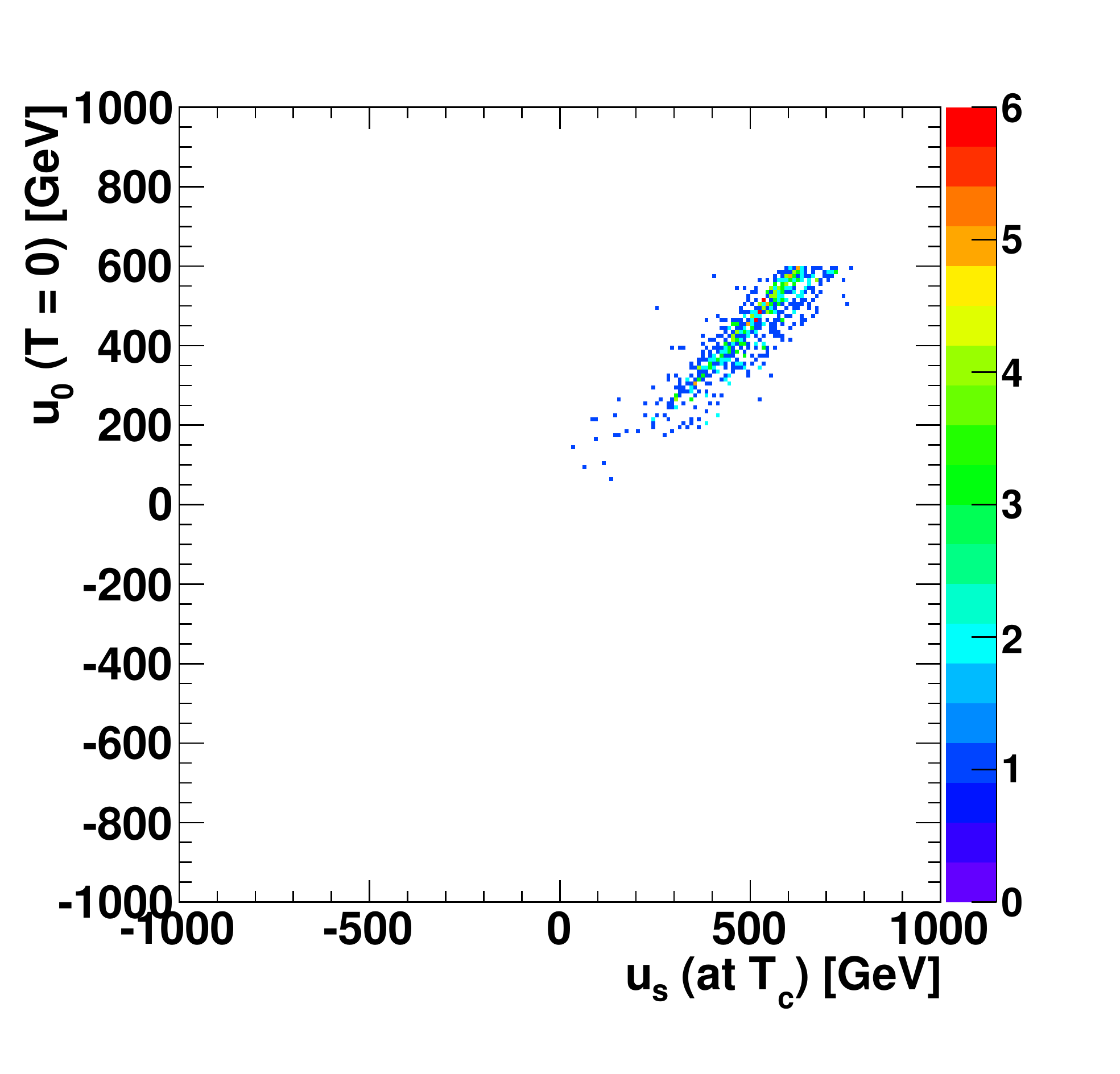} 
\includegraphics[width=0.3\textwidth]{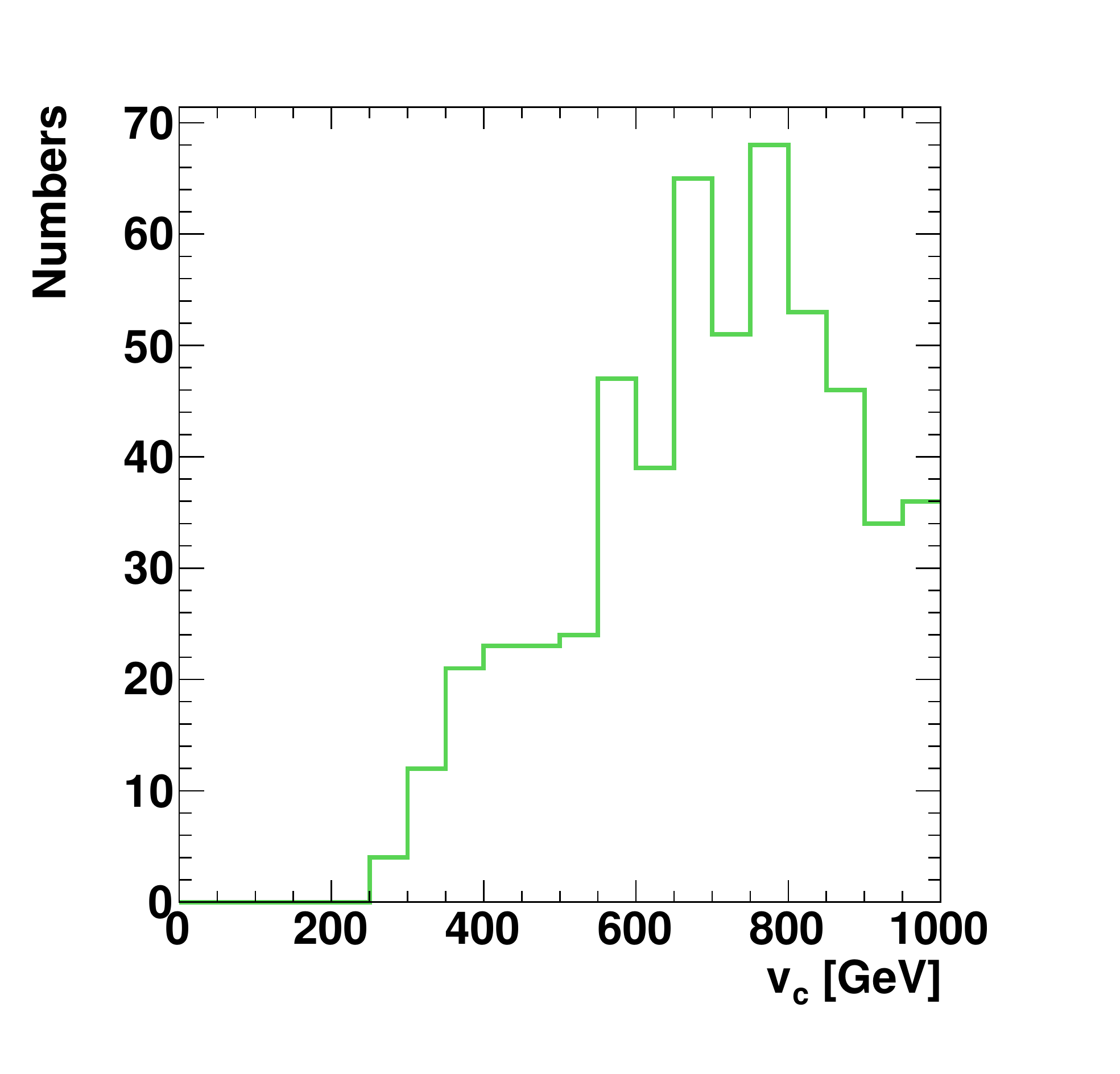} \\
\includegraphics[width=0.3\textwidth]{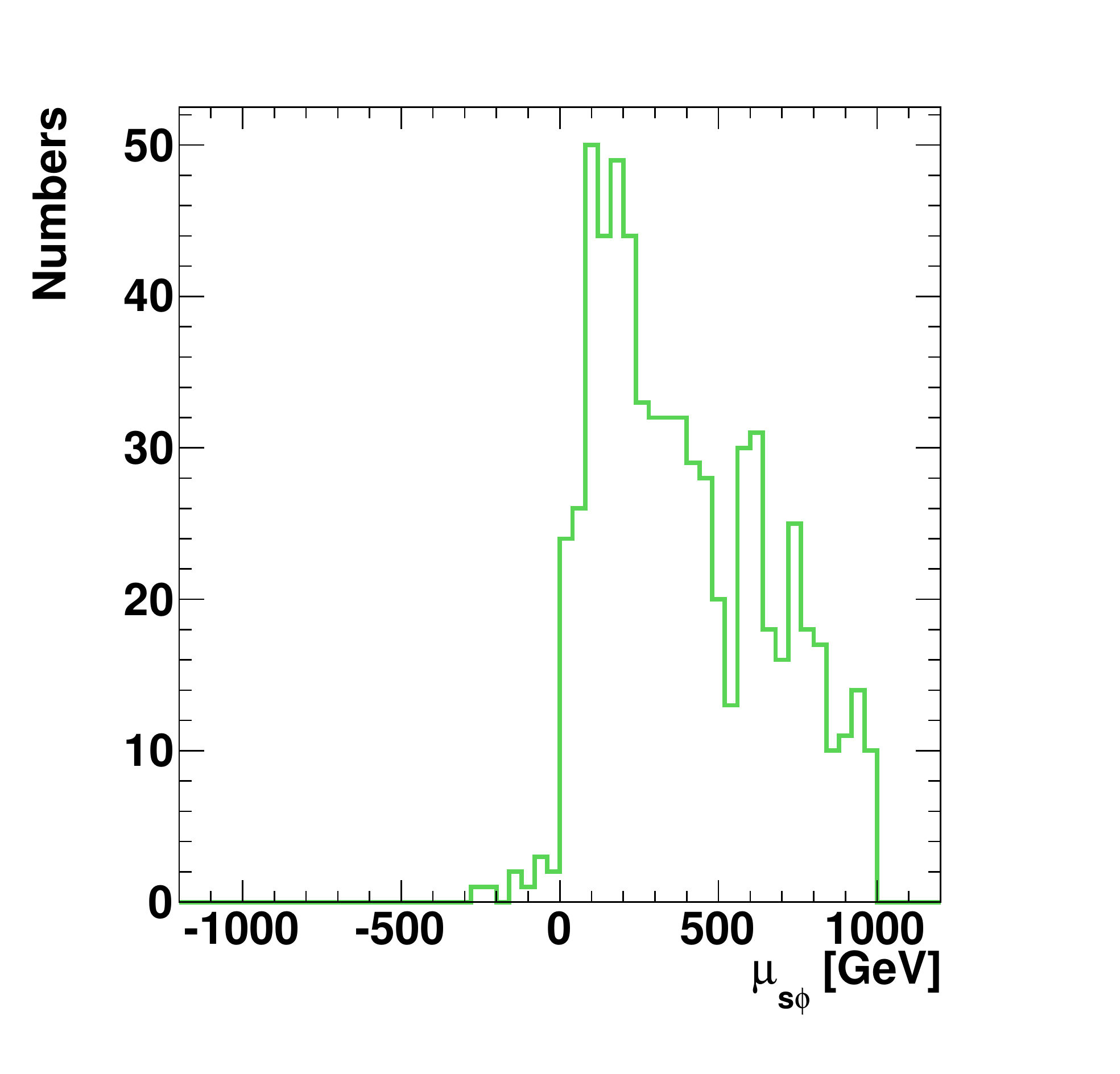}
\includegraphics[width=0.3\textwidth]{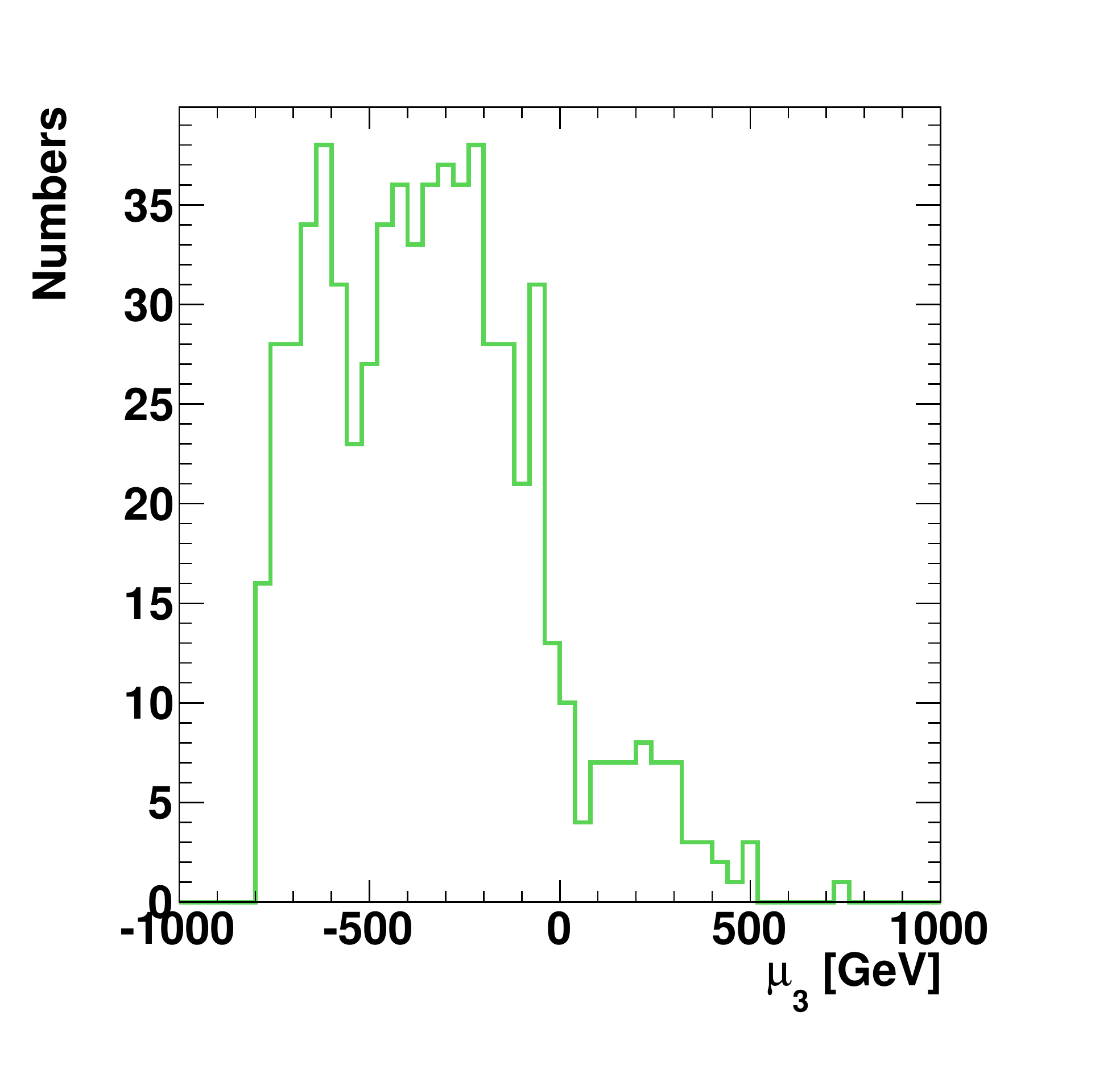} 

\caption{\small In the case IIIb with $u_0 > 0$,  the allowed values of the model parameters ($u_0$, $u_b$), $v_c$, $\mu_{s\phi}$, and $\mu_3$ are shown. The color palette on the right shows the density of the scatter points in one GeV interval. 
}
\label{fig:scan4musph4}
\end{center}
\end{figure}

In our investigation of the phase transition, we focused on looking for the critical temperature when the $\phi$ vev jumps from a non-zero value to 0. We didn't investigate whether there was a jump below this temperature. Nevertheless, we learn from the correlation between the zero temperature vev $u_0$ and the critical temperature vev $u_b$: if there was any big difference between them, we would expect another phase transition. In our scan, we find a rough linear relation between $u_0$ and $u_s$ (shown in Figure~\ref{fig:scan4musph4}), implying that the zero temperature vacuum and the above-critical temperature vacuum are on the same branch.

\begin{figure}[!htb]
\begin{center}
\includegraphics[width=0.3\textwidth]{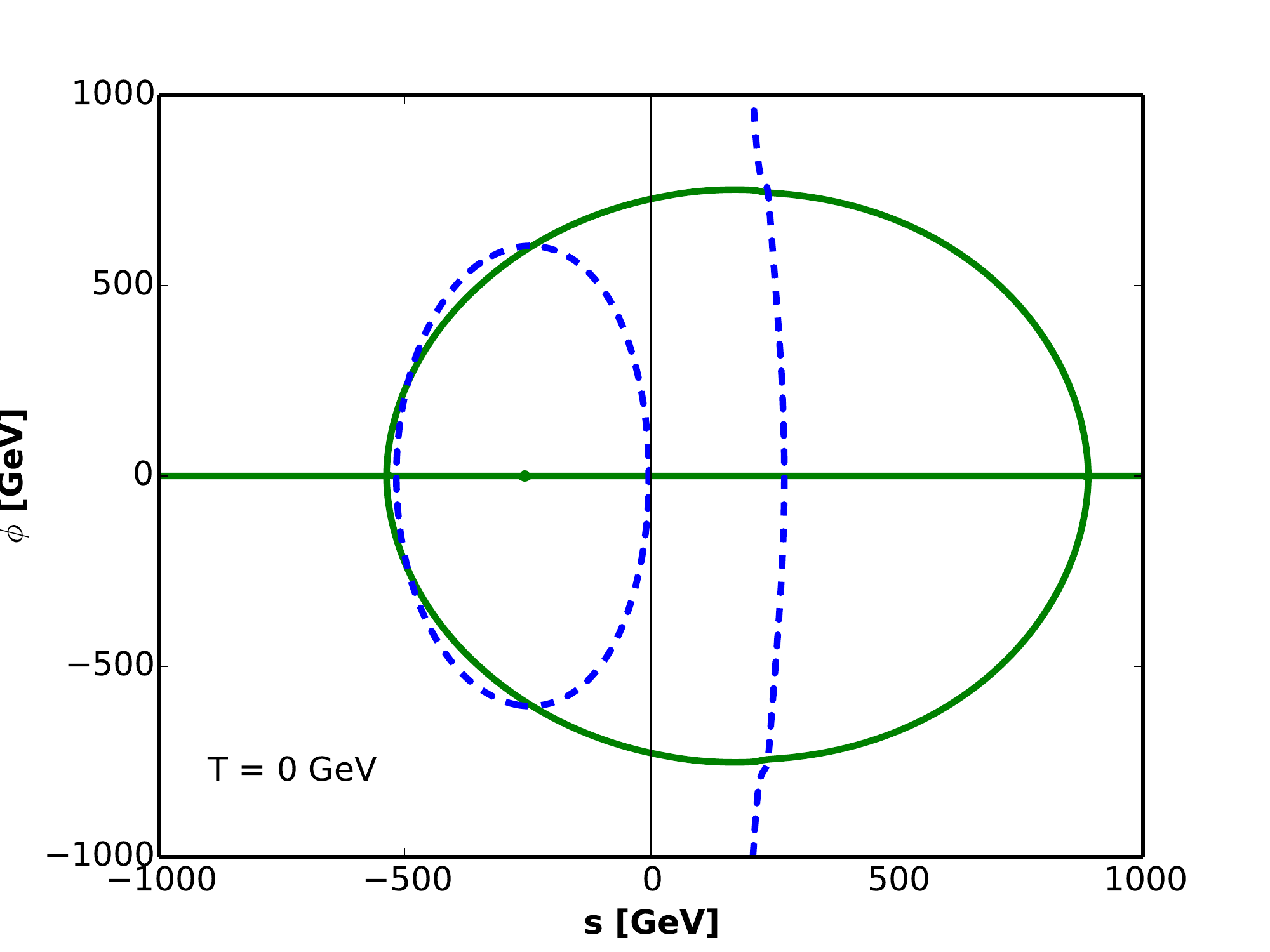} 
\includegraphics[width=0.3\textwidth]{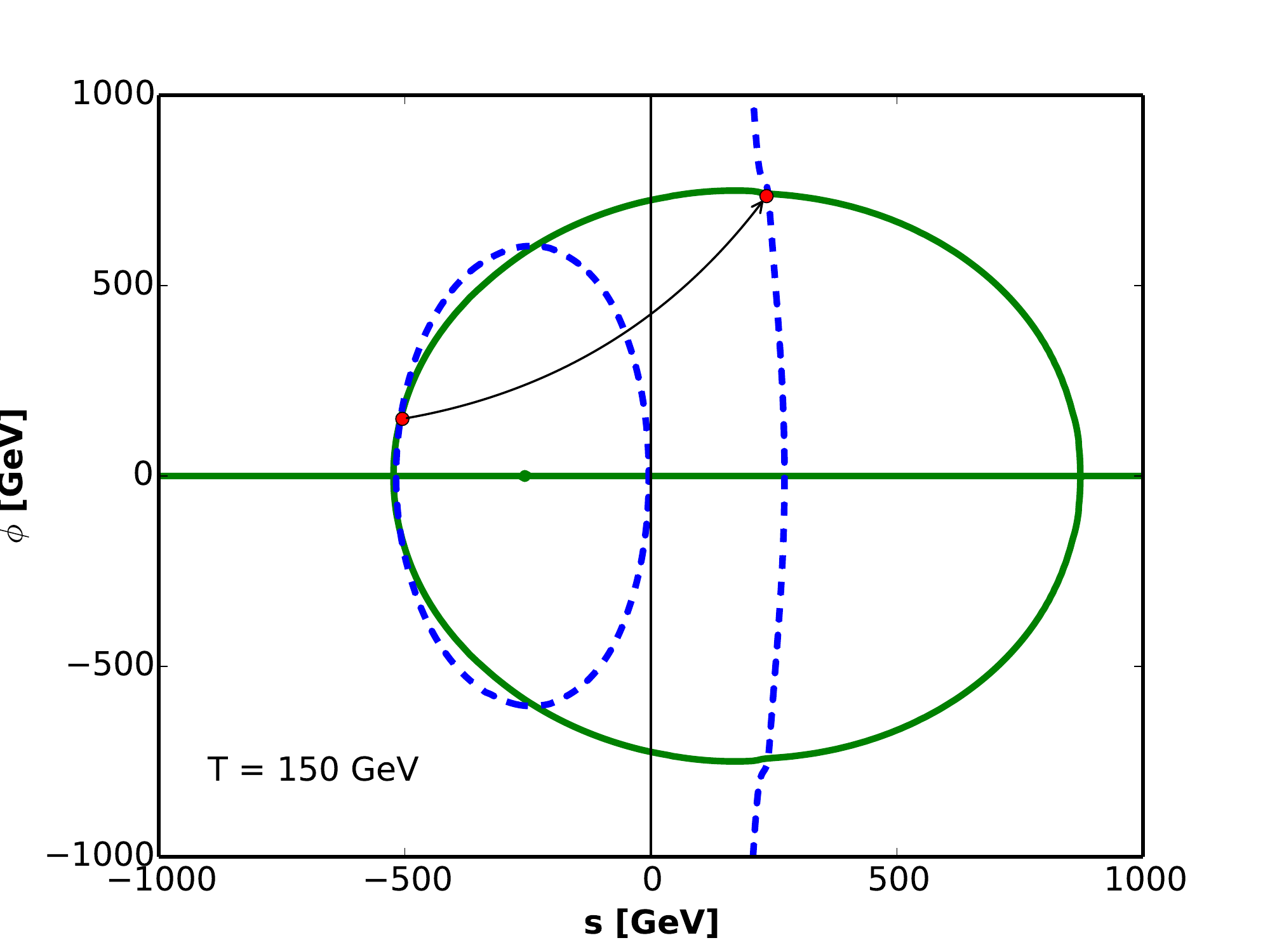} \\
\includegraphics[width=0.3\textwidth]{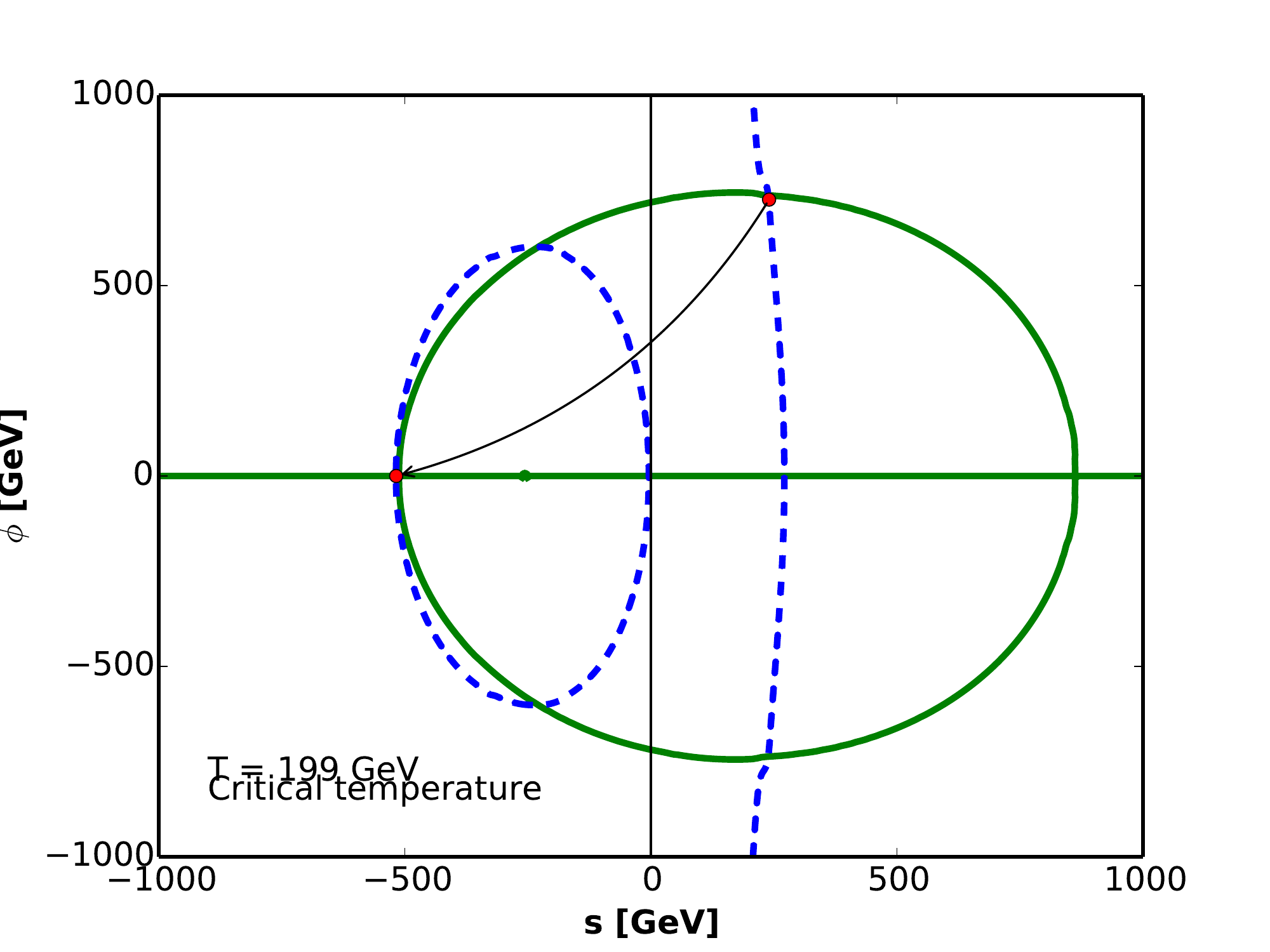} 
\includegraphics[width=0.3\textwidth]{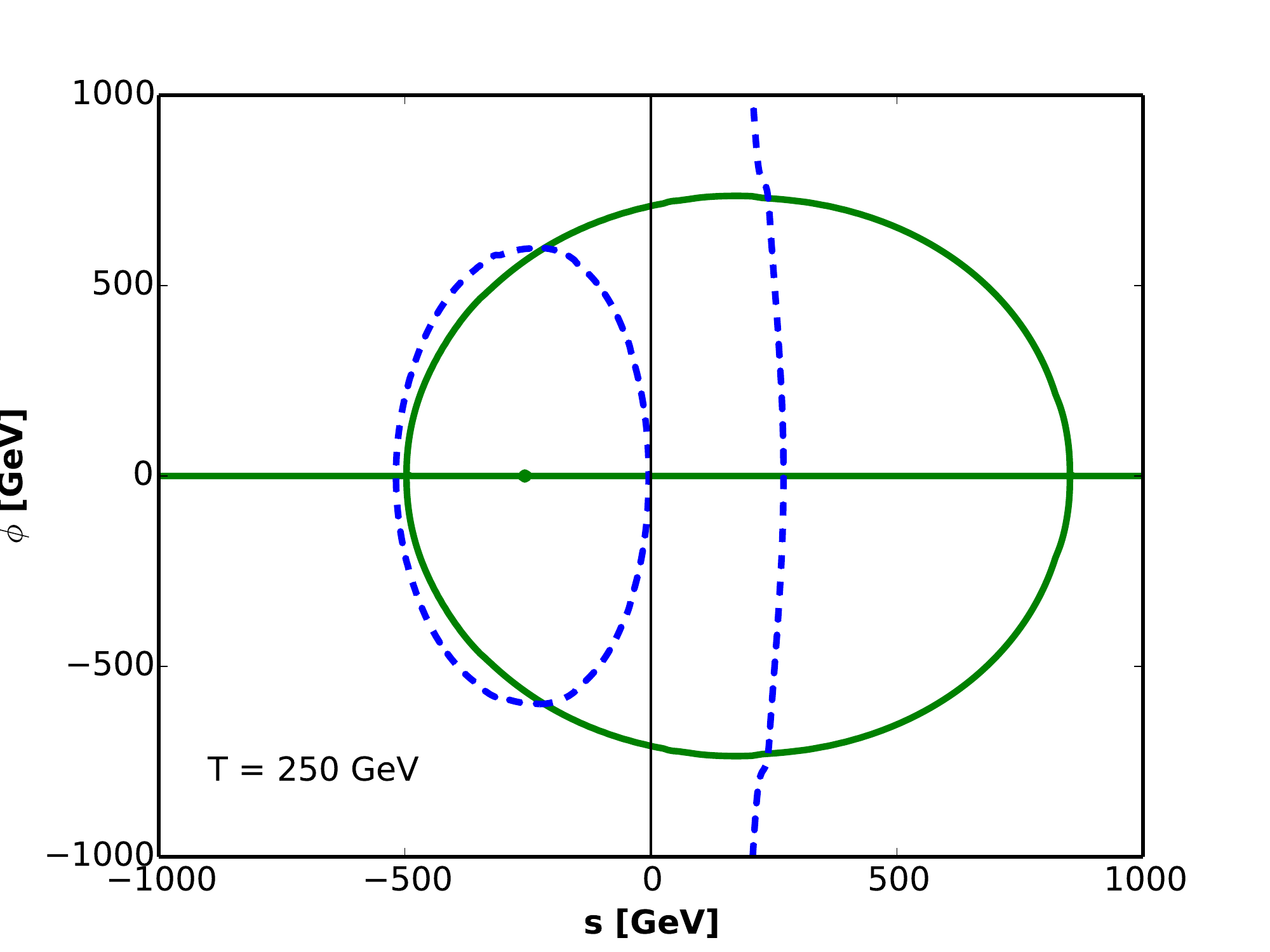}
\caption{\small In the phase transition pattern IIIb, the curves are same as the Figure~\ref{fig:phase1}. }
\label{fig:phase4}
\end{center}
\end{figure}

There are many possible ways of realizing this. One of them would be that a elliptical $\phi$ curve is placed across multiple branches of the $s$ curve, forming several broken minima, as shown in Figure~\ref{fig:phase4}. The global minimum may transit from one branch to another, and jump back later, causing the $u_0,u_b,u_s$ correlation previously described. 

Unlike other cases, one could find more events with small values of $\mu_{s\phi}$ in this pattern, indicating a small $s_*$ and thus a $\phi$ curve centered near $(0,0)$, which may be more likely to intersect with multiple branches of the $s$ curve. The signs of the parameters can be inferred from the comparison of potentials among the three points: negative $\mu_3$ is preferred for a final minimum at $(0,u_s > 0)$, and positive $\mu_{s\phi}$ is preferred for a minimum at $(v_c, u_b<0)$. Figure~\ref{fig:scan4musph4} precisely shows these features.

As the Figure~\ref{fig:TcvcFOPT} shows, events concentrate at the region $T_c<v_c<246$, with $T_c$ typically around 100GeV. In this pattern IIIb, however, we found that the critical temperature tends to be large, and $v_c$ is even larger, which is very different from the other patterns. Large critical temperature leaves sufficient space for a second phase transition below it, but may be harmful for a strong enough CP violation strength because it appears in the denominator of the strength with 8 powers. Our estimation of the upper bound for $T_c$ pretty much rule out this pattern as candidate of EW baryogenesis.
Another interesting feature is that this pattern becomes more and more likely as $u_0$ becomes larger. Test scans showed that a larger range of $u_0$ would lead to dramatically more events with pattern IIIb. In order to evade the interference from these "complex" situations, we chose a smaller range of $u_0$ for the scan.

In summary, the above four cases are the typical patterns in our results of the parameter scan. 
The first two cases are the most concerned when $u_0$ is within the typical range of energy scale for new physics, in that they account for 90\% of the data from our scan.
The pattern IIIa is overall very rare, but the pattern IIIb would become favored at large $u_0$. 

\subsection{Numerical Results}

Given the analysis about the transition patterns, we would like to present our numerical results on physical parameters.
Since we performed a random scan on the parameter space with flat prior, 
the distributions of the parameters should reflect the preference in the model to have the strong first order phase transition. 
%
%


\begin{figure}[!htb]
\begin{center}
\includegraphics[width=0.3\textwidth]{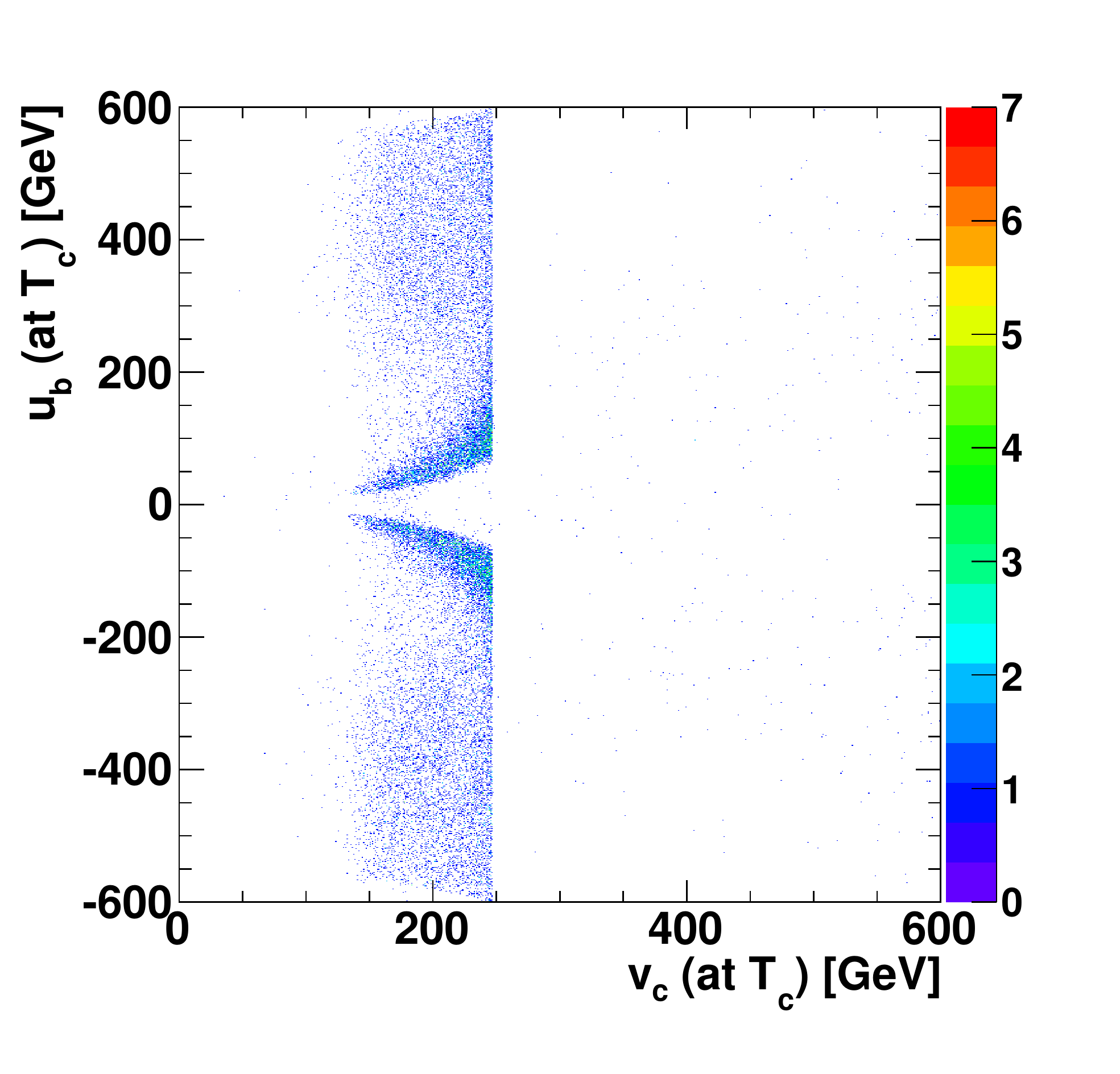} 
\includegraphics[width=0.3\textwidth]{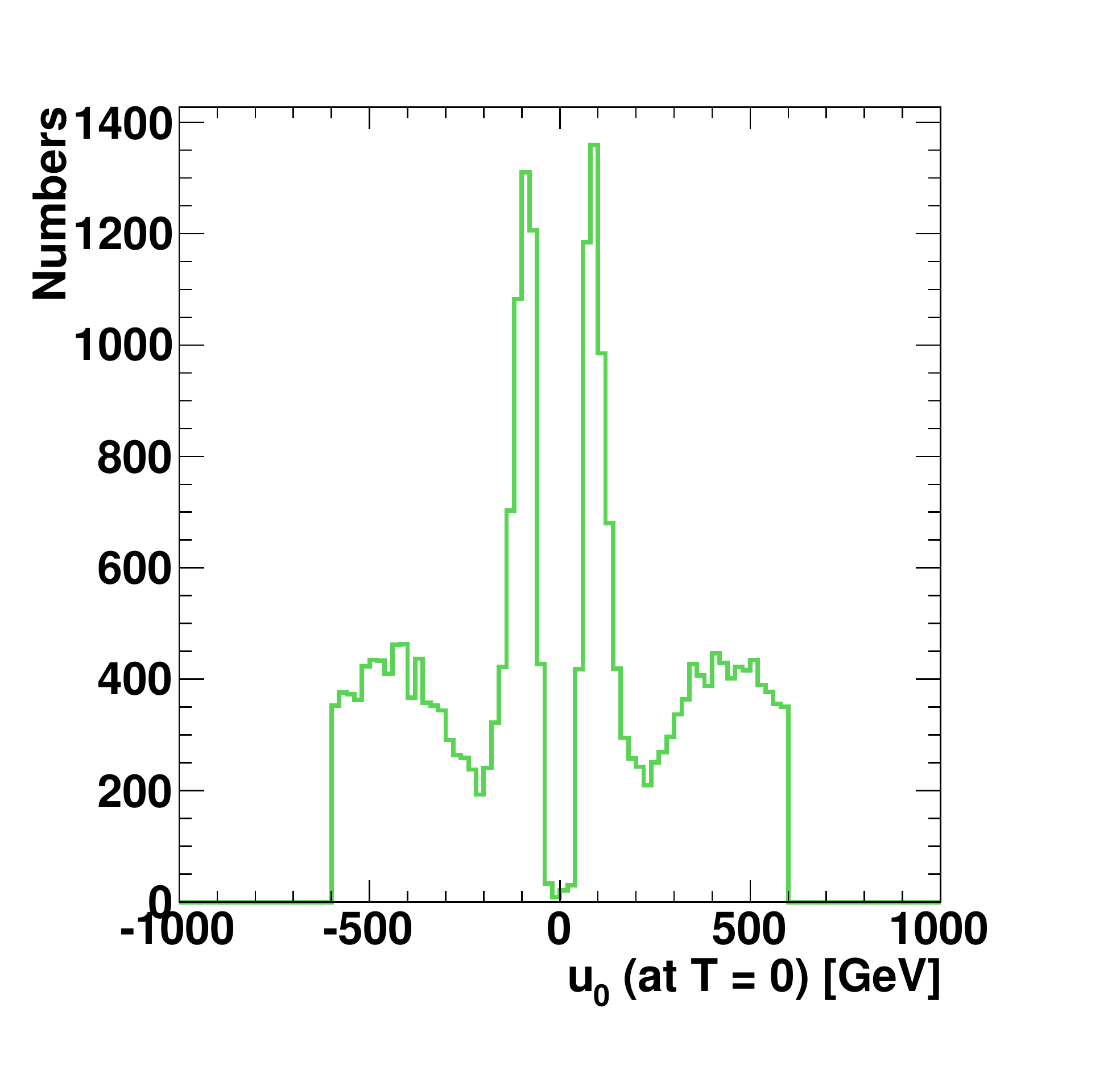} 
\includegraphics[width=0.3\textwidth]{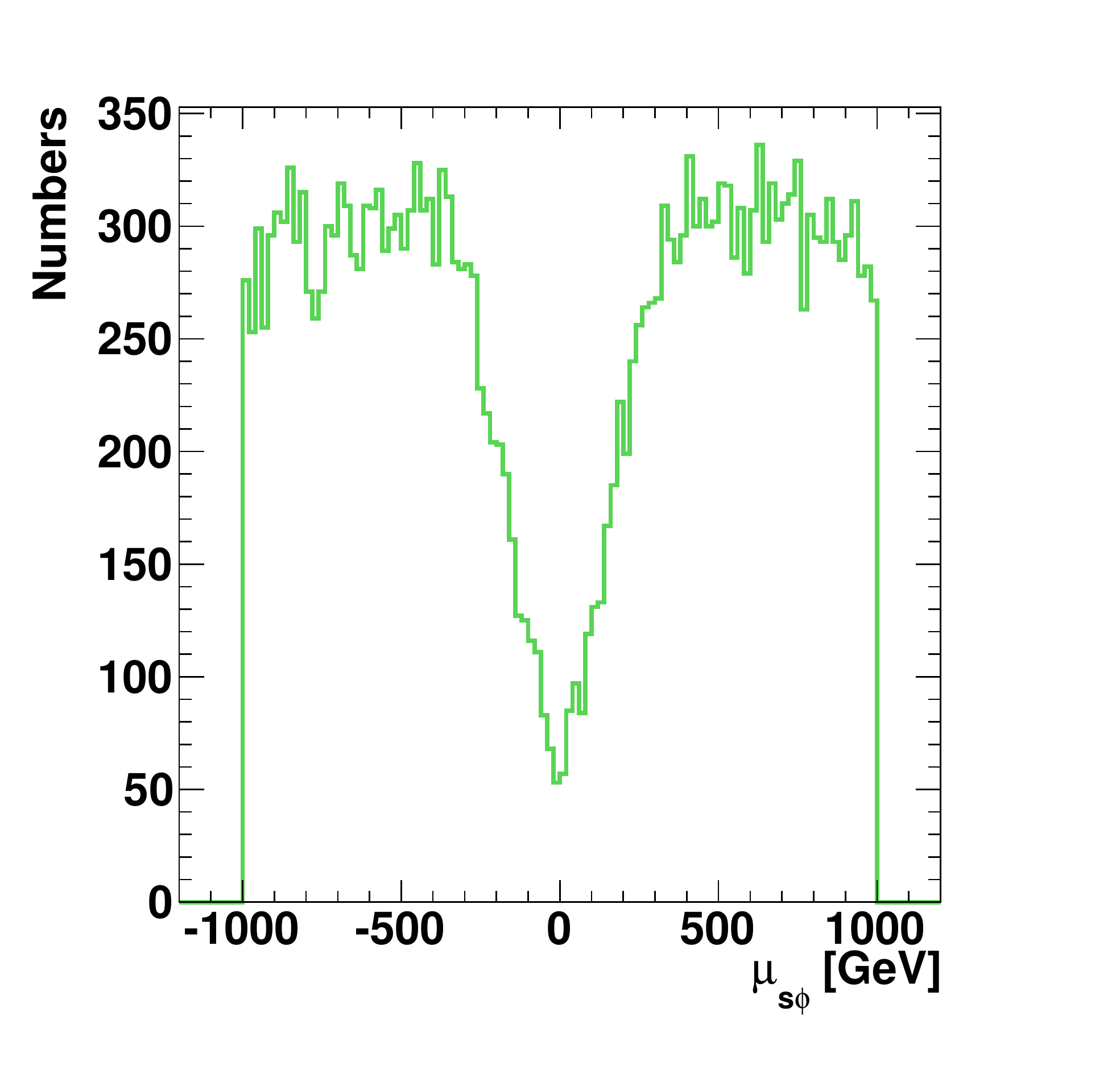} \\ 
\includegraphics[width=0.3\textwidth]{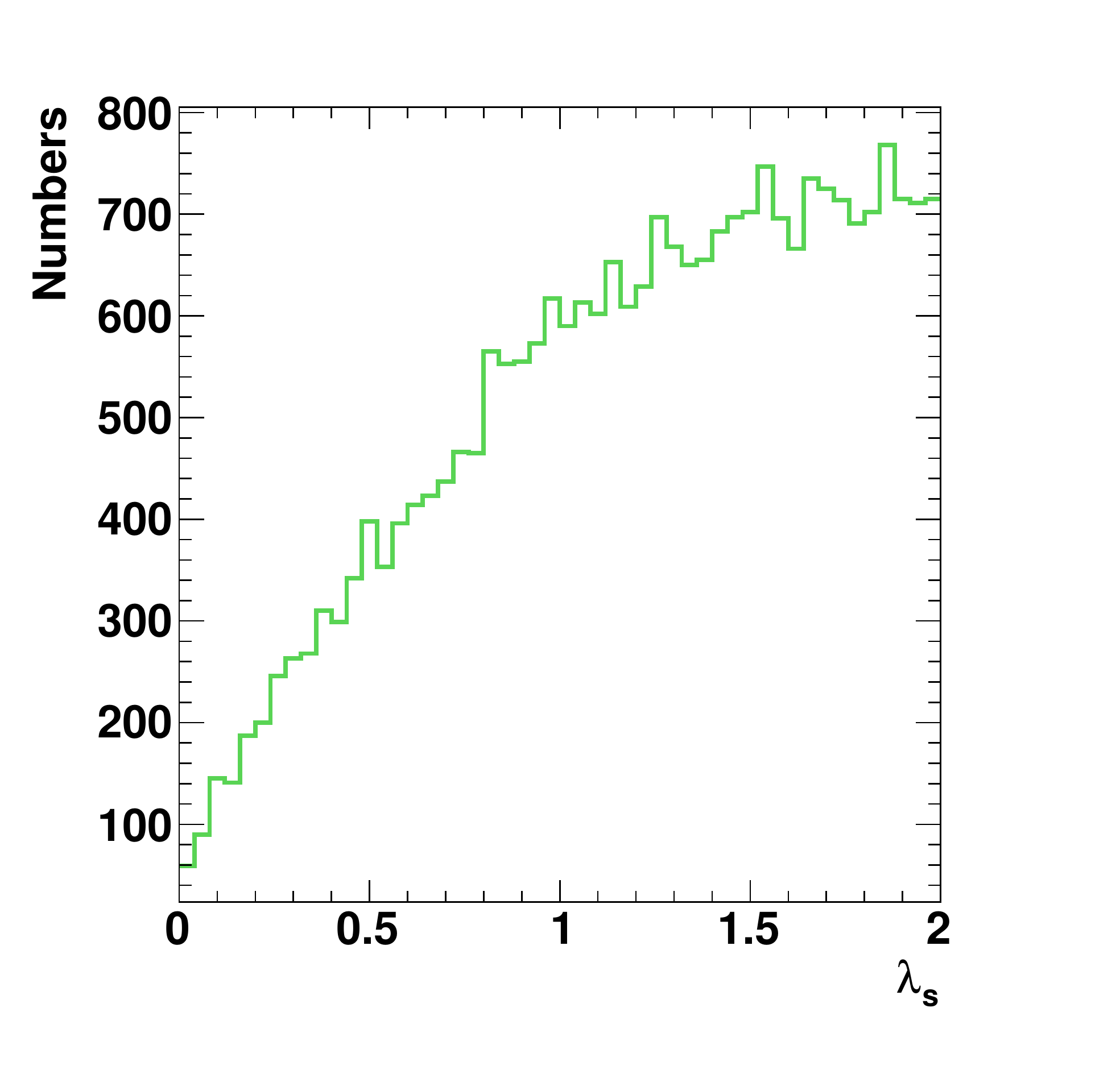} 
\includegraphics[width=0.3\textwidth]{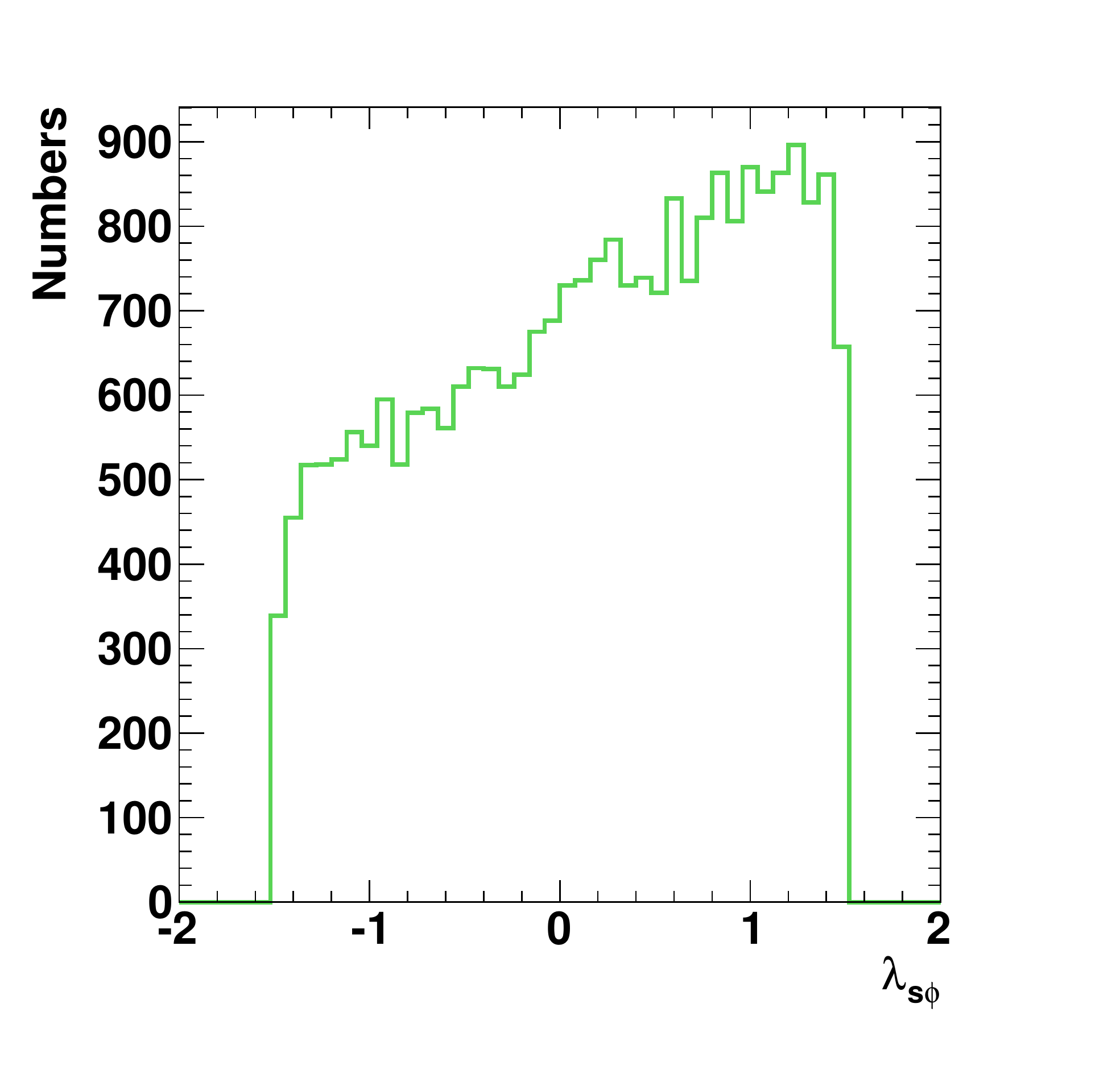}
\includegraphics[width=0.3\textwidth]{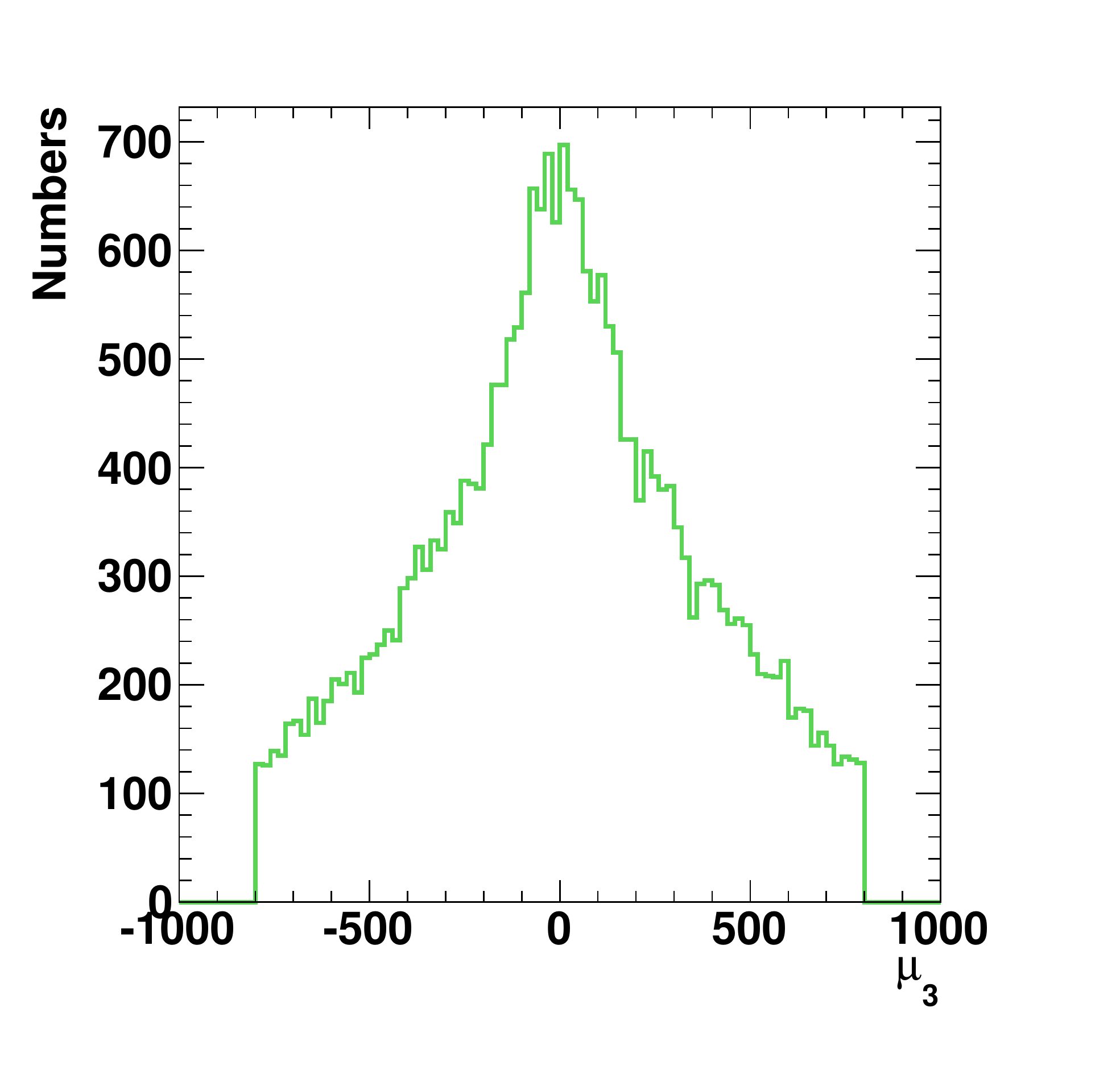}\\
\includegraphics[width=0.3\textwidth]{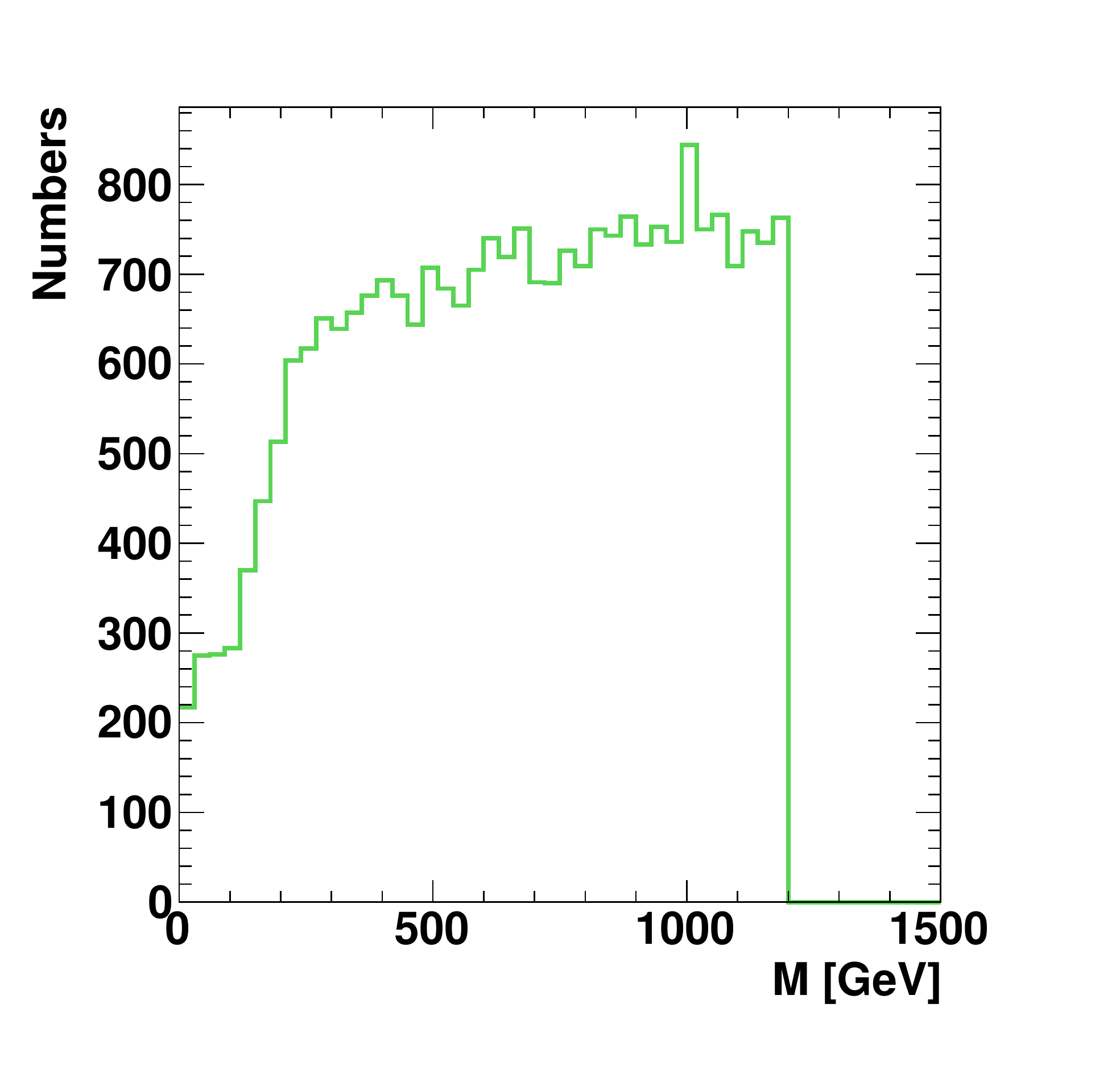} 
\includegraphics[width=0.3\textwidth]{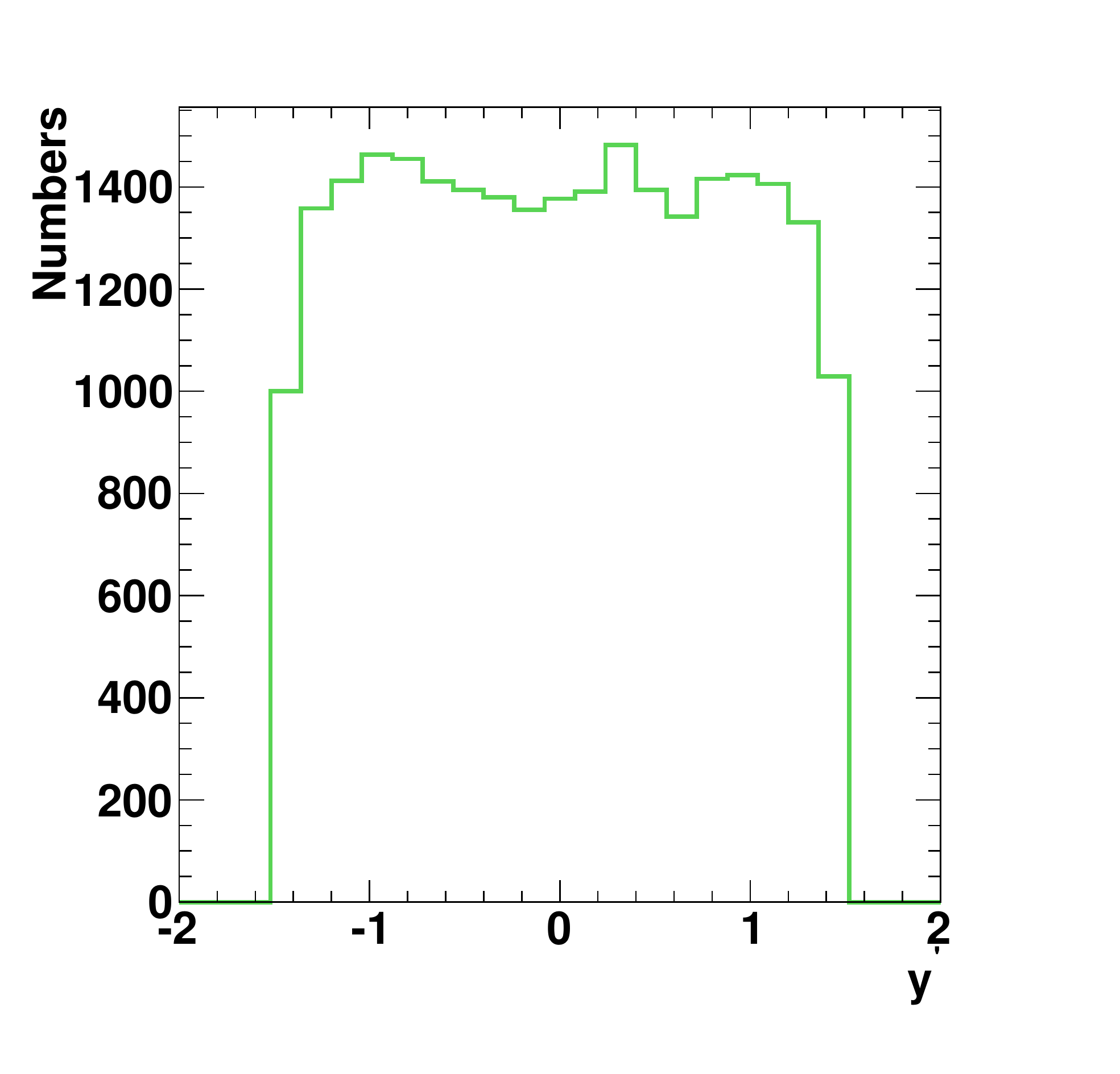} 
\includegraphics[width=0.3\textwidth]{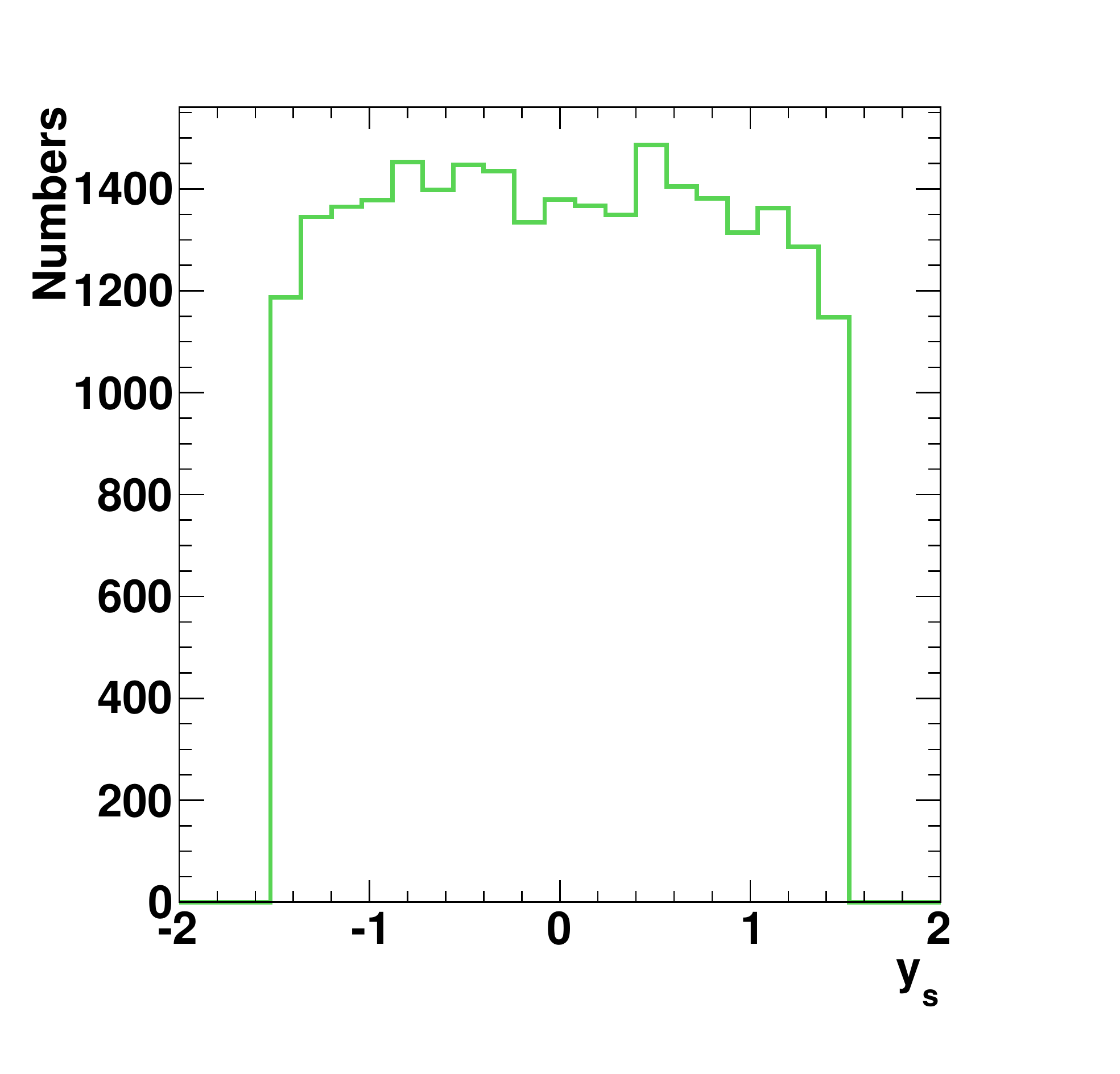}
\caption{\small The allowed values of the broken minimum $(v_c, u_b)$, and all eight input parameters allowed by the strongly 
first order phase transition. }
\label{fig:scaninput}
\end{center}
\end{figure}

Figure~\ref{fig:scaninput} shows the allowed values of the broken minimum $(v_c, u_b)$  and 
all eight input parameters.
This is basically a summary of the detailed discussion in the previous subsection.
To complete our discussion, Figure~\ref{fig:scaninput} also shows the parameters in the fermion sector. 
We learned that the fermion sector is almost irrelavent to the strong first order phase transition.

\begin{figure}[!htb]
\begin{center}
\includegraphics[width=0.3\textwidth]{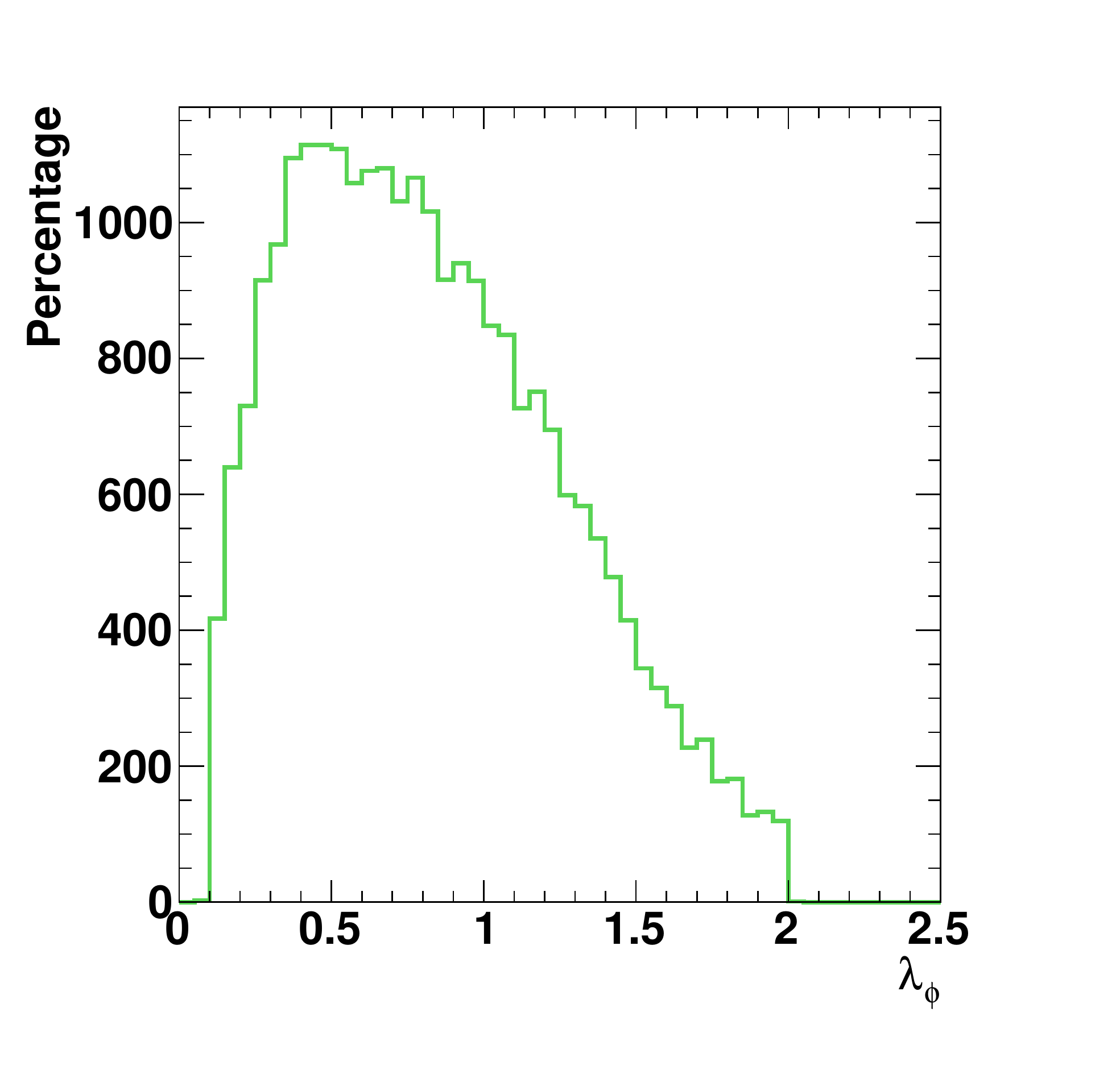} 
\includegraphics[width=0.3\textwidth]{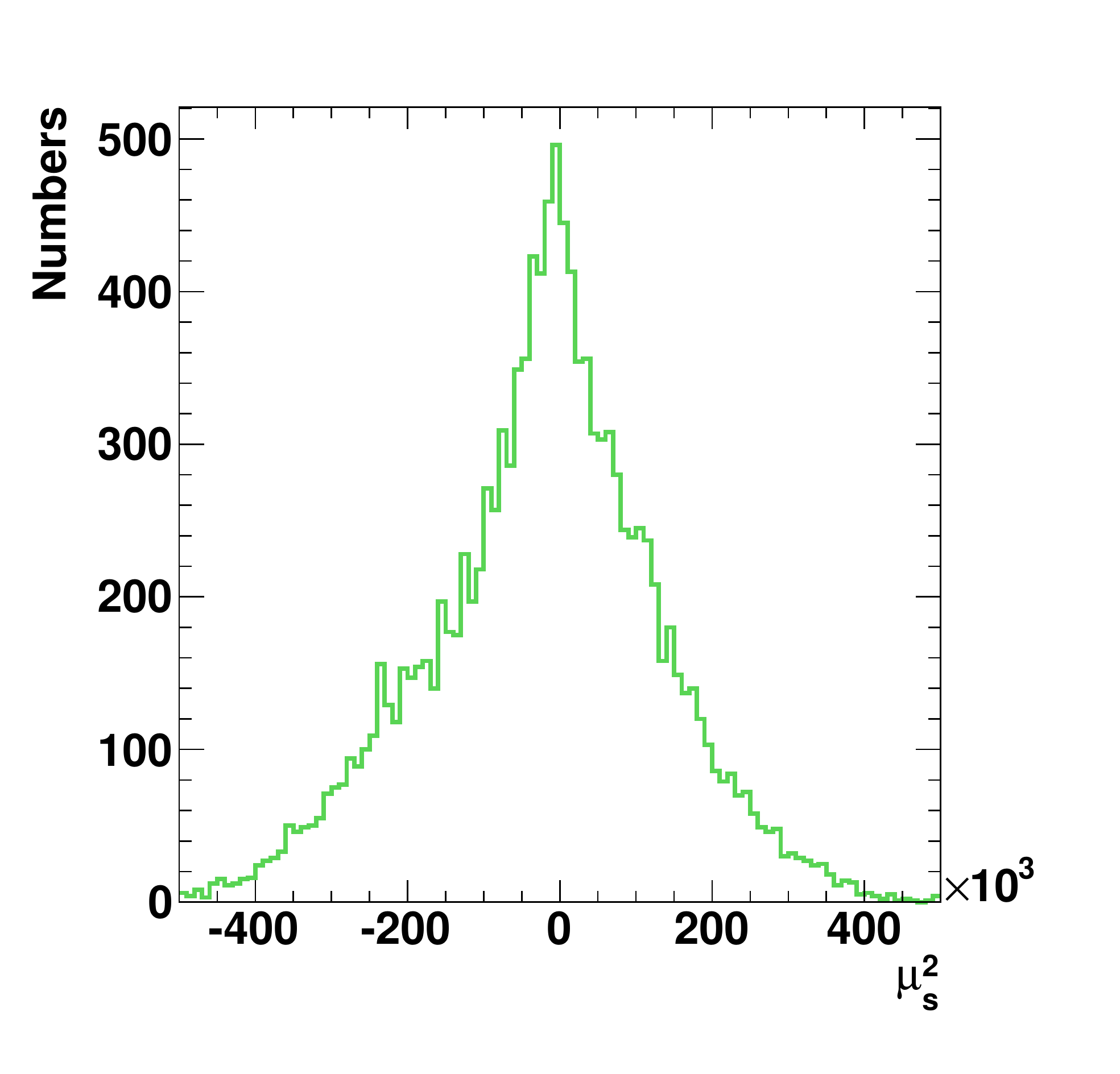}
\includegraphics[width=0.3\textwidth]{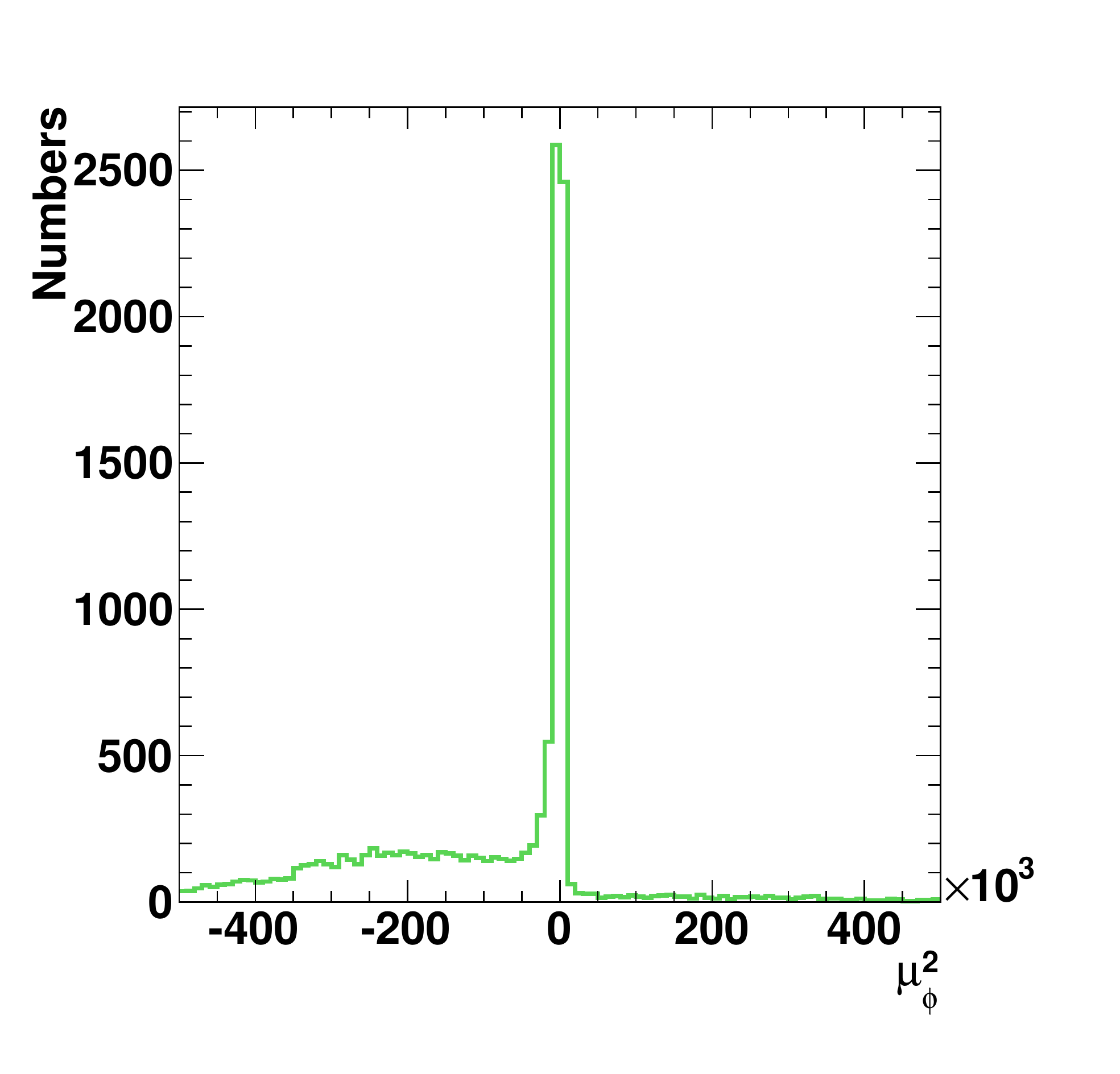}
\caption{\small The allowed values of the derived parameters $\lambda_\phi$, $\mu_s^2$ and $\mu_\phi^2$ in GeV. }
\label{fig:scanoutput1}
\end{center}
\end{figure}

We also show the allowed region for the derived parameters in Figure~\ref{fig:scanoutput1}.
We found that the value of the $\lambda_\phi$ is bounded from a minimum value of about 0.1, and peaks at around 0.5. 
This indicates that a Higgs self-coupling stronger than that in the SM is expected in the singlet scalar extended model.
The $\mu_s^2$, as a controller of the intersection between the $s$ curve and the $s$ axis, has been discussed in the last section. Obviously, for patterns I and II, it tends to be negative and positive respectively. Interestingly, Figure~\ref{fig:scanoutput1} shows that it has no specific sign preference over all. 
The $\mu_\phi^2$ parameter characterize the intersection between the $\phi$ curve and the $\phi$ axis. There are non-zero intersecions only when $\mu_\phi^2>0$. The high peak in Figure~\ref{fig:scanoutput1} represents the pattern I in which $\phi$ curve roughly go across the origin $(0,0)$. The large tail on the negative side represents pattern II, in which the $\phi$ curve is usually far away from the $\phi$ axis.

\begin{figure}[!htb]
\begin{center}
\includegraphics[width=0.3\textwidth]{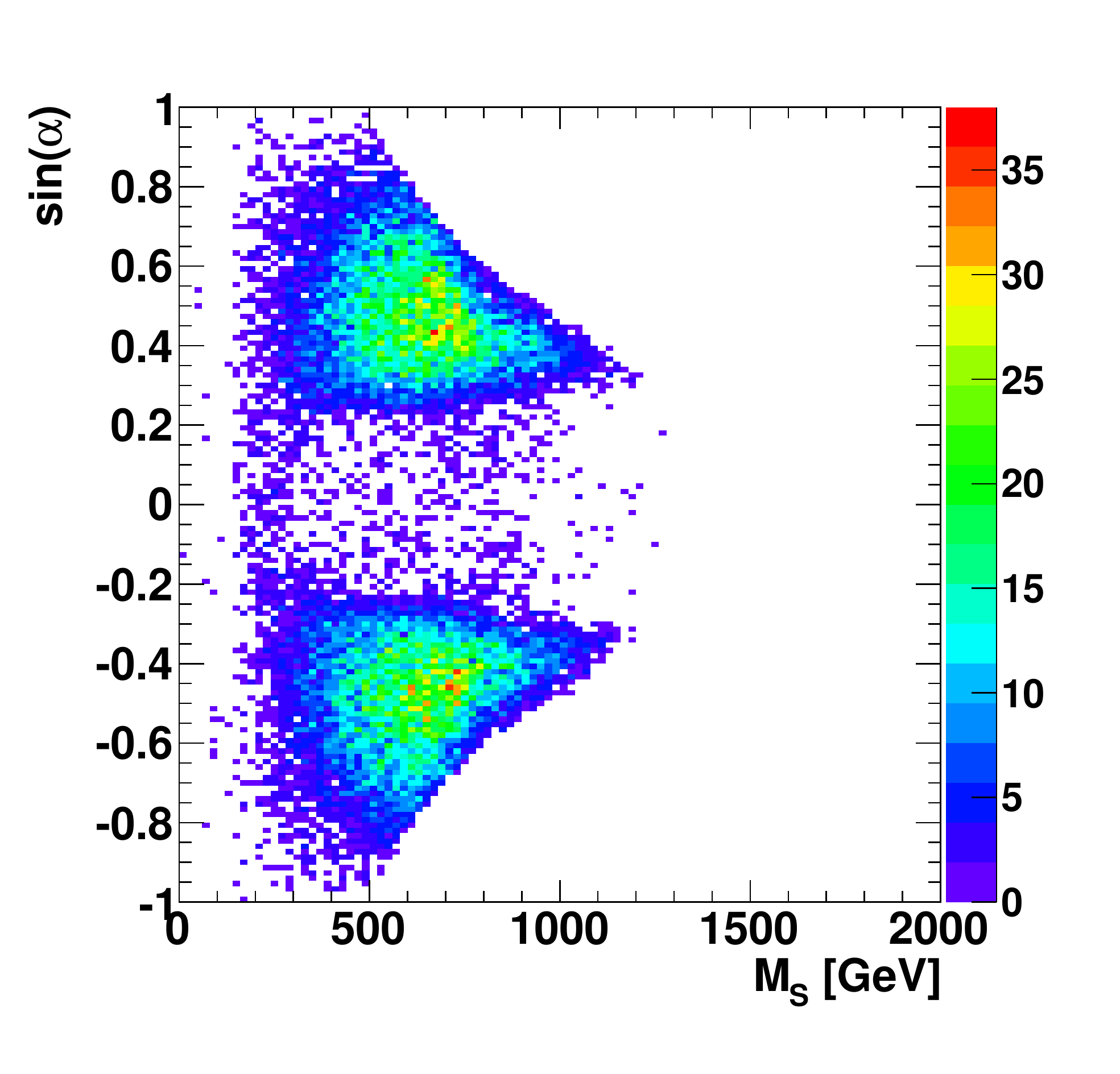} 
\includegraphics[width=0.3\textwidth]{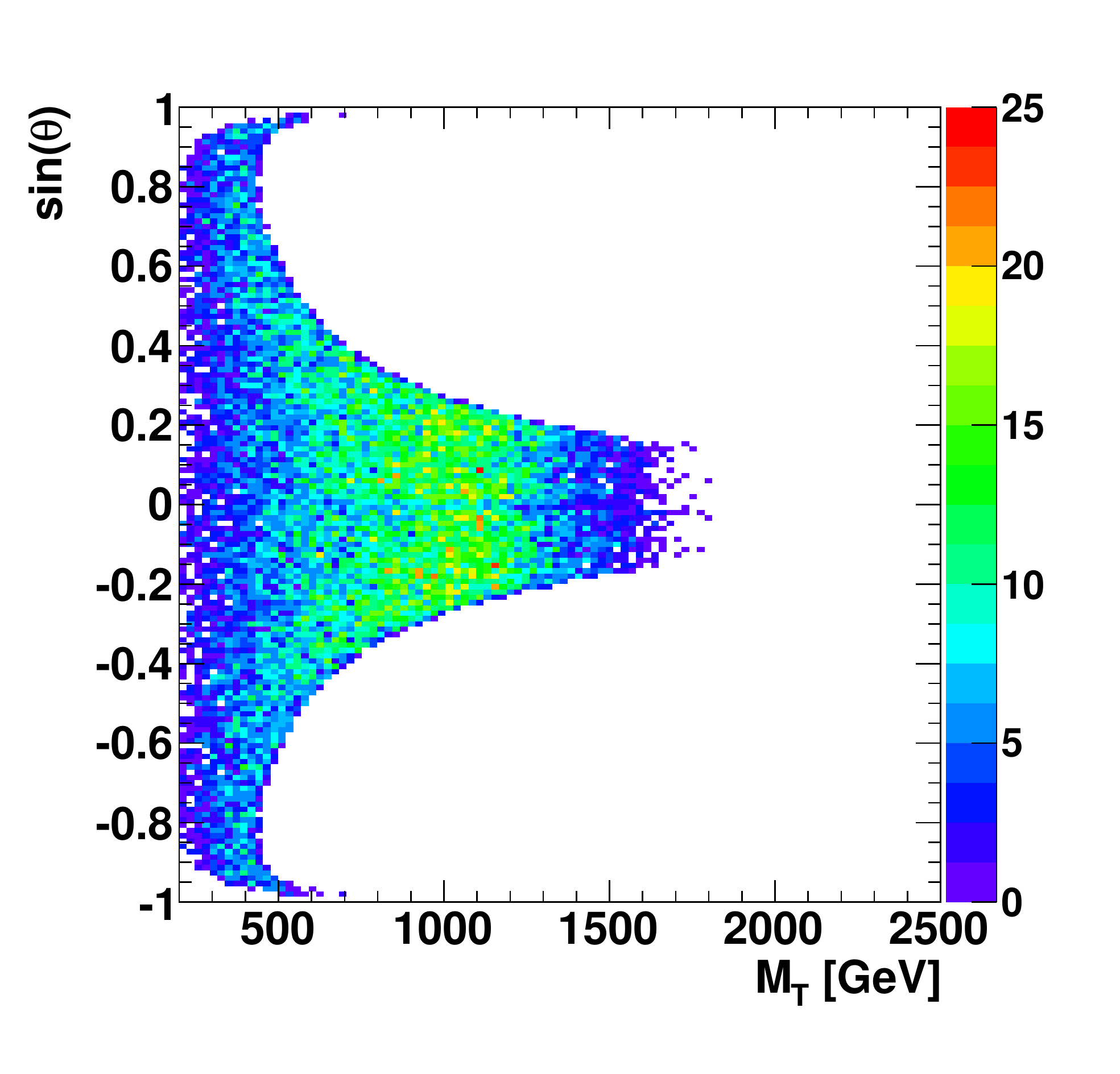}
\includegraphics[width=0.3\textwidth]{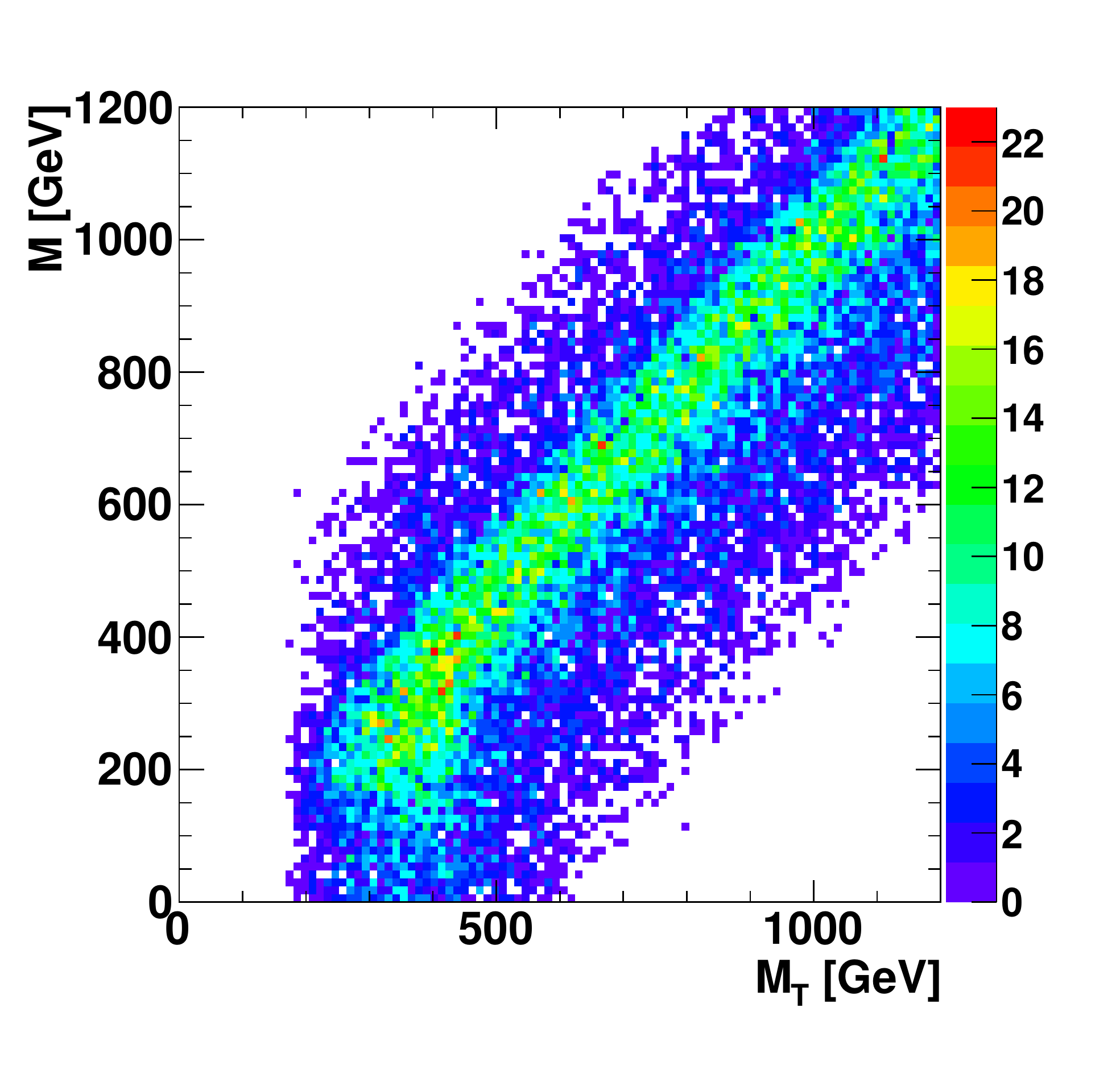} 
\caption{\small The allowed values of the contours of the physical mass and couplings  $(m_S, \sin\varphi)$ in the scalar sector, and $(m_T, \sin\theta_L)$ and $(M, m_T)$ in the fermion sector.  }
\label{fig:scanoutput2}
\end{center}
\end{figure}

Besides the input parameters in the model, we also obtain the favored region of the physical observables, like the masses and mixing angles of the new particles, in light of the strong first order phase transition.
In Figure~\ref{fig:scanoutput2}, we present the two dimensional contours of the physical parameters $(m_S, \sin\alpha)$ in the scalar sector, and those of $(m_T, \sin\theta)$ and $(M, m_T)$ in the fermion sector.
It is shown that the scalar with its mass around 500 - 1000 GeV and a medium mixing angle is favored. 
We recognized the feature that small mixing angle are disfavored, as expected from the fact that the scalar needs to couple with the Higgs boson to render strong first order phase transition. 
This favored region is compatible with that allowed by vacuum stability criteria~\cite{Xiao:2014kba}.
Unlike the scalar mixing angle, the fermionic mixing angle can be very small, which indicates the decoupling between 
the new fermion and the phase transition criteria. 
We expect that the precision data and the Higgs data will put stronger constraints on the fermion sector.
Finally, the strong correlation in the last panel implies that the mass of the new fermion is mainly controlled by the Dirac mass term, thus $M+y_su_0 \approx M$, which may result from a relatively small region of $u_0$ that we chose.

%
%


\section{Implications at the LHC}
\label{sec:lhc}

From the numerical results, we found that different transition patterns exist, among which parameter preference are different. Regarding to 
the physical parameters, we note that a new scalar boson with $500 \sim 800 $ GeV mass and  medium mixing angle are favored. 

In this section,  we check
whether the strongly FOPT parameter region is still allowed by the current experimental constraints, such as the oblique corrections $S, T$, Higgs coupling measurements, and direct collider searches.

The oblique corrections $S, T$ put the tightest constraints on the fermion mass and mixing angle. 
%
For a singlet vector-like fermion, the $Zb\bar{b}$ measurement is less stringent than the oblique correction $T$.
Here we collect the results~\cite{Xiao:2014kba} on the oblique parameters into boson-loop contributions $T_S,\ S_S$ and 
fermion-loop contributions $T_F,\ S_F$.
For the fermion-loop contributions, the NP effect is only involved in the vacuum polarization amplitudes 
 where the top quark and heavy top quark are in the loop.
Subtracting the SM contributions due to the third generation quarks
\bea
	T^{\textrm{SM}}_F &=& \frac{3 m_t^2}{4\pi e^2 v^2} , \\
	S^{\textrm{SM}}_F &=& \frac1{2\pi} \left( 1 - \frac13 \log\frac{m_t^2}{m_b^2} \right) ,
\eea
we arrive at the expressions from the fermion contributions
\bea
\label{eq:deltaTf}
\Delta T_F&=&T^{\textrm{SM}}s_L^2\Big[-(1+c_L^2) + s_L^2\frac{m_T^2}{m_t^2} + c_L^2\frac{2m_T^2}{m_T^2-m_t^2}\ln\frac{m_T^2}{m_t^2}\Big],\\
\Delta S_F&=&-\frac{s_L^2}{6\pi}\Big[(1-3c_L^2)\ln\frac{m_T^2}{m_t^2}+5c_L^2 - \frac{6c_L^2m_t^4}{(m_T^2-m_t^2)^2}\Big(\frac{2m_T^2}{m_t^2}-\frac{3m_T^2-m_t^2}{m_T^2-m_t^2}\ln\frac{m_T^2}{m_t^2}\Big)\Big],
\eea
Similarly the contributions from the Higgs and scalar loop are 
\bea
\Delta T_S &=&  s_{\varphi}^2 \Big[ T_s(m_S^2) - T_s(m_h^2)  \Big], \\
\Delta S_S &=&  s_{\varphi}^2 \Big[ S_s(m_S^2) - S_s(m_h^2)  \Big],
\eea
where the functions are defined as
\bea
\label{eq:oblqf}
T_s(m)&=&-\frac{3}{16\pi c_W^2}\bigg[\frac{1}{(m^2-m_Z^2)(m^2-m_W^2)} \Big(m^4\ln m^2-s_W^{-2}(m^2-m_W^2)m_Z^2\ln m_Z^2 \nn\\
&&+ s_W^{-2}c_W^2(m^2-m_Z^2)m_W^2\ln m_W^2\Big)-\frac{5}{6}\bigg], \\
S_s(m)&=&\frac{1}{12\pi}\bigg[\ln m^2-\frac{(4m^2+6m_Z^2)m_Z^2}{(m^2-m_Z^2)^2} + \frac{(9m^2+m_Z^2)m_Z^4}{(m^2-m_Z^2)^3}\ln\frac{m^2}{m_Z^2}-\frac{5}{6}\bigg].
\eea
We note that the scalar contributions are much smaller than the fermion contributions.
Thus the constraints from $S, T$ on the scalar mass and mixing angle are quite weak.

On the other hand, the Higgs coupling measurements  at the LHC put the tightest constraints on the scalar mixing angle.
In our model, due to mixing between the Higgs boson and the heavy scalar, all the 
tree-level Higgs couplings are modified as
\bea
	g_{h ff}^{\tbox{NP}} = \cos\varphi g_{h ff}^{\tbox{SM}}, 
	\quad  g_{h VV}^{\tbox{NP}} = \cos\varphi g_{h VV}^{\tbox{SM}}.
\eea
At the same time, the loop-induced Higgs couplings to the photon and the gluon are also modified 
by the new contribution from the vector-like fermion loop. 
So the Higgs couplings to the photon and the gluon are
\bea
	g_{h gg}^{\tbox{NP}}          &=&
	 \frac{g_s^2}{16\pi^2} \left( \sum_{f} \frac{g_{hff}}{m_f} A_{1/2}(\tau_f) + \frac{g_{hTT}}{m_T} A_{1/2}(\tau_T)\right), \\
    g_{h\gamma\gamma}^{\tbox{NP}} &=& \frac{e^2}{16\pi^2}\left[\frac{g_{\tbox{hWW}}}{m_W^2} A_{1}(\tau_W) 
	+ \sum_{f} 2 N_{c}^{f} Q_f^2 \frac{ g_{hff} }{m_f} A_{1/2}(\tau_f)   +  \frac83 \frac{g_{hTT}}{m_T} A_{1/2}(\tau_T) \right],
	\label{eq:couphignew}
\eea
where $\tau_T = \frac{4 m_T^2}{m_h^2}$, and $g_{htt}, g_{hTT}$ are the Higgs couplings to the fermions, given in Ref.~\cite{Xiao:2014kba}.
In the Higgs to photon and gluon process, there is almost no fermion mass dependence if the heavy vector-like fermion 
is much heavier than the top quark. 
Therefore, the Higgs coupling measurements put constraints on the scalar and fermion mixing angles. 

The direct searches on the new fermion and new scalar boson at the LHC 
also put tight constraints on their masses and couplings. 
At the LHC, the vector-like quark can be produced in pair through QCD production $pp\to T\bar{T}$, 
or be singly produced via electroweak process $pp \to T\bar{b}$. 
In our paper, the up-type vector-like quark predominantly couple to the third-generation quarks
through $T \to tZ, th, Wb$. 
From an updated CMS analysis~\cite{Chatrchyan:2013uxa} which uses the 8 TeV data collected 
up to integrated luminosity of $19.5 $ fb$^{-1}$,
the lower limits on the mass $m_T$ are set to be around $687 - 782$ GeV.

The heavy scalar $S$ is CP-even and has the same quantum number as the SM Higgs boson. 
The search limits on the high mass Higgs boson at the LHC
could be recasted to put constraints on the mass and the coupling of the heavy scalar.
The production mechanism is quite similar to the SM Higgs boson.  
The dominant channel is the gluon fusion channel $ gg \to S$. 
The decay channels of the $S$ boson include 
\bea
	S \to WW, \quad S \to ZZ, \quad S \to hh, \quad S \to t\bar{t},
\eea
and  $S \to t T$ only if the heavy top is much lighter than the scalar $S$.
Other decay channels, such as $S \to \gamma \gamma/gg$, $S \to f\bar{f}$, 
where $f$ is the fermion other than the top quark, are negligible. 
Using the updated analysis~\cite{Chatrchyan:2013yoa} in the $S \to WW$ and $S \to ZZ$ decay channels from CMS, 
the mass range between $145$ GeV and $1000$ GeV has been investigated.
According to the analysis, if the high mass Higgs boson has the same coupling as the SM, 
the mass range between $145$ GeV and $710$ GeV are excluded at the 95\% CL. 
In our model, we recast the exclusion limits to the constraints on the scalar mass and mixing angles.

\begin{figure}[!htb]
\begin{center}
\includegraphics[width=0.47\textwidth]{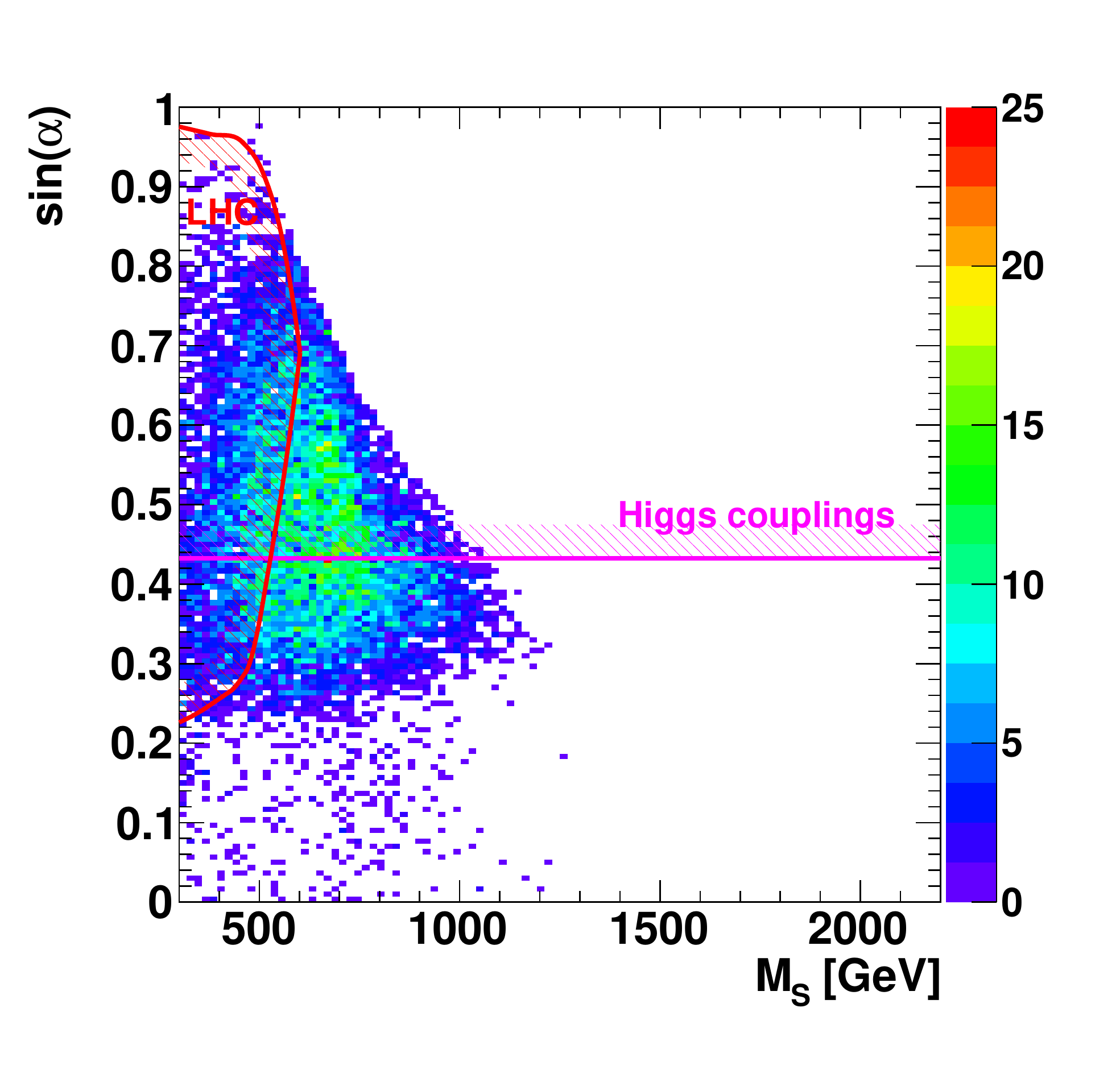} 
\includegraphics[width=0.47\textwidth]{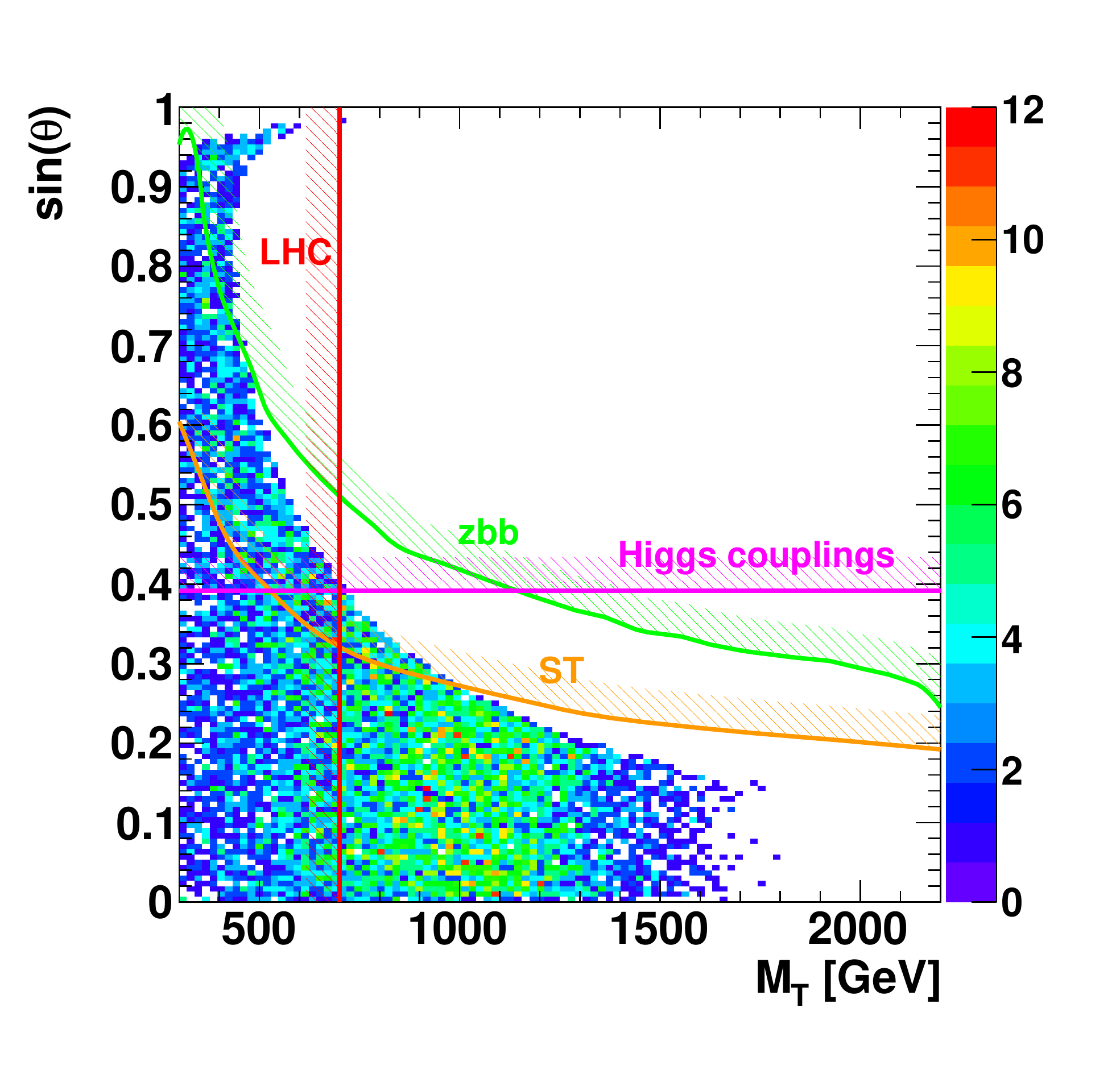}
\caption{\small The allowed parameter contour $(m_S, \sin\varphi)$ (left) and $(m_T, \sin\theta_L)$ (right) in light of the strong first order phase transition. 
The constraints from the $S, T$ parameters, Higgs coupling measurements, and direct LHC searches are shown as the exclusion lines.
}
\label{fig:scanoutput3}
\end{center}
\end{figure}

The numerical results on the constraints from the $S, T$ parameters, Higgs coupling measurements, and direct LHC searches are shown in Figure~\ref{fig:scanoutput3}.
From the Figure~\ref{fig:scanoutput3} (left), the parameter region with the mixing angle $\sin\alpha$ greater than 0.43 has been ruled out by the current Higgs coupling measurements. 
The direct LHC searches also exclude the scalar boson with mass less than 500 GeV.
We expect that the future Higgs coupling data put more stringent constraints on the mixing angle $\sin\varphi$, and thus put stronger limit on the 
favored region by the strong first order phase transition.
On the right panel of the Figure~\ref{fig:scanoutput3}, we note that 
the $S, T$ parameters can only exclude very small region which is favored by the strong first order phase transition.
The direct LHC searches could exclude the top partner with mass less than 700 GeV. 
There are large available parameter regions in the fermion sector. 
Therefore, the CP violation rate from the fermion sector is adequate to generate the needed baryon number asymmetry.

Finally, we expect that the future Higgs data could explore the parameter region on the $(m_S, \sin\varphi)$ contour. 
Furthermore, from the Figures~\ref{fig:scaninput} and~\ref{fig:scanoutput1}, we note that  
the large scalar coupling $\lambda_{s\phi}$ and moderate $\lambda_\phi$  is strongly favored. 
We should be able to explore the scalar trilinear coupling $\lambda_{s\phi}$ and $\lambda_\phi$  at the high luminosity LHC.
If the trilinear couplings are enhanced compared to the SM Higgs self-coupling, the 
Higgs pair production cross section should be larger than the SM value. 
Through the Higgs pair production process $pp \to h/S \to h h$, we could extract out the trilinear couplings from the production cross section measurements.


\section{Conclusions}
\label{sec:conclusion}

We investigated the necessary conditions to realize the electroweak baryogenesis 
in a scalar-assisted vector-like fermion model.  
In the fermion sector, the extended CKM matrix provides additional sources of the 
CP violation effects, parametrized by Jarlskog-like invariant. 
We found that the CP violation rate is greatly enhanced by the heavy mass of the new fermion. 
With the flavor constraints on the extended $4 \times 3$ CKM matrix considered,
we estimated the CP violation strength, which turns out to be adequate for the baryon number asymmetry.

We focused on the one-loop, finite-temperature effective potential in our model and its implications on the electroweak phase transition. 
Unlike the case of the SM, 
the new scalar extends the field space in which the phase transition occurs. In the two-dimensional field space, we have more possible ways of constructing barriers between minima. 
We utilized the shape of the derivative of the potential: $s$ curve and the $\phi$ curve, as a tool to analyse 
the  two-dimensional effective potential. 

The first order phase transition occurs in the form of bubble nucleation of the symmetry broken phase.
The sphaleron decoupling criteria  $\xi = \frac{v_c}{T_c} \ge 1$ is used in this model, to prevent the baryon asymmetry generated in the symmetry broken phase from being washed out by sphalerons.
We performed a parameter scan over the 8 independent model parameters, and 
obtained the allowed parameter region which could have strong first order phase transition.

According to the different regions in the $(u_b, u_s)$ contour at the critical temperature,
transition patterns are classified into four patterns: single-branch barrier transition (pattern I or II, with or without the existence of multiple relevant branches in the $s$ curve), 
inter-branch barrier transition (pattern IIIa) and multi-step transition (pattern IIIb). 
For the single-branch barrier transition, the large trilinear mass term $\mu_{s\phi}$ is favored because the width of the barrier is strongly related to it. 
However, small $\mu_{s\phi}$ is preferred in the patterns IIIa and IIIb. 
The preferences of parameters we have got from the scan results can be justified by analysing the shapes of the $s$ curve and the $\phi$ curve and the intersections between the two curves in different patterns.
We also note that all the patterns prefer large quartic scalar couplings and moderate mixing angle between the Higgs and the scalar.

Finally we combine the constraints from strong first order phase transition and the experimental limits on the $S, T$ parameters, Higgs coupling measurements, and direct LHC searches. 
We found that there is still a significant amount of parameter region for the fermion mass and couplings to satisfy all the constraints, and have adequate CP violation strength to realize the baryon asymmetry at the same time.
%
%
The new scalar with mass around 500 - 1100 GeV and mixing angle $\sin\varphi$ around $0.25 - 0.42$ are still allowed and favored by the strong first order phase transition.
%
%
The future Higgs coupling measurements and the Higgs boson pair production cross section will be able to further explore the allowed parameter space. 
%

\section*{Acknowledgements}

We would like to thank Daniel Chung, Jacques Distler, Vadim Kaplunovsky, Can Kilic, David Morrissey, Michael Ramsey-Musolf for helpful discussions, and especially thank Willy Fischler for his insightful comments. 
The research was supported by the National Science Foundation under Grant Numbers PHY-1315983 and PHY-1316033.


\appendix

%
%
%
%
%



\section{Details of The Effective Potential}
\label{sec:appen1}

\subsection{The Effective Potential in the On-shell Scheme}

In this appendix, we first review how the effective potential is written in on-shell scheme. Then we extend the discussion on SM Higgs boson in Ref.~\cite{Anderson:1991zb, Delaunay:2007wb} to the Higgs-new scalar mixing case.

We start from the one-loop effective potential in the Landau gauge, with the dimensional-regularization applied,
\bea
\label{eq:0TVeff}
V_{\eff}(T=0) &=& V_{\tree} + V_{CW} \nn\\
&=& V_{\tree} + \frac{1}{64\pi^2}\STr M_{\varphi}^4(\phi, s)\Big(\log\frac{M_{\varphi}^2(\phi, s)}{\mu^2}  - \frac32 - 
C_{\varphi}^{\rm UV}\Big),	
\eea
where the super-trace is taken among all the dynamical fields $\varphi$ that have $(\phi,s)$ dependent masses $M_{\varphi}$. 
The UV divergent piece $C_{\varphi}^{\rm UV}$ are defined as
\bea
	C_{\varphi}^{\rm UV} = 
	\begin{cases}
	\frac{1}{2-\frac{n}{2}} -\gamma_E + {\rm log} 4\pi & \varphi = \text{scalar and fermion}\\
	(n-1)\left(\frac{1}{2-\frac{n}{2}} -\gamma_E + {\rm log} 4\pi\right) &\varphi = \text{gauge boson}
	\end{cases}
\eea
The UV divergence has to be absorbed by the counterterms.
We introduce the following counterterms
\bea
\label{eq:ct}
\Delta V &=&  A(\phi^2-v^2) + B(\phi^2-v^2)^2 +  C(s-u) + D (s-u)^2 + E(\phi^2-v^2)(s-u) \nn\\
&& + F (s-u)^3 + G (s-u)^4 + H (\phi^2-v^2)(s-u)^2.
\eea
The renormalization conditions are needed to fix the above counterterms. 
In the $\overline{MS}$ renormalization scheme, the renormalization conditions consists in subtracting the term proportional to
$\frac{1}{2-\frac{n}{2}} -\gamma_E + {\rm log} 4\pi $ in the regularized potential.
We will choose the on-shell renormalization scheme. 
The effective potential  can be expanded using the one-particle irreducible (IPI)
Green function $\Gamma^{(n)}$ at zero external momentum:
\bea
	V_{\rm eff}(\Phi) = -\sum_{n=0}^{\infty} (\Phi- \Phi_{\rm VAC})^n \Gamma^{(n)}(p = 0),
\eea 
where $\Phi$ denotes the scalar fields $(\phi, s)$ and $\Phi_{\rm VAC}$ denotes the vacuum $(v, u)$.
Therefore, we define the renormalized mass of the scalar field as the negative
inverse propagator at zero momentum
\bea
\left.
	M_{ij}^2 = - \Gamma^{(2)}(p = 0) =  \frac{\partial^2 V_{\rm eff}}{\partial\Phi_i\partial\Phi_j}\right|_{\phi=v,s=u}.
\eea
Of course, we could also define the renormalized couplings as the four-point IPI Green function $\Gamma^{(4)}$.
However, the renormalization conditions on the couplings are not unique. 
Since we want to keep our discussion as general as possible, we only impose the tadpole conditions and mass conditions, as follows 
%
\eq{
\label{eq:OS}
&\left.\frac{\partial}{\partial \Phi_i}(V_{CW} + \Delta V)\right|_{\phi=v,s=u} = 0,\\
&\left.\frac{\partial^2}{\partial\Phi_i\partial\Phi_j}(V_{CW} + \Delta V)\right|_{\phi=v,s=u} = 0,\\
&\Phi_i = {\phi,s}.
}
The five conditions fixes the tree level VEVs and scalar masses to be the physical ones, regardless of the couplings.
As all the renormalization conditions are evaluated at point $\phi = v$ and $s = u$, the only relevant variables here are only $A,B,C,D,E$, which can be uniquely determined, while the other three are totally arbitrary~\footnote{
One may use the following counterterms
\eq{
\Delta V = A s + B s^2 + C s^3 + D s^4 + E \phi^2 + F \phi^2 s + G \phi^2 s^2 + H \phi^4.
}
However, in this parametrization, if we only apply the five conditions, it is not enough to determine certain counterterms. 
One has to use three renormalization conditions on the couplings to determine them uniquely.
Due to the arbitrariness on the renormalization conditions on the couplings, the counterterms could be quite different.  
}.

Although we can solve for these 5 coefficients using the 5 equations, but we found an easy way to do it systematically. 
The trick is to make use of the following function~\cite{Anderson:1991zb}:
\eq{
\label{eq:CWOS}
\tilde{V}_{CW}^{OS} = \frac{1}{64\pi^2}\STr\left[M_{\varphi,m}^4\left(\ln\frac{M_{\varphi,m}^2}{M_{\varphi,phy}^2} - \frac{3}{2}\right) + 2M_{\varphi,phy}^2M_{\varphi,m}^2\right]
}
where the mass matrices $M_{\varphi,m}^2$ are supposed to be diagonal, in the basis of mass eigenstates, and $M_{\varphi,phy}^2$ are $M_{\varphi,m}^2$ evaluated at the vacuum point $\phi=v,\ s=u$. It's easy to verify that the function satisfies all the 5 renormalization conditions that we apply, regardless of the details of the mass matrices\footnote{
By adding higher powers of $M_{\varphi,m}^2-M_{\varphi,phy}^2$ with appropriate coefficients, one can construct functions that satisfy higher-order on-shell conditions. For instance, if we want to have all the Lagrangian couplings to be the on-shell values, we simply add terms up to the 4th power of $M_{\varphi,m}^2-M_{\varphi,phy}^2$ to it:
\eq{
\label{eq:CWOSfull}
\tilde{V}_{CW}^{OS} &= \frac{1}{64\pi^2}\STr\Big[M_{\varphi,m}^4\left(\ln\frac{M_{\varphi,m}^2}{M_{\varphi,phy}^2} - \frac{3}{2}\right) + 2M_{\varphi,phy}^2M_{\varphi,m}^2\\
&\qquad - \frac{1}{2M_{\varphi,m}^2}(M_{\varphi,m}^2-M_{\varphi,phy}^2)^3 + \frac{1}{2M_{\varphi,m}^4}(M_{\varphi,m}^2-M_{\varphi,phy}^2)^4\Big]
}}. 

The problem is that whether it can be achieved from the original Coleman-Weinberg potential through adding counterterms like Eq.~\ref{eq:ct}. The answer is, luckily, yes for theories without mixing particles, but is no for the model we are dealing with, where both scalar sector and top quark sector may have large mixing. 

To see this, let's take the difference
\eq{
\Delta\tilde{V} = \tilde{V}_{CW}^{OS} - V_{CW} = \frac{1}{64\pi^2}\STr\left[M_{\varphi}^4\ln\frac{\mu^2}{M_{\varphi,phy}^2} + 2M_{\varphi,phy}^2M_{\varphi}^2\right].
}
For mixed fields, though the mass matrix elements in gauge eigenbasis are usually polynomials of the scalar fields, which are allowed in the counterterms, in the mass eigenbasis they are typically irrational expressions. Specifically, $\Tr M_{\varphi,m}^2$ and $\Tr M_{\varphi,m}^4$ can still be expressed in terms of the coefficients of the characteristic polynomial of the mass matrix, and hence are polynomial of the scalar fields, but $\Tr DM_{\varphi,m}^4$ and $\Tr DM_{\varphi,m}^2$ are not, where $D$ is a diagonal matrix with non-degenerate eigenvalues. The two terms in the above expressions are exactly in this form.

What we do is to expand $\Delta\tilde{V}$ at the vacuum point like in Eq.~\ref{eq:ct}, and truncate the expression at the order as we like. For instance, we can retain the terms of $A',B',E',F',H'$, and throw away all the other terms, so that the 5 renormalization conditions are still satisfied. The coefficients we get this way are unambiguous, which must be the same as what people get by any other methods. In our calculation, we retained all the 8 terms that are allowed in the counterterms, thus recover the full form of $\Delta V$, and throw away the higher order terms. The coefficients $C',D',G'$, however, are ambiguous, which depend on the additional renormalization conditions that people may add to the scheme.

Suppose that after truncation, we get $\Delta V$ out of $\Delta \tilde{V}$, therefore the final on-shell potential is
\eq{
V_{eff}^{OS} = V_{eff} + \Delta V
}
in which the coefficients are
\eq{
A &= \frac{\partial}{\partial s}\Delta\tilde{V}\Big|_{\phi=v,s=u}\\
B &= \frac{1}{2}\frac{\partial^2}{\partial s^2}\Delta\tilde{V}\Big|_{\phi=v,s=u}\\
E &= \frac{1}{2v}\frac{\partial}{\partial \phi}\Delta\tilde{V}\Big|_{\phi=v,s=u}\\
F &= \frac{1}{2v}\frac{\partial^2}{\partial s\partial \phi}\Delta\tilde{V}\Big|_{\phi=v,s=u}\\
H &= \frac{1}{8v^2}\frac{\partial^2}{\partial \phi^2}\Delta\tilde{V}\Big|_{\phi=v,s=u}\\
}

\subsection{Goldstone Infrared Divergence}

There are two problems for the above potential even before the truncation. First, by definition, the potential is defined at scale $p^2=0$, and the second derivatives don't give pole masses but renormalized masses at scale $\mu=0$. Second, in the Goldstone contribution to the Coleman-Weinberg potential, there is IR divergence from $\ln m_G^2(v,u)$, since the field-dependent mass of the Goldstone boson
\eq{
\label{eq:mG}
m_G^2(\phi,s) = \lambda_{\phi}(\phi^2-v^2) +\frac{\lambda_{s\phi}}{2}(s^2-u^2) + \mu_{s\phi}(s-u) ,
}
is zero at the vacuum point. In this section, we will show that these two effects cancel each other, according to the discussion in \cite{Espinosa:2011ax}.

Let's find out the relation between the zero momentum parameters appearing in the effective potential and the physical observables that we need in an OS scheme. In general, we have the vertex functions
\begin{align}
\Gamma(\phi_1, \phi_2, \cdots, \phi_n;p_i) = \Gamma^X_r + \Gamma_L(p_i) + \Gamma^X_{ct}
\end{align}
where $\Gamma_r$ is the tree-level renormalized coupling, or the inverse propagator in the case of $n=2$. $\Gamma_L$ is the loop contributions, which depends on the external momenta. $\Gamma^X_{ct}$ comes from the counter terms defined in scheme $X$. 

As we are working in the OS scheme, we have
\begin{equation}
\Gamma(\phi_1, \phi_2, \cdots, \phi_n;p_i) = \Gamma_{phy} + \Gamma_L(p_i) + \Gamma^{OS}_{ct}
\end{equation}
while
\begin{equation}
\Gamma_L(OS) + \Gamma^{OS}_{ct} = 0
\end{equation}
where the $OS$ inside the parenthesis indicates the on-shell momenta, instead of on-shell scheme, as $\Gamma_L$ is scheme independent. Thus we have the $p^2=0$ values for the vertex functions:
\begin{equation}
\Gamma(\phi_1, \phi_2, \cdots, \phi_n; p_i^2=0) = \Gamma_{phy} + \Gamma_L(p_i^2=0) - \Gamma_L(OS) \equiv \Gamma_{phy} - \Delta\Gamma_L
\end{equation}
where $\Delta\Gamma_L$ is defined to be the difference between on-shell loop contribution and zero-momenta loop contributions.

In this spirit, the renormalization conditions Eq.~\ref{eq:OS} should be modified as
\eq{
\label{eq:OS_1}
&\left.\frac{\partial}{\partial \Phi_i}(V_{CW}^{\overline{DR}} + \Delta V)\right|_{\phi=v,s=u} = 0\\
&\left.\frac{\partial^2}{\partial\Phi_i\partial\Phi_j}(V_{CW}^{\overline{DR}} + \Delta V)\right|_{\phi=v,s=u} = -\Delta\Sigma\\
}
where $\Sigma$ is the loop contribution to the mass matrix. Tadpole term is not changed because the tadpole loop does not depend on external momentum. For couplings, we have conditions like
\begin{equation}
\label{eq:IR coupling}
\frac{\partial^4V_{\eff}}{\partial s^2 \partial\phi^2} \Big|_{\phi=v, s=u} = \lambda_{s\phi} + \frac{\partial^4V_{CW}}{\partial s^2 \partial\phi^2} \Big|_{\phi=v, s=u} = \lambda_{m,phy} + \frac{\partial^4V_{CW}}{\partial s^2 \partial\phi^2} \Big|_{\phi=v, s=u} - \Delta\Gamma_m .
\end{equation}
Note that, unlike the case for masses, the Coleman-Weinberg potential does have contributions to the couplings (unless we use the more complete form Eq.~\ref{eq:CWOSfull}).

One may notice that the off-diagonal element of $\Sigma$ is not well-defined in OS scheme. In addition, some couplings like $s^3$ also don't have natural on-shell definition. Here we assume that the $\Delta\Gamma_L$'s are not sensitive to the tricky details of how we define the OS renormalization conditions. Here we only focus on the IR divergence from the Goldstone loops, for which we use $m_\chi^2 = m_G^2(v,u)$ as an IR regulator, and choose a convenient but inexact form of the IR-finite part.

Thus we only retain the IR divergent terms, and replace all the parameters for physical quantities we need in an OS scheme. These terms are
\begin{equation}
\label{eq:VIR}
\begin{split}
V_{\textrm{IR}} =& \frac{\Sigma_\phi(0)}{8v^2}(\phi^2-v^2)^2 + \frac{\Sigma_{s\phi}(0)}{2v}(\phi^2-v^2)(s-u) + \frac{\Sigma_s(0)}{2}(s-u)^2 + \frac{\Gamma^{(m)}_L(0)}{4}(\phi^2-v^2)(s-u)^2 \\
& + \frac{\Gamma^{(3)}_L(0)}{3!}(s-u)^3 + \frac{\Gamma^{(4)}_L(0)}{4!}(s-u)^4
\end{split}
\end{equation}
while the $\ln m_\chi^2$ order IR divergences are:
\bea
\Sigma_\phi(0)    &\sim& \frac{3}{2}\times(2\lambda_{\phi}v)^2\times \frac{1}{16\pi^2}\ln \frac{m_\chi^2}{m_H^2}, \\
\Sigma_s(0)       &\sim& \frac{3}{2}\times(\lambda_{s\phi}u+\mu_{s\phi})^2\times \frac{1}{16\pi^2}\ln \frac{m_\chi^2}{m_H^2}, \\
\Sigma_{s\phi}(0) &\sim& \frac{3}{2}\times2\lambda_{\phi}v(\lambda_{s\phi}u+\mu_{s\phi})\times \frac{1}{16\pi^2}\ln \frac{m_\chi^2}{m_H^2},
\eea
\bea
\Gamma^{(m)}_L(0) &\sim& \frac{3}{2}\times2\lambda_{s\phi}\lambda_{\phi}\times \frac{1}{16\pi^2}\ln \frac{m_\chi^2}{m_H^2} + \textrm{power-law IR divergence}, \\
\Gamma^{(3)}_L(0) &\sim& \frac{9}{2}\times\lambda_{s\phi}(\lambda_{s\phi}u+\mu_{s\phi})\times \frac{1}{16\pi^2}\ln \frac{m_\chi^2}{m_H^2} + \textrm{power-law IR divergence}, \\
\Gamma^{(4)}_L(0) &\sim& \frac{9}{2}\times\lambda_{s\phi}^2\times \frac{1}{16\pi^2}\ln \frac{m_\chi^2}{m_H^2} + \textrm{power-law IR divergence},
\eea
where $\ln m_H^2$ is for the IR finite contributions which, as already explained, are not exact, but errors are negligible. The power-law IR divergences in the couplings will be cancelled by the Coleman-Weinberg contributions (such as the last two term in Eq.~\ref{eq:CWOSfull}, which we don't prove explicitly here.

Plugging them into Eq.~\ref{eq:VIR}, together with the knowledge of Eq.~\ref{eq:mG}, we get
\begin{equation}
V_{\textrm{IR}} = \frac{3}{64\pi^2}m_G^4(\phi,s)\ln \frac{m_\chi^2}{m_H^2},
\label{eq:vcwir}
\end{equation}
which neatly cancels the IR divergence of the Coleman-Weinberg term. Thus we only need to replace the Goldstone pole mass in the logarithm in Eq.~\ref{eq:CWOS} with the higgs mass to fix the IR divergence problem. Again, there are corrections and small subtleties from momentum-shift effect, such as the $\Delta\Sigma$ contribution to the counterterms $B',E',H'$, which are computable but we neglected.



\subsection{Field-dependent Masses}

In effective potential, the particles running in the loop are 
the particles in the model
with the following degrees of freedom in the Landau gauge:
\bea 
n_{W}=6, \  n_{Z}=3, \  n_{\pi}=3, \  n_{h} = n_{S} = 1, \  n_{t}=-12, \ n_{T}=-12.
\eea
The field-dependent masses of the top quark, gauge bosons and Goldstone bosons at zero temperature are given by
\bea   
m_{W}^2(\phi) &=& \frac{g^2}{4} \phi^2, \ 
m_{Z}^2(\phi) = \frac{g^2+g'^2}{4} \phi^2,\\  
m_{\pi}^2(\phi, s) &=& \lambda_{\phi} \phi^2 - \mu_\phi^2 + \frac12\lambda_{s\phi} s^2 + \mu_{s\phi} s. 
\label{masses}
\eea
The field-dependent masses of the scalars $h$ and $S$ are obtained as
\bea
	m^2_{h,S}(\phi,s) &=&  \frac{1}{2} \left(m^2_{\phi\phi}(\phi,s) + m^2_{ss}(\phi,s)\right) \mp 
	\frac12\sqrt{  \left(m^2_{\phi\phi}(\phi,s) - m^2_{ss}(\phi,s)\right)^2 +4 m^4_{s\phi }(\phi,s) },
\eea
where the field-dependent quantities 
are
\bea
	m^2_{\phi\phi}(\phi,s) &=& 3 \lambda_\phi \phi^2 - \mu_\phi^2+ \frac{\lambda_{s\phi}}{2} s^2 + \mu_{s\phi} s ,\\
	m^2_{s\phi}(\phi,s) &=& m^2_{\phi s}(\phi,s) =  (\lambda_{s\phi} s + \mu_{s\phi}) \phi,\\
	m^2_{ss}(\phi,s) &=& 3 \lambda_s s^2 + 2 \mu_3 s - \mu_s^2 + \frac{\lambda_{s\phi}}{2}\phi^2.
\eea
The field-dependent masses of the top quark and heavy vector-like top quark $T$ are obtained as
\bea
	m^2_{t,T}(\phi,s) &=&  \frac{1}{2} \left(m^2_{tt}(\phi,s) + m^2_{TT}(\phi,s)\right) \mp 
	\frac12\sqrt{  \left(m^2_{tt}(\phi,s) - m^2_{TT}(\phi,s)\right)^2 +4 m^4_{tT}(\phi,s) },
\eea
where the field-dependent quantities 
are
\bea
	m^2_{tt}(\phi,s) &=& \frac{1}{2}(y_t^2 + y'^2)\phi^2 ,\\
	m^2_{tT}(\phi,s) &=& m^2_{Tt}(\phi,s) = \frac{1}{\sqrt2} y' \phi (y_s s + M),\\
	m^2_{TT}(\phi,s) &=& (y_s s + M)^2.
\eea

The finite-temperature potential needs to be corrected by the thermal field-dependent masses. 
The thermal field-dependent masses is calculated by adding the Debye masses, calculated from the the quadratically divergent bubbles and Daisy resummation. 
This leads to a shift of the bosonic field-dependent masses $m_i^2(\phi,s)$ to the thermal field-dependent masses (Debye masses)
\bea
	m_i^2(\phi,s, T) \equiv m_i^2(\phi,s)+ \Pi_i (\phi,s, T),
\eea 
where $\Pi_i (\phi,s, T)$ is the self-energy of the bosonic field $i$ in the IR limit.
In particular, the longitudinal and transversal polarizations of the gauge bosons have to be taken into account separately: only the longitudinal components get a thermal mass correction and the transversal ones will not.
Since the ring diagrams will only contribute significantly at high-temperature,
only the zero-mode of the Matsubara frequency behave as a massless degree of freedom and generate IR-divergences at high-temperature,
while other modes lead to subdominant contributions.
For the SM bosonic contributions, the gauge boson thermal self energy  is 
\bea
	m_{\rm V}^2(h,s,T)  = m_{\rm V}^2(h,s) + \Pi_{\rm V},
\eea
where
\bea
m_{\rm V}^2 &=&  \phi^2\left(
                 \begin{array}{cccc}
                   \frac{g^2  }{4} & 0 & 0 & 0 \\
                   0 & \frac{g^2  }{4} & 0 & 0 \\
                   0 & 0 & \frac{g^2  }{4} & -\frac{g g'  }{4} \\
                   0 & 0 & -\frac{g g'  }{4} & \frac{g'^2  }{4} \\
                 \end{array}
               \right)
\label{mgb}
\eea
and
\bea
\Pi_{\rm V} &=& {\rm diag} \left[\frac{11}{6} g^2 T^2, \; \frac{11}{6} g^2 T^2,
 \; \frac{11}{6} g^2 T^2, \; \frac{11}{6} g'^2 T^2 \right]
\eea
The Goldstone boson is
\bea
\Pi_{\pi} &=&
\left(\frac{3}{16} g^2 + \frac{1}{16} g'^2 + \frac{\lambda_{_H}}{2}  + \frac{y_t}{4} + \frac{\lambda_{_{HS}}}{3} \right)  T^2
\label{thermal masses}
\eea

For the new scalar bosons, 
\bea
	{\cal M}^2(\phi,s, T) = {\cal M}^2(\phi,s) +  \begin{pmatrix}
	c_{\phi} & 0 \\
	0 & c_s
	\end{pmatrix} T^2, 
\eea
where
\bea
	c_\phi &=& \frac{\lambda_\phi}{2} + \frac{\lambda_{s\phi}}{24} + \frac{3g^2 + g^{'2}}{16} 
	+ \frac{y_t^2}{4}, \\
	c_s &=& \frac{\lambda_s}{4} + \frac{\lambda_{s\phi}}{6}. 
\eea
After diagonalization, we obtain the Deybe squared mass $m_{h,S}(\phi,s,T)$.
For the SM bosonic particles, we obtain
\bea
	\Pi_{W_L} &=& \frac{11}{6} g^2 T^2, \quad \Pi_{W_T} = 0, \\
	\Pi_{Z_L} &=& \frac{11}{6} (g^2 + g^{'2}) T^2,  \quad \Pi_{Z_T} = 0, \\
	\Pi_{\pi} &=& c_\phi T^2.
\eea


\begin{thebibliography}{99}
	


\bibitem{Ade:2015xua} 
  P.~A.~R.~Ade {\it et al.}  [Planck Collaboration],
  arXiv:1502.01589 [astro-ph.CO].


\bibitem{Sakharov:1967dj} 
  A.~D.~Sakharov,
  Pisma Zh.\ Eksp.\ Teor.\ Fiz.\  {\bf 5}, 32 (1967)
  [JETP Lett.\  {\bf 5}, 24 (1967)]
  [Sov.\ Phys.\ Usp.\  {\bf 34}, 392 (1991)]
  [Usp.\ Fiz.\ Nauk {\bf 161}, 61 (1991)].



\bibitem{Kuzmin:1985mm} 
  V.~A.~Kuzmin, V.~A.~Rubakov and M.~E.~Shaposhnikov,
  Phys.\ Lett.\ B {\bf 155}, 36 (1985).

\bibitem{Shaposhnikov:1986jp} 
  M.~E.~Shaposhnikov,
  JETP Lett.\  {\bf 44}, 465 (1986)
  [Pisma Zh.\ Eksp.\ Teor.\ Fiz.\  {\bf 44}, 364 (1986)].

\bibitem{Shaposhnikov:1987tw} 
  M.~E.~Shaposhnikov,
  Nucl.\ Phys.\ B {\bf 287}, 757 (1987).
   
\bibitem{Cohen:1993nk} 
  A.~G.~Cohen, D.~B.~Kaplan and A.~E.~Nelson,
  Ann.\ Rev.\ Nucl.\ Part.\ Sci.\  {\bf 43}, 27 (1993)
  [hep-ph/9302210].
  
\bibitem{Trodden:1998ym} 
  M.~Trodden,
  Rev.\ Mod.\ Phys.\  {\bf 71}, 1463 (1999)
  [hep-ph/9803479].
  A.~Riotto,
  hep-ph/9807454.
  A.~Riotto and M.~Trodden,
  Ann.\ Rev.\ Nucl.\ Part.\ Sci.\  {\bf 49}, 35 (1999)
  [hep-ph/9901362].
  
\bibitem{Quiros:1999jp} 
  M.~Quiros,
  hep-ph/9901312.
  
  
\bibitem{Bernreuther:2002uj} 
  W.~Bernreuther,
  Lect.\ Notes Phys.\  {\bf 591}, 237 (2002)
  [hep-ph/0205279].
  
  
\bibitem{Cline:2006ts} 
  J.~M.~Cline,
  hep-ph/0609145.
    
\bibitem{Morrissey:2012db} 
  D.~E.~Morrissey and M.~J.~Ramsey-Musolf,
  New J.\ Phys.\  {\bf 14}, 125003 (2012)
  [arXiv:1206.2942 [hep-ph]].
  

\bibitem{Bochkarev:1990gb} 
  A.~I.~Bochkarev, S.~V.~Kuzmin and M.~E.~Shaposhnikov,
  Phys.\ Rev.\ D {\bf 43}, 369 (1991).
  
  
\bibitem{Bochkarev:1987wf} 
  A.~I.~Bochkarev and M.~E.~Shaposhnikov,
  Mod.\ Phys.\ Lett.\ A {\bf 2}, 417 (1987).
  
  




\bibitem{Choi:1993cv} 
  J.~Choi and R.~R.~Volkas,
  Phys.\ Lett.\ B {\bf 317}, 385 (1993)
  [hep-ph/9308234].
  
  

\bibitem{Cline:1996mga} 
  J.~M.~Cline and P.~A.~Lemieux,
  Phys.\ Rev.\ D {\bf 55}, 3873 (1997)
  [hep-ph/9609240].

\bibitem{Ham:2004cf} 
  S.~W.~Ham, Y.~S.~Jeong and S.~K.~Oh,
  J.\ Phys.\ G {\bf 31}, 857 (2005)
  [hep-ph/0411352].
  
\bibitem{Ahriche:2007jp} 
  A.~Ahriche,
  Phys.\ Rev.\ D {\bf 75}, 083522 (2007)
  [hep-ph/0701192].

  

\bibitem{Profumo:2007wc} 
  S.~Profumo, M.~J.~Ramsey-Musolf and G.~Shaughnessy,
  JHEP {\bf 0708}, 010 (2007)
  [arXiv:0705.2425 [hep-ph]].

  
\bibitem{Espinosa:2011ax} 
  J.~R.~Espinosa, T.~Konstandin and F.~Riva,
  Nucl.\ Phys.\ B {\bf 854}, 592 (2012)
  [arXiv:1107.5441 [hep-ph]].

  
\bibitem{Li:2014wia} 
  T.~Li and Y.~F.~Zhou,
  JHEP {\bf 1407}, 006 (2014)
  [arXiv:1402.3087 [hep-ph]].

\bibitem{Fuyuto:2014yia} 
  K.~Fuyuto and E.~Senaha,
  Phys.\ Rev.\ D {\bf 90}, 015015 (2014)
  [arXiv:1406.0433 [hep-ph]].
  
\bibitem{Profumo:2014opa} 
  S.~Profumo, M.~J.~Ramsey-Musolf, C.~L.~Wainwright and P.~Winslow,
  arXiv:1407.5342 [hep-ph].



\bibitem{Xiao:2014kba} 
  M.~L.~Xiao and J.~H.~Yu,
  Phys.\ Rev.\ D {\bf 90}, no. 1, 014007 (2014)
  [Phys.\ Rev.\ D {\bf 90}, no. 1, 019901 (2014)]
  [arXiv:1404.0681 [hep-ph]].


\bibitem{ATLAS13}
  The ATLAS Collaboration, ``Search for resonances decaying to photon pairs in 3.2 fb$^{-1}$ of pp collisions at $\sqrt{s}=13$ TeV with the ATLAS detector'', Tech. Rep. ATLAS-CONF-2015-081, CERN, Geneva, Dec, 2015.

\bibitem{CMS13}
  The CMS Collaboration, ``Search for new physics in high mass diphoton events in proton-proton collisions at 13 TeV'', Tech. Rep. CMS-PAS-EXO-15-004, CERN, Geneva, 2015.  


\bibitem{Franceschini:2015kwy} 
  R.~Franceschini {\it et al.},
  arXiv:1512.04933 [hep-ph].
    D.~Buttazzo, A.~Greljo and D.~Marzocca,
    arXiv:1512.04929 [hep-ph].
  W.~Chao, R.~Huo and J.~H.~Yu,
  arXiv:1512.05738 [hep-ph].
    A.~Falkowski, O.~Slone and T.~Volansky,
    arXiv:1512.05777 [hep-ph].
    W.~Altmannshofer, J.~Galloway, S.~Gori, A.~L.~Kagan, A.~Martin and J.~Zupan,
    arXiv:1512.07616 [hep-ph].

\bibitem{ArkaniHamed:2002qy} 
  N.~Arkani-Hamed, A.~G.~Cohen, E.~Katz and A.~E.~Nelson,
  JHEP {\bf 0207}, 034 (2002)
  [hep-ph/0206021].
  
  
\bibitem{Kaplan:1983sm} 
  D.~B.~Kaplan, H.~Georgi and S.~Dimopoulos,
  Phys.\ Lett.\ B {\bf 136}, 187 (1984).
  K.~Agashe, R.~Contino and A.~Pomarol,
  Nucl.\ Phys.\ B {\bf 719}, 165 (2005)
  [hep-ph/0412089].
 
\bibitem{Randall:1999ee} 
  L.~Randall and R.~Sundrum,
  Phys.\ Rev.\ Lett.\  {\bf 83}, 3370 (1999)
  [hep-ph/9905221].
  T.~Appelquist, H.~C.~Cheng and B.~A.~Dobrescu,
  Phys.\ Rev.\ D {\bf 64}, 035002 (2001)
  [hep-ph/0012100].
  
\bibitem{Jarlskog:1985ht} 
  C.~Jarlskog,
  Phys.\ Rev.\ Lett.\  {\bf 55}, 1039 (1985).
  
\bibitem{Bernabeu:1986fc} 
  J.~Bernabeu, G.~C.~Branco and M.~Gronau,
  Phys.\ Lett.\ B {\bf 169}, 243 (1986).
  
	
\bibitem{Hou:2008xd} 
  W.~-S.~Hou,
  Chin.\ J.\ Phys.\  {\bf 47}, 134 (2009)
  [arXiv:0803.1234 [hep-ph]].
%
  W.~S.~Hou, Y.~Y.~Mao and C.~H.~Shen,
  Phys.\ Rev.\ D {\bf 82}, 036005 (2010)
  [arXiv:1003.4361 [hep-ph]].
 
\bibitem{delAguila:1997vn} 
  F.~del Aguila, J.~A.~Aguilar-Saavedra and G.~C.~Branco,
  Nucl.\ Phys.\ B {\bf 510}, 39 (1998)
  [hep-ph/9703410].
  
  
\bibitem{Branco:1998yk} 
  G.~C.~Branco, D.~Delepine, D.~Emmanuel-Costa and F.~R.~Gonzalez,
  Phys.\ Lett.\ B {\bf 442}, 229 (1998)
  [hep-ph/9805302].



\bibitem{Djouadi:2012ae} 
  A.~Djouadi and A.~Lenz,
  Phys.\ Lett.\ B {\bf 715}, 310 (2012)
  [arXiv:1204.1252 [hep-ph]].
  
\bibitem{Eberhardt:2012gv} 
  O.~Eberhardt, G.~Herbert, H.~Lacker, A.~Lenz, A.~Menzel, U.~Nierste and M.~Wiebusch,
  Phys.\ Rev.\ Lett.\  {\bf 109}, 241802 (2012)
  [arXiv:1209.1101 [hep-ph]].
  
\bibitem{Botella:2012ju} 
  F.~J.~Botella, G.~C.~Branco and M.~Nebot,
  JHEP {\bf 1212}, 040 (2012)
  [arXiv:1207.4440 [hep-ph]].
  
\bibitem{Alok:2015iha} 
  A.~K.~Alok, S.~Banerjee, D.~Kumar, S.~U.~Sankar and D.~London,
  arXiv:1504.00517 [hep-ph].

\bibitem{Baker:2006ts} 
  C.~A.~Baker, D.~D.~Doyle, P.~Geltenbort, K.~Green, M.~G.~D.~van der Grinten, P.~G.~Harris, P.~Iaydjiev and S.~N.~Ivanov {\it et al.},
  Phys.\ Rev.\ Lett.\  {\bf 97}, 131801 (2006)
  [hep-ex/0602020].

\bibitem{Liao:2000re} 
  Y.~Liao and X.~Li,
  Phys.\ Lett.\ B {\bf 503}, 301 (2001)
  [hep-ph/0005063].






\bibitem{Patel:2011th} 
  H.~H.~Patel and M.~J.~Ramsey-Musolf,
  JHEP {\bf 1107}, 029 (2011)
  [arXiv:1101.4665 [hep-ph]].



\bibitem{Parwani:1991gq} 
  R.~R.~Parwani,
  Phys.\ Rev.\ D {\bf 45}, 4695 (1992)
  [Phys.\ Rev.\ D {\bf 48}, 5965 (1993)]
  [hep-ph/9204216].

\bibitem{Arnold:1992rz} 
  P.~B.~Arnold and O.~Espinosa,
  Phys.\ Rev.\ D {\bf 47}, 3546 (1993)
  [Phys.\ Rev.\ D {\bf 50}, 6662 (1994)]
  [hep-ph/9212235].


\bibitem{Sanchez:2006tt} 
  A.~Sanchez, A.~Ayala and G.~Piccinelli,
  Phys.\ Rev.\ D {\bf 75}, 043004 (2007)
  [hep-th/0611337].



\bibitem{Anderson:1991zb} 
  G.~W.~Anderson and L.~J.~Hall,
  Phys.\ Rev.\ D {\bf 45}, 2685 (1992).
  
    
\bibitem{Delaunay:2007wb} 
  C.~Delaunay, C.~Grojean and J.~D.~Wells,
  JHEP {\bf 0804}, 029 (2008)
  [arXiv:0711.2511 [hep-ph]].
  
  
\bibitem{Barger:2011vm} 
  V.~Barger, D.~J.~H.~Chung, A.~J.~Long and L.~T.~Wang,
  Phys.\ Lett.\ B {\bf 710}, 1 (2012)
  [arXiv:1112.5460 [hep-ph]].
  
  
\bibitem{Kozaczuk:2015owa} 
  J.~Kozaczuk,
  arXiv:1506.04741 [hep-ph].
 
  

\bibitem{James:1975dr} 
  F.~James and M.~Roos,
  Comput.\ Phys.\ Commun.\  {\bf 10}, 343 (1975).
  
  

  
  

\bibitem{Chatrchyan:2013uxa} 
  S.~Chatrchyan {\it et al.}  [CMS Collaboration],
  Phys.\ Lett.\ B {\bf 729}, 149 (2014)
  [arXiv:1311.7667 [hep-ex]].
  
  
\bibitem{Chatrchyan:2013yoa} 
  S.~Chatrchyan {\it et al.}  [CMS Collaboration],
  Eur.\ Phys.\ J.\ C {\bf 73}, 2469 (2013)
  [arXiv:1304.0213 [hep-ex]].

  

\end{thebibliography}
\end{document}